\documentclass[fleqn,usenatbib]{mnras}
\usepackage{newtxtext,newtxmath}
\usepackage[T1]{fontenc}

\DeclareRobustCommand{\VAN}[3]{#2}
\let\VANthebibliography\thebibliography
\def\thebibliography{\DeclareRobustCommand{\VAN}[3]{##3}\VANthebibliography}

\usepackage{graphicx}	% Including figure files
\usepackage{amsmath}	% Advanced maths commands
\title[Stable accretion: EX Lupi and TW Hya]{Stable accretion in young stars: The cases of EX Lupi and TW Hya}

\author[Sicilia-Aguilar et al.]{
A. Sicilia-Aguilar$^{1}$\thanks{E-mail: asiciliaaguilar@dundee.ac.uk},
J. Campbell-White$^{2,1}$,
V. Roccatagliata$^{4,5,6}$,
J. Desira$^{1}$, 
S.G. Gregory$^{1}$, A. Scholz$^{3}$,
\newauthor  
M. Fang$^{7}$, F. Cruz-Saenz de Miera$^{8,9}$, \'{A}. K\'{o}sp\'{a}l$^{8,9,10,11}$, S. Matsumura$^{1}$, and P. \'{A}brah\'{a}m$^{8,9,10}$
\\
$^{1}$SUPA, School of Science and Engineering, University of Dundee, Nethergate, DD1 4HN, Dundee, UK\\
$^{2}$European Southern Observatory, Karl-Schwarzschild-Strasse 2, 85748 Garching bei M\"unchen, Germany\\
$^{3}$SUPA, School of Physics and Astronomy, University of St Andrews, North Haugh, St Andrews KY16 9SS, Scotland, UK\\
$^{4}$INAF-Osservatorio Astrofisico di Arcetri, Largo E. Fermi 5, 50125 Firenze, Italy\\
$^{5}$Department of Physics ``E. Fermi'', University of Pisa, Largo Bruno Pontecorvo 3, 56127 Pisa, Italy\\
$^{6}$INFN, Sezione di Pisa, Largo Bruno Pontecorvo 3, 56127 Pisa, Italy\\
$^{7}$Purple Mountain Observatory, Chinese Academy of Sciences, 10 Yuanhua Road, Nanjing 210023, China\\
$^{8}$Konkoly Observatory, HUN-REN Research Centre for Astronomy and Earth Sciences, Konkoly-Thege Mikl\'os \'ut 15-17, 1121 Budapest, Hungary\\
$^{9}$CSFK, MTA Centre of Excellence, Konkoly-Thege Mikl\'os \'ut 15-17, 1121 Budapest, Hungary\\
$^{10}$ELTE E\"otv\"os Lor\'and University, Institute of Physics, P\'azm\'any P\'eter s\'et\'any 1/A, 1117 Budapest, Hungary\\
$^{11}$Max Planck Institute for Astronomy, K\"onigstuhl 17, 69117 Heidelberg, Germany\\
}

\date{Accepted 2023 October 03. Received 2023 October 03; in original form 2023 June 16}

\pubyear{2023}

\begin{document}
\label{firstpage}
\pagerange{\pageref{firstpage}--\pageref{lastpage}}
\maketitle

\begin{abstract}
We examine the long-term spectroscopic and photometric variability of EX~Lupi and TW~Hya, studying the presence of stable accretion and the role it plays in the observed variability. Analysing the velocity modulations of the emission lines with STAR-MELT, we obtain information on the structure of the accretion columns and the disk-star connection. The emission line radial velocities reveal that TW Hya, like EX Lupi, has a remarkably stable or slow-varying accretion column footprint, locked to the star for several years. The line-emitting regions are non-polar for both EX Lupi and TW Hya, and species with different energies differ in position. In contrast, the continuum emission as observed in the photometry is very variable and can be modelled by hot spot(s) that change over time in phase, shape, temperature, size, and location with respect to the emission line region. The continuum emission region may not be limited to the stellar surface, especially during episodes of high accretion.  The broad line emission observed in EX Lupi during episodes of increased accretion reveals a further structure, which can be fitted by non-axisymmetric disk in Keplerian rotation inwards of the corotation radius. Since the radial velocity modulation due to accretion footprints is so stable, we used it to search for further velocity modulations. While no residual modulation (other than caused by stellar rotation) is found in these objects, a similar analysis could help to identify young planets/companions. Therefore, determining whether stable accretion footprints are common among young stars is a key to detect young planets.
\end{abstract}


%%%%%%%%%%%%%%%%%%%%%%%%%%%%%%%%%%%%%%%%%%%%%%%%%%

%%%%%%%%%%%%%%%%% BODY OF PAPER %%%%%%%%%%%%%%%%%%

\section{Introduction}

Young, pre-main sequence stars have spectra rich in emission lines. These lines have been, since the beginning, one of the defining characteristics of T Tauri stars \citep[TTS;][]{joy45}. They are highly variable, especially in accreting stars (classical T Tauri stars, CTTS). CTTS lines include forbidden and permitted transitions and typically display broad and narrow components (BC and NC) that originate in different structures around the object, including accretion columns and their hot spots footprints, and winds \citep{hamann_emission-line_1992,hamann_emission-line_1994,beristain_permitted_1998,dodin12}. The variability in the line profiles reveals the structure and distribution of the accretion columns around the star \citep{alencar_spectral_2001}, their stability over time \citep{kurosawa_spectral_2013}, and the spatial scales of the disk wind \citep{fang_gw_2014}. 

The last two decades have seen substantial work concentrated on modelling the Hydrogen and Helium emission lines to constrain accretion rates \citep[e.g.][]{muzerolle_emission-line_1998,lima10,alencar_accretion_2012,alcala_x-shooter_2014,alcala17}, although the metallic lines also provide information about the processes taking place on the surface of the star \citep{beristain_iron_2000,sicilia-aguilar_accretion_2015}, including chromospheric activity \citep{ingleby11}. In particular, the typically narrow, metallic emission lines are very useful to track the location of the footprints of accretion columns and their time variability in CTTS \citep{sicilia-aguilar_accretion_2015,campbellwhite21}. 
These metallic lines tend to be less saturated and not so much affected by self-absorption and/or extended emission from the accretion columns themselves or from winds. 
Different metallic lines often originate in regions with different temperatures and densities, thus being very useful to study the post-shock accretion region and to track the accreting material.
The study of the NC has revealed that, at least in some objects, the line-emitting region is surprisingly stable over several years \citep{sicilia-aguilar_accretion_2015,campbellwhite21}. This stability of accretion structures can explain the difficulties detecting periodic signals from companions or exoplanets, since radial velocity modulations due to accretion can pervade the data over decades \citep{kospal_radial_2014}.

The main limitation to use the NC metallic lines velocities is that the post-shock region is very close to the stellar surface \citep{dodin12} and, if located at high latitudes, the modulation caused by the rotation of the star spans only a fraction of the projected rotational velocity, v$sini$ \citep{gahm_face_2013,sicilia-aguilar_accretion_2015,mcginnis_magnetic_2020}. We thus need to study many high-resolution spectra over time and extract with high accuracy as many lines as possible. This is now feasible thanks to the STAR-MELT code \citep{campbellwhite21}, which enables a quick extraction and analysis of the emission lines. 

Here, we present the STAR-MELT analysis of EX Lupi and TW Hya, two well-known young stars for which there are large numbers of high-resolution spectra available. Both objects have unmistakable signatures of different types of variability, ranging from year-long accretion bursts \citep[EX Lupi;][]{herbig07} to hours-timescale events \citep[TW Hya;][]{dupree_tw_2012}. They thus represent two very different cases among young stars, despite their similar spectral type and mass.
We take a first step expanding the study of the stability of accretion columns, combining the emission line analysis with ground-based and TESS photometry to improve our understanding of accretion around very different systems. We also examine the causes behind the photometric variability, and use the spectroscopic data to explore further periodic or quasi-periodic signals over longer timescales. 
The NC of EX Lupi have been already thoroughly examined \citep{sicilia-aguilar_accretion_2015,campbellwhite21}, so we concentrate on the BC and the photometry. TW Hya is one of the most observed young stars, and thus offers a great opportunity to test the capabilities of STAR-MELT and the feasibility of using line velocities to find the rotational period in objects with irregular photometry. Section \ref{obs} discusses the available observations. Section \ref{analysis} presents the analysis of both the spectroscopy and photometry data. The results are discussed in Section \ref{discussion} and summarised in Section \ref{conclu}.

\begin{figure*}
    \centering
    \includegraphics[width=15cm]{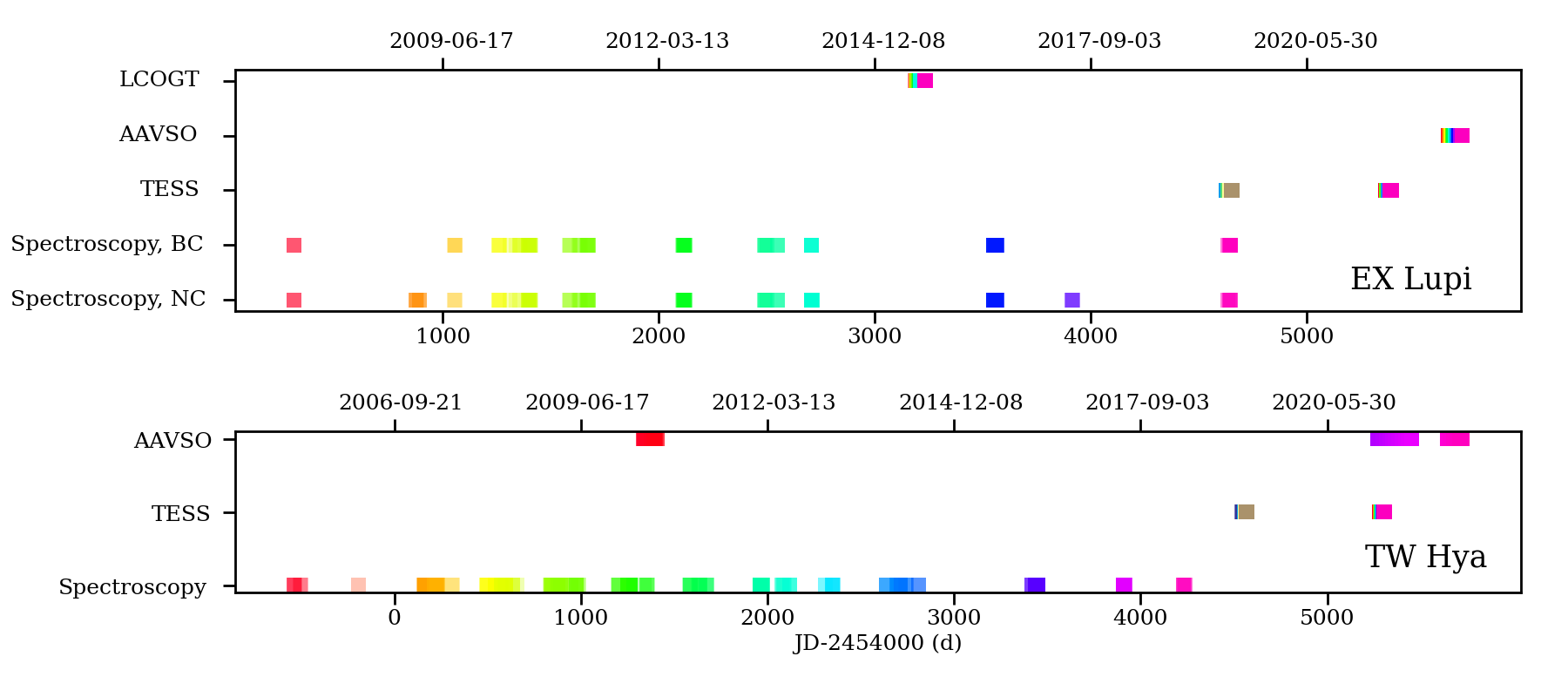}
    \caption{ Timeline of the observations of EX Lupi (top) and TW Hya (bottom), showing the date-dependent colour-code used in the rest of figures.}
    \label{timeline-fig}
\end{figure*}

\section{Presenting the data and observations}\label{obs}

In this section, we briefly describe the sources and available data. Figure \ref{timeline-fig} presents the data available and temporal coverage.

\subsection{A brief summary of source properties \label{sourceprops}}
 
Since some source properties are important for the analysis, and some of them are further revised in our work, we summarise them here. 

EX Lupi is the prototype of the EXor class, low-mass stars with recurrent accretion outburst with timescales between months to years \citep{herbig77,herbig01,herbig08,audard14,fischer22}. As such, it has been typically studied during outburst. Although EX Lupi can increase in brightness by several magnitudes during strong outbursts \citep{herbig77,jones08}, it also shows milder accretion variations on a continuous basis \citep{lehmann95} and bursts \citep{cruz23}, which alter the number, strength, and width of the many emission lines in its spectrum \citep{sicilia-aguilar_accretion_2015}.
The Gaia DR3 parallax for EX Lupi is 6.463$\pm$0.015 mas \citep[RUWE=1.115;][]{gaia16,gaia21}, giving distance of 154.7$\pm$0.4 pc.
EX Lupi has a 7.417d rotation period, first determined via the radial velocities of its emission lines \citep{sicilia-aguilar_accretion_2015} and later confirmed using TESS data from sector 12 \citep{campbellwhite21}. This rotation period, together with the wealth of emission lines and line-dependent veiling, is responsible for the radial velocity modulation of its photospheric spectrum \citep{kospal_radial_2014,sicilia-aguilar_accretion_2015}. According to our previous NC work \citep{sicilia-aguilar_accretion_2015} and to inner disk modeling \citep{sipos09}, the system has a low-to-intermediate inclination ($\sim$ 20-40$^{\circ}$), also confirmed for the outer disk with ALMA \citep[32$^{\circ}$;][]{white20}. The matter content in the inner disk is known to change as result of the outbursts, and the inner disk structures are non-axisymmetric \citep{kospal11,goto11,sicilia-aguilar_optical_2012,banzatti15}.

 TW Hya is one of the best studied CTTS thanks to its privileged distance, 60.14$\pm$0.05pc \citep[Gaia DR3, RUWE=1.175, parallax 16.629$\pm$0.015 mas;][]{gaia16,gaia21}.
The system is viewed at low inclination \citep{qi04,huang18}. If the inclination of the star is similar to the inclination of the inner disk, which is not too different from the outer disk inclination, the value would be 4.3$\pm$1.0$^{\circ}$ \citep{pontoppidan08}, although the stellar inclination may be somewhat higher \citep[$\sim$15$^{\circ}$][]{donati11} and there is evidence of misaligned structures \citep{debes23}. Since the radial velocities of spectral lines at the stellar surface are very small, TW Hya also tests the capabilities of STAR-MELT.

TW Hya has a rotational period of 3.56~d \citep{huelamo08}, albeit not as clear as the one of EX Lupi. \citet{huelamo08} report photometric variations with periods of the order of 6.1 and 6.5~d, while \citet{siwak14,siwak18} detected oscillations with 4.18~d and 3.7-3.8d periods using MOST on two different epochs, \citet{rucinski08} found considerable noise and a 3.7~d peak, and \citet{dupree_tw_2012} retrieved a photometric period of 4.774~d. The difficulty in determining the rotational period via photometry arises from the very irregular, bursting lightcurve. At very low inclinations, significant changes in the visibility of photometric spots are rare, but we also note that the day-to-day variability in TW Hya is nearly as large as that of EX Lupi in quiescence, albeit not periodic. TW Hya is a relatively fast rotator and has a non-negligible projected rotational velocity \citep[$vsini$=6.2$\pm$1.3 km/s or 7$\pm$3 km/s,][respectively]{weise10,venuti19}. The accretion rate is low, $\sim$3 $10^{-9}$ M$_\odot$/yr \citep{weise10,venuti19}, as expected for a relatively evolved disk in a $\sim$8 Myr old association \citep{donaldson16}.

\subsection{Spectroscopy }\label{sec:spec}
The spectroscopy data were obtained from public programs via the European Southern Observatory (ESO) Science Portal\footnote{http://archive.eso.org/scienceportal/home} and the Canada-France-Hawaii Telescope archive\footnote{https://www.cadc-ccda.hia-iha.nrc-cnrc.gc.ca/en/search/}, see Table \ref{tab:Spectra} for a summary of observations. The data comprise observations from the high-resolutions spectroscopes HARPS \citep[High Accuracy Radial-Velocity Planet Searcher; R$\sim$115,000, wavelength coverage 3780-6710~\AA;][]{mayor03}, FEROS \citep[Fiber-fed Extended Range Optical Spectrograph, R$\sim$48,000, wavelength coverage 3500-9200~\AA\ with some gaps;][]{kaufer99}, and ESPaDOnS \citep[Echelle Spectro Polarimetric Device for the Observation of Stars;  intensity spectra with R$\sim$65,000, wavelength coverage 3700-10500~\AA\ with some gaps;][]{donati97,moutou15}. They were reduced by the corresponding instrument pipelines, which are well suited for our purposes \citep{campbellwhite21}. For FEROS, the barycentric correction was further revised following \citet{mueller13}.

\begin{figure*}
    \centering
    \includegraphics[width=14cm]{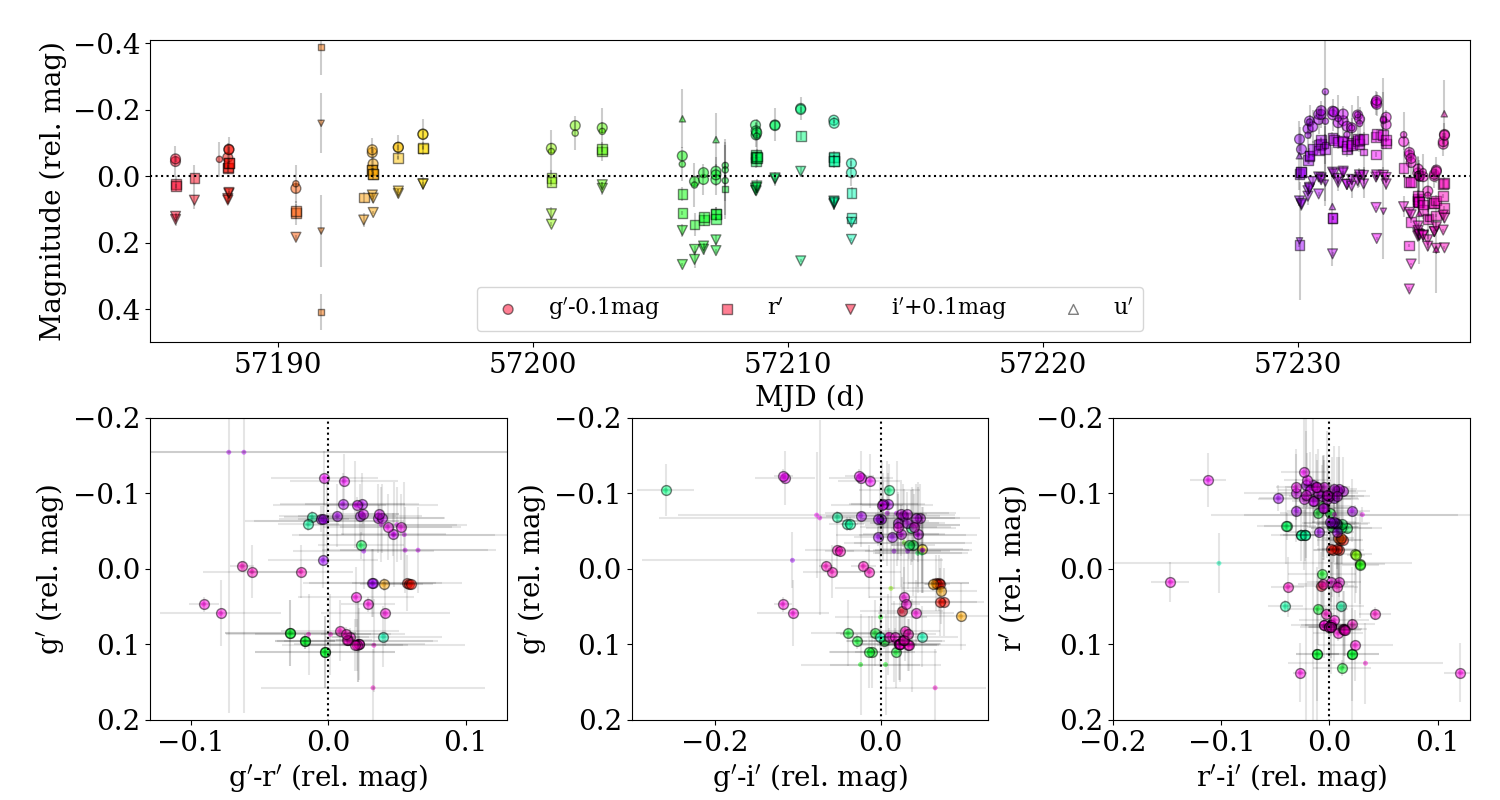}
    \caption{Lightcurve (top) and colour variability (bottom) for EX Lupi as observed with the LCOGT. All magnitudes are shown relative to the mean value, and an extra offset (-0.1 mag for g', +0.1 mag for i') has been added in the top panel for better visualisation.  The colours of the symbols are assigned depending on date (see Fig. \ref{timeline-fig}). Small symbols are used to denote measurements that have a large errors and are thus less accurate than the rest.}
    \label{lcogt-curve}
\end{figure*}

TW Hya was observed using  FEROS (183 spectra), HARPS (34 spectra), and ESPaDOnS (213 spectra), taken between 2005 and 2018. 
EX Lupi was observed between 2007 and 2019. Here, we concentrated on the BC, which are seen in the strongest lines during periods of increased accretion \citep[see][]{sicilia-aguilar_accretion_2015}. The spectra displaying significant BC were selected after visualization with STAR-MELT. This resulted in 62 observations for EX Lupi, taken between 2007 and 2014 \citep[excluding the 2008 outburst data, which was analysed by][]{sicilia-aguilar_optical_2012} where the BC were strong enough to be fitted and extracted. Not all the lines are visible in all the spectra, and the BC of lines such as Ca~II and Fe~II are mostly visible during periods of increased accretion.

\subsection{LCOGT photometry}

We obtained photometric data for EX Lupi with the automated 1m telescopes from Las Cumbres Observatory Global Telescope Network \citep[LCOGT;][]{brown13}, between June and August 2015, a phase or relative quiescence. Observations were taken at variable cadences between hours to days, with a higher cadence during the last 5 days. 
Each set of observations consisted of 10s and 100s exposures (for g',r',i') and a 300s exposure for the u' band. LCOGT provides the basic data calibration, including bias, flat field and an astrometry solution. We performed aperture photometry using Python $photutils$. The automated astrometry solution had numerous failures throughout the dataset,  so we used $astropy$ \citep{astropy_collaboration_astropy_2013, astropy_collaboration_astropy_2018} and $astroquery$  task $AstrometryNet$ to refine the coordinate solution \citep{lang10}. A relative calibration was obtained using stars with similar magnitudes within the field of view \citep[following][]{sicilia13,sicilia-aguilar_2014-2017_2017}. The calibration failed when it had less than 5 comparison stars or when the relation was non-linear. A large number of u' band images failed due to their few sources. Excluding the failed data, this resulted in 8 photometry points for u', 107 data points each for g' and r', and 106 for i'. The lightcurve is shown in Fig.  \ref{lcogt-curve} and the relative photometry is listed in Table \ref{tab:LCOGTphoto}. 

To study the colour variations, measurements from different bands were considered simultaneous if taken within 30~min. Given the 7.417d rotational period \citep{sicilia-aguilar_accretion_2015} and what we see in TESS data, the uncertainties arising from this time lapse are minimal. When more than one measurement was available within 30~min, all were used to derive the colour, so the spread gives information on the uncertainty. We also removed points that deviate by more than 1.5 sigma from those taken within the same 2h (a range in which the lightcurve does not show significant variations).

\subsection{AAVSO photometry}

\begin{figure*}
    \centering
    \begin{tabular}{c}
    {\bf TW Hya}\\
    \includegraphics[width=13.5cm]{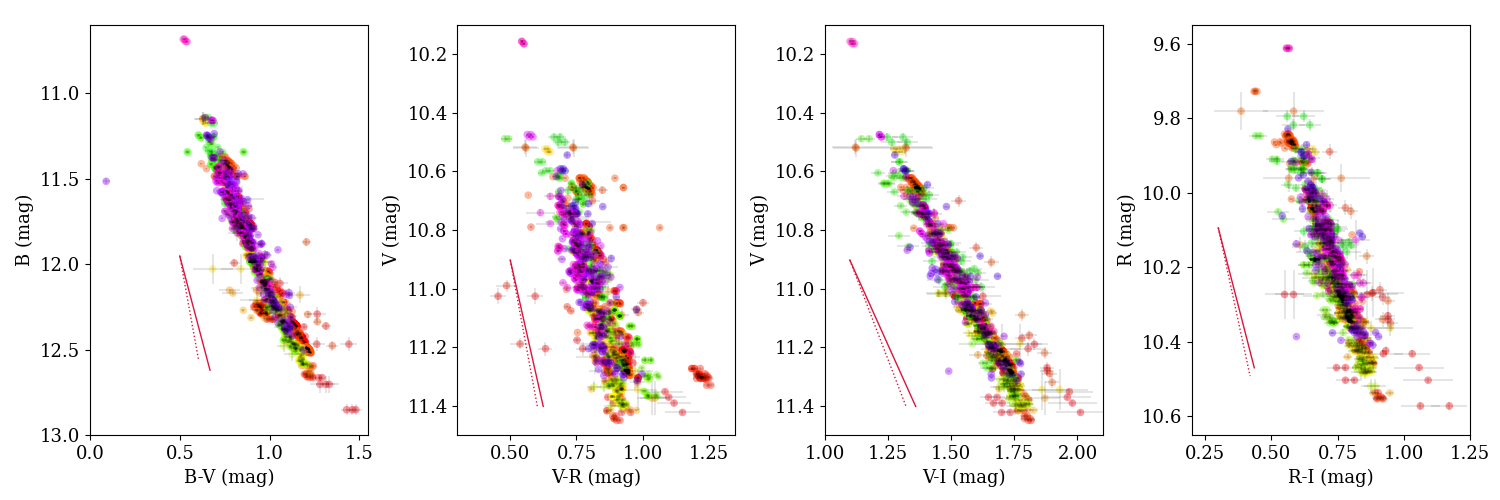} \\
    {\bf EX Lupi}\\
     \includegraphics[width=13.5cm]{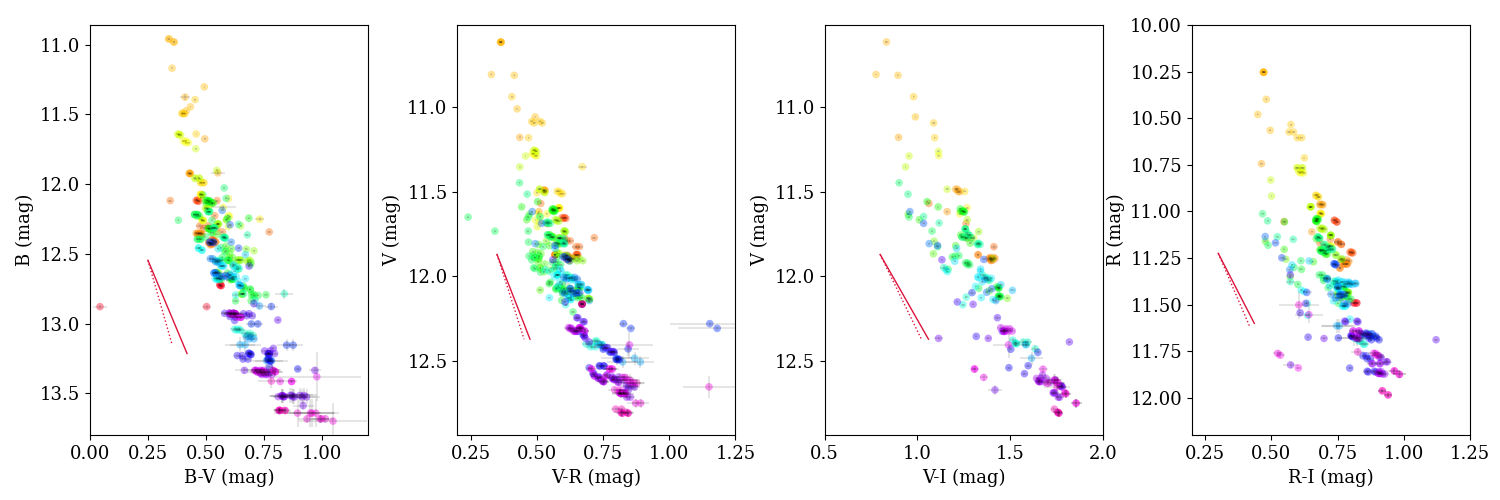} \\
    \end{tabular}
    \caption{Colour-magnitude diagrams showing the variability for TW Hya (top) and EX Lupi (botton) in the AAVSO data. 
    The data points are coloured from red to purple according to the epoch when they were observed  (see Fig. \ref{timeline-fig}), so points that are close in time have similar colours. Extinction vectors for A$_V$=0.5 mag and R$_V$=3.1 (red bold line) and
    R$_V$=5 (dotted red line) are also displayed for comparison, following \citet{cardelli89}. For EX Lupi, these data correspond to the evolution out of a minor outburst or burst in 2022, so that it is substantially different from what we observe with the LCOGT during quiescence. }
    \label{twhyaexlupi-colors}
\end{figure*}

To offer a similar multi-band photometry view for TW Hya, we used data from the American Association of Variable Stars Observers \citep[AAVSO;][]{kafka21,kloppenborg22}, available via the AAVSO database\footnote{See https://www.aavso.org}. The data we use comprised verified observations in Johnsons filters B,V,R and I, contributed by different observers, and obtained between JD 2459269 and JD 2459724, comprising 847 (B), 949 (V), 777 (R) and 686 (I) datapoints. Many observations are nearly-simultaneous (taken within less than 6 min), and were used to explore colour variations. Older data and other filters had worse sampling and higher uncertainties, so they are not considered. The data are displayed in Fig.  \ref{twhyaexlupi-colors}.

The AAVSO also contains data for EX Lupi (see Fig.  \ref{twhyaexlupi-colors}, bottom), acquired during and after the 2008 outburst (visual filter only) and during a recent significant outburst (Johnsons filters B,V,R, and I), so they are not directly comparable with our quiescence LCOGT data. On the other hand, the multiwavelength dataset offers a very good view of the photometric changes during episodes of increased accretion, displaying the contrast between quiescence and outburst, unlike the LCOGT and TESS data. We thus chose to include the 2022 burst data for our analysis, non-uniformly sampled between JD 2459651 and JD 2459720, and containing 248, 316, 242, and 190 datapoints in B, V, R and I, respectively.

\subsection{TESS photometry} 

Both EX Lupi and TW Hya have been observed by the the Transiting Exoplanet Survey Satellite \citep[TESS;][]{ricker15}. We use the TESS data to verify the rotational period and to identify other types of variability that may affect the spectral lines, such as extinction events or buster behaviour.

EX Lupi was observed by TESS in sectors 12 (2019-05-21 to 2019-06-19, 30~min cadence) and 39 (2021-05-26 to 2021-06-24, 10~min cadence). The lightcurve data were obtained from the Barbara A. Mikulski Archive for Space Telescopes \citep[MAST\footnote{https://mast.stsci.edu/portal/Mashup/Clients/Mast/Portal.html};][]{powell22}.
For TW Hya, the TESS observations correspond to Sector  9 (2019-Feb-28 to 2019-Mar-26, 30~min cadence) and Sector 36 (2021-Mar-07 to 2021-Apr-02, 10~min cadence). Since the lightcurves were not available, we used $eleanor$\footnote{https://pypi.org/project/eleanor/} \citep{feinstein19,brasseur19} to extract them. Following our previous experience \citep{campbellwhite21}, we used the principal component analysis (PCA) to detrend the data. There were no significant differences when using various types of apertures, and the MAST lightcurves for EX Lupi were essentially identical if extracted with $eleanor$, demonstrating their robustness. 

Both objects show flares and burster behaviour (especially, TW Hya and, to a lesser extent, EX Lupi during 2021), but there is no evidence of dips or occultations by circumstellar material. Those are not seen in the ground-based photometry either, suggesting that the observed variability is dominated by rotation in objects with hot and/or cold spots and accretion variability.

\section{Analysis }\label{analysis}

\begin{figure*}
    \centering
    \begin{tabular}{cccc}
    & {\bf He II 4686\AA} & {\bf TESS} & \\
         &  
          \includegraphics[height=4.5cm]{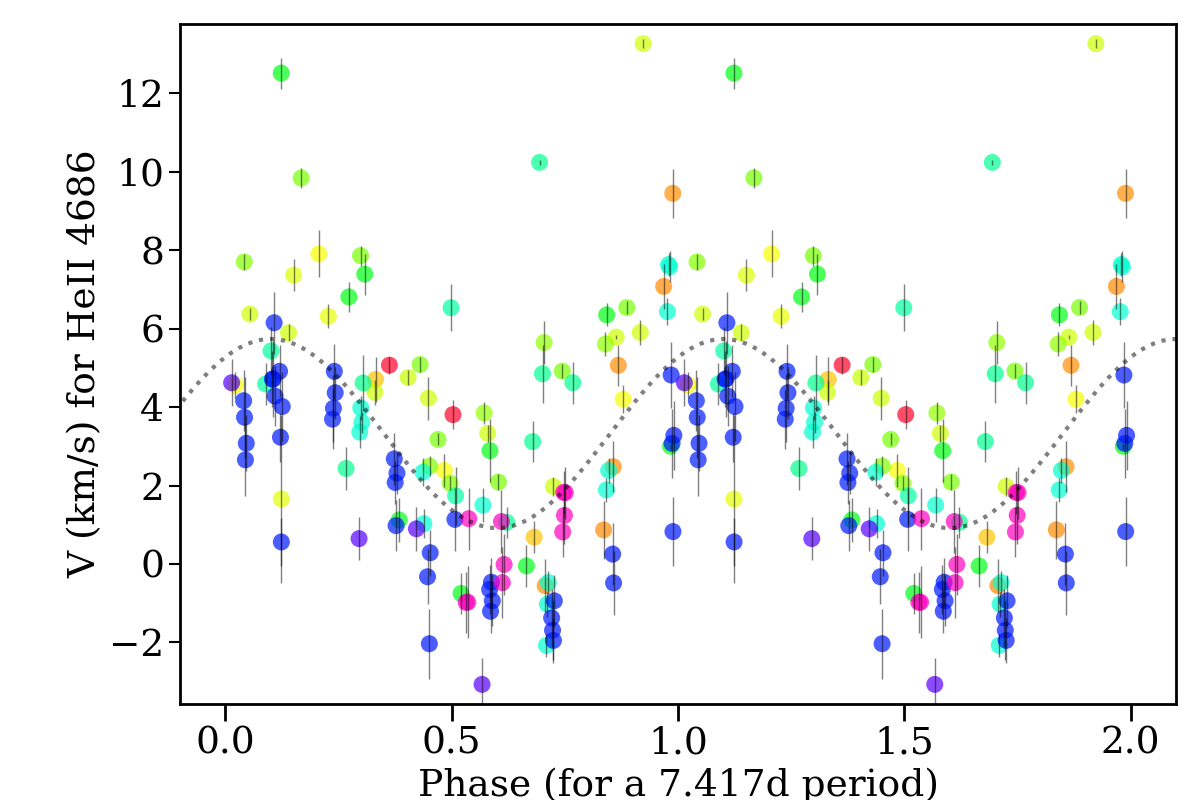} &
          \includegraphics[height=4.5cm]{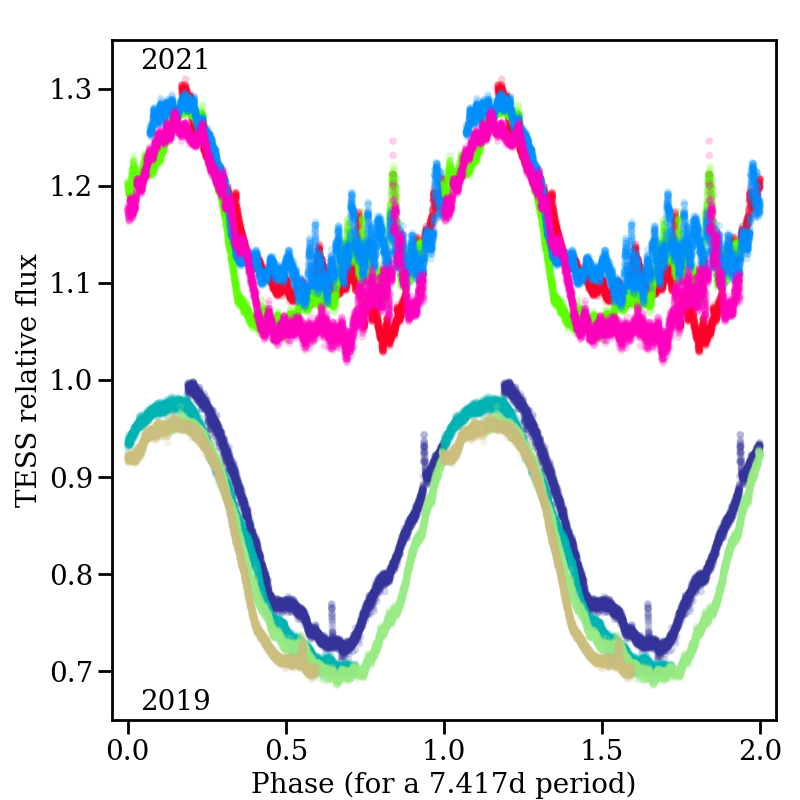} &
         \\
    \end{tabular}
    \begin{tabular}{c}
    {\bf LCOGT}\\
    \includegraphics[height=4cm]{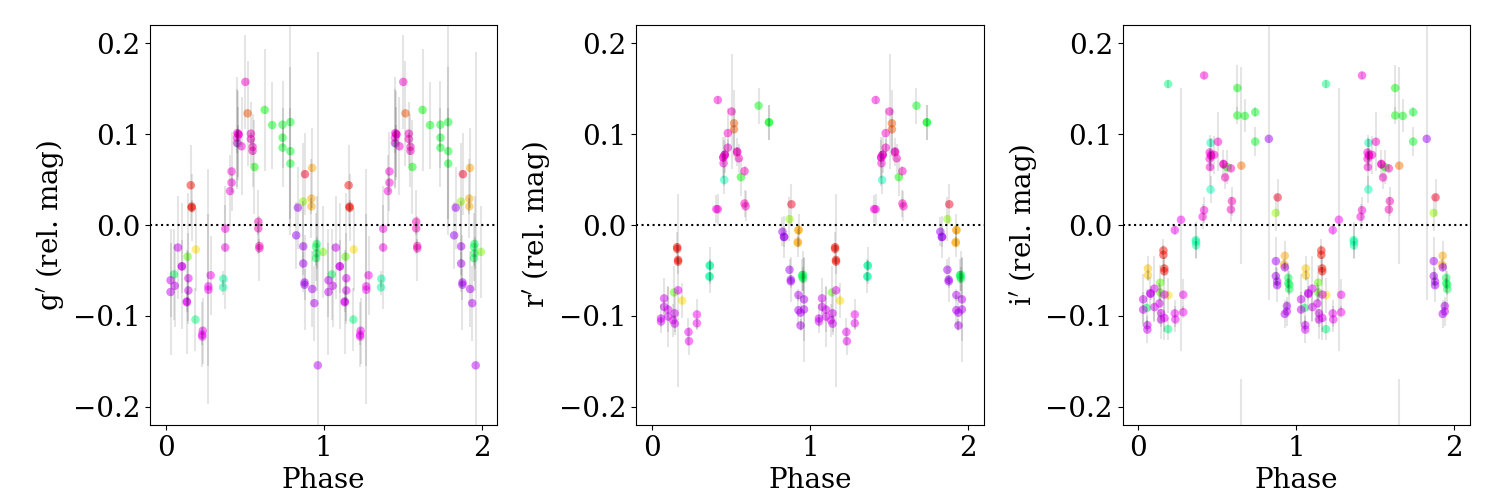}\\
     \includegraphics[height=4cm]{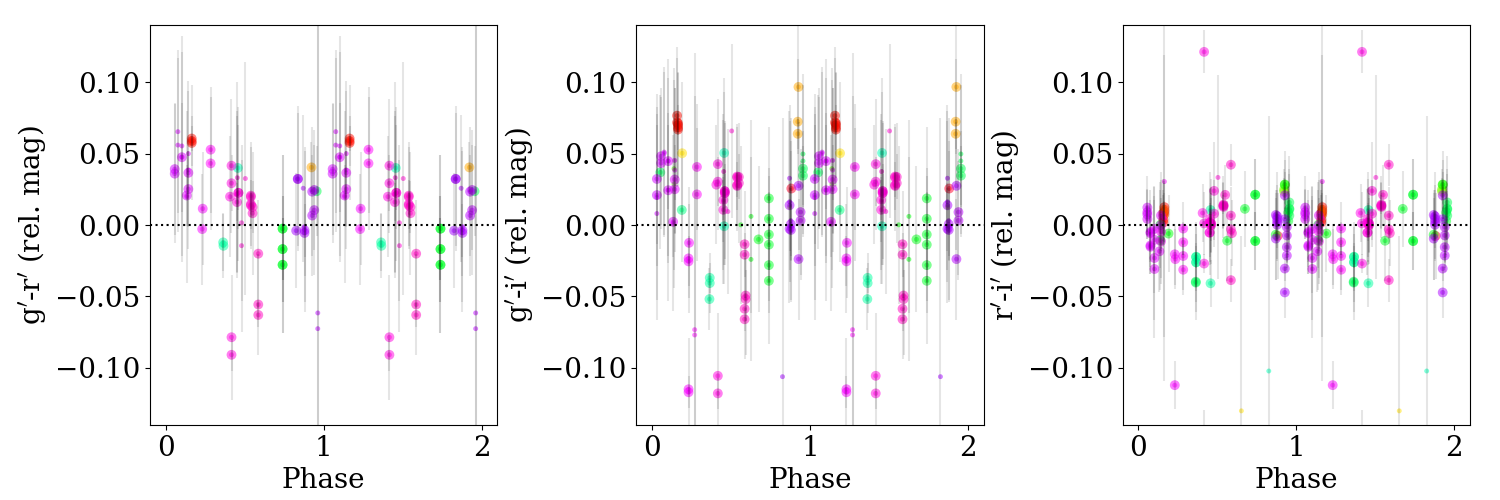}\\
     {\bf AAVSO} \\
      \includegraphics[height=4cm]{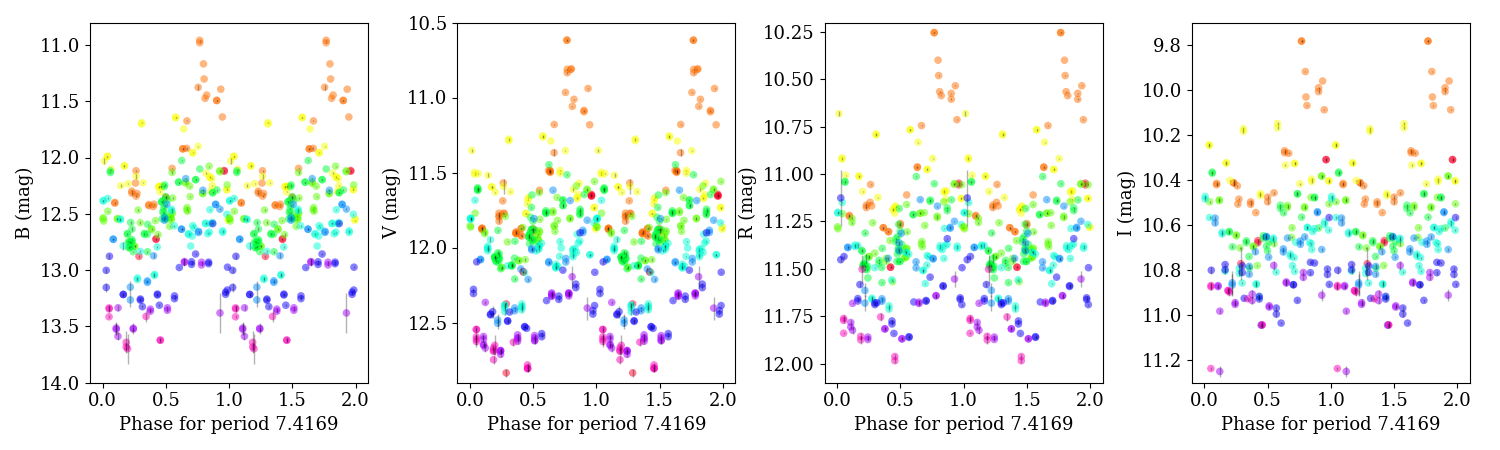}\\
       \includegraphics[height=4cm]{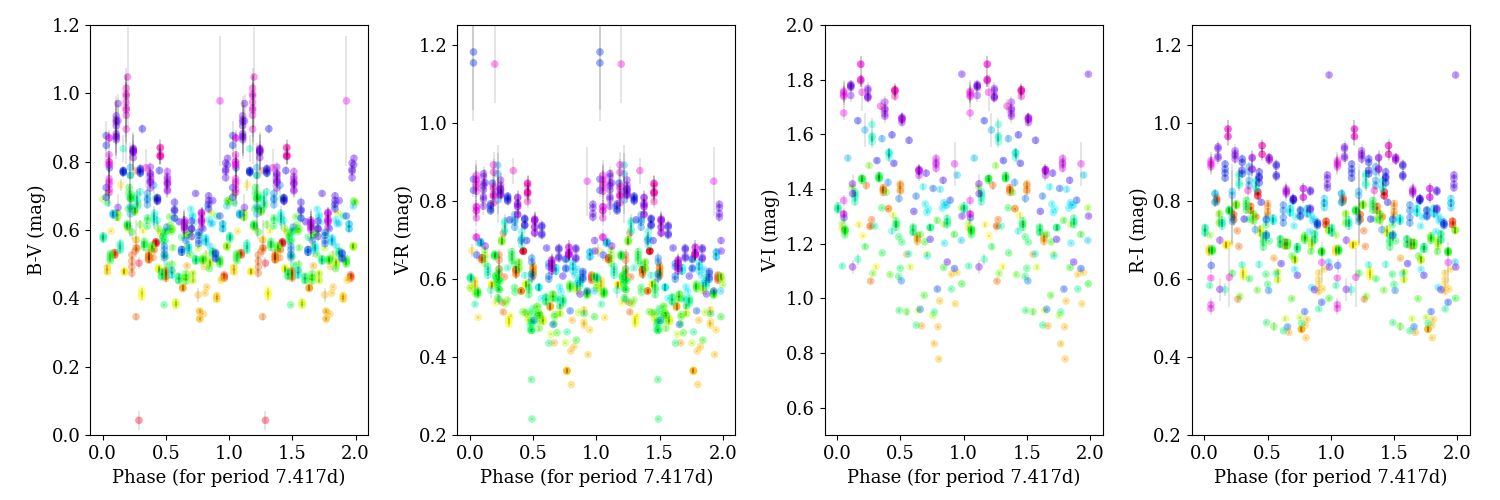}\\
       \end{tabular}
    \caption{Phase-folded data showing the phase variations observed in EX Lupi in spectral lines vs magnitudes and colours. All the data here are wrapped with the same period as obtained from spectroscopy, 7.4169~d, and taking as reference MJD=54309.1147~d \citep[starting point of data in][]{campbellwhite21}. In each figure, the data are coloured from red to purple according to the epoch, taking as limits the beginning and end of each dataset (see Fig. \ref{timeline-fig}). The top plot shows the He~II 4686\AA\ line with it sinusoidal fit (dashed line) from \citet{campbellwhite21}, next to the TESS data, noting that the rest of spectral lines display the same behaviour with differences only up to $\sim$10~$^{\circ}$. Below, from top to bottom, we display the  phase-folded LCOGT magnitudes, the LCOGT colours, the AAVSO magnitudes, and the AAVSO colours. }
    \label{exlup-phase}
\end{figure*}

\begin{figure*}
    \centering
    \begin{tabular}{ccc}
    {\bf HeI 5016\AA} & {\bf FeII 5018} & {\bf TESS} \\
         \includegraphics[height=4cm]{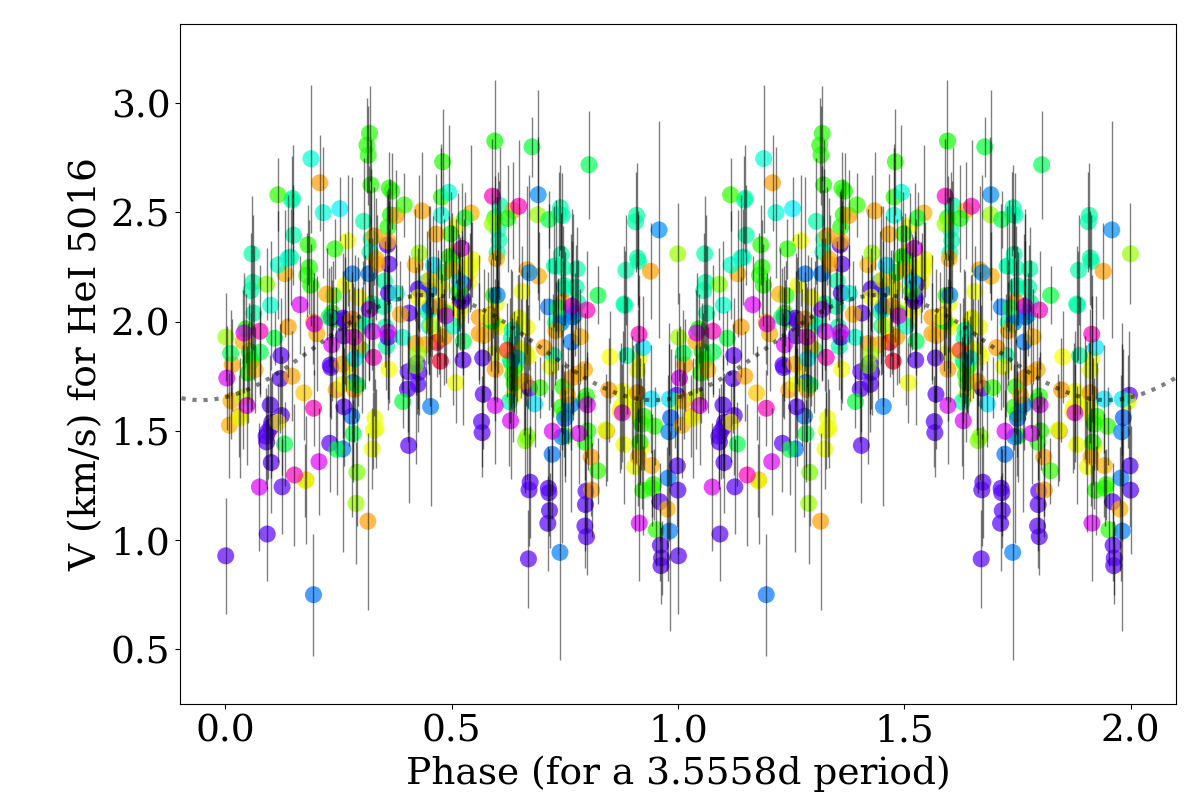} &
         \includegraphics[height=4cm]{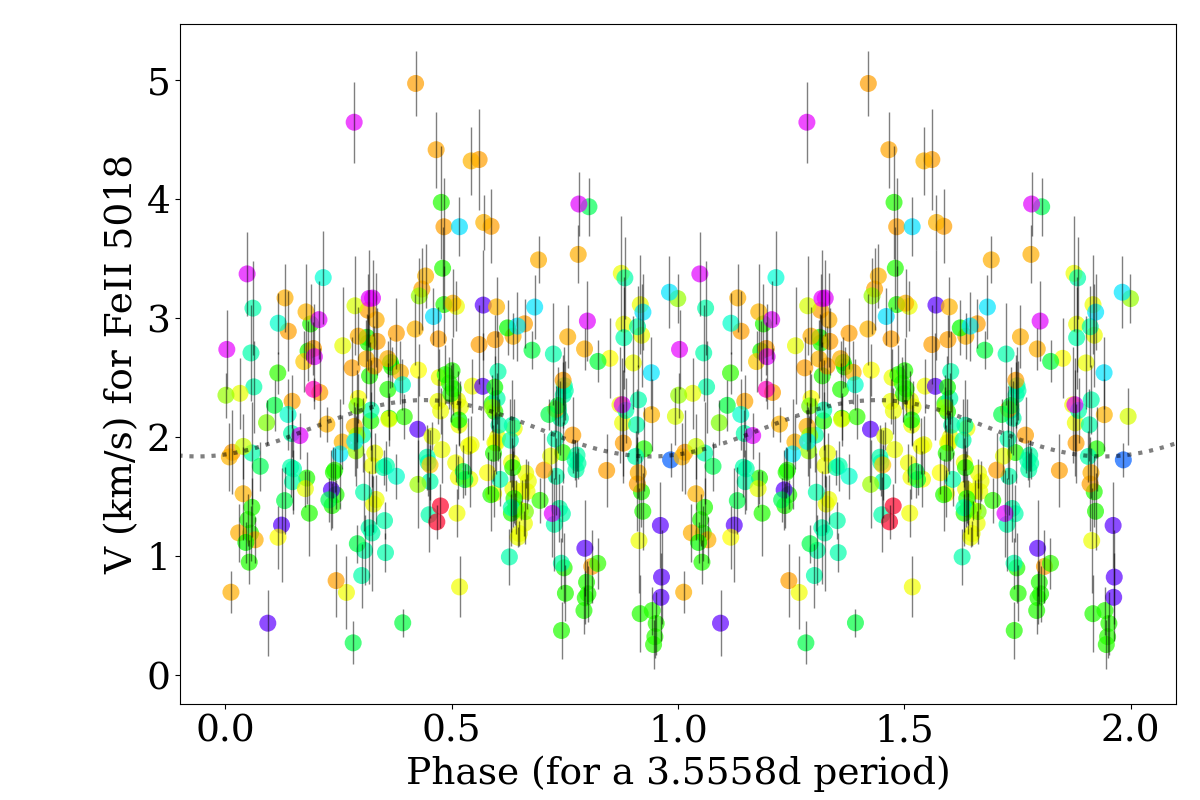} &
         \includegraphics[height=4cm]{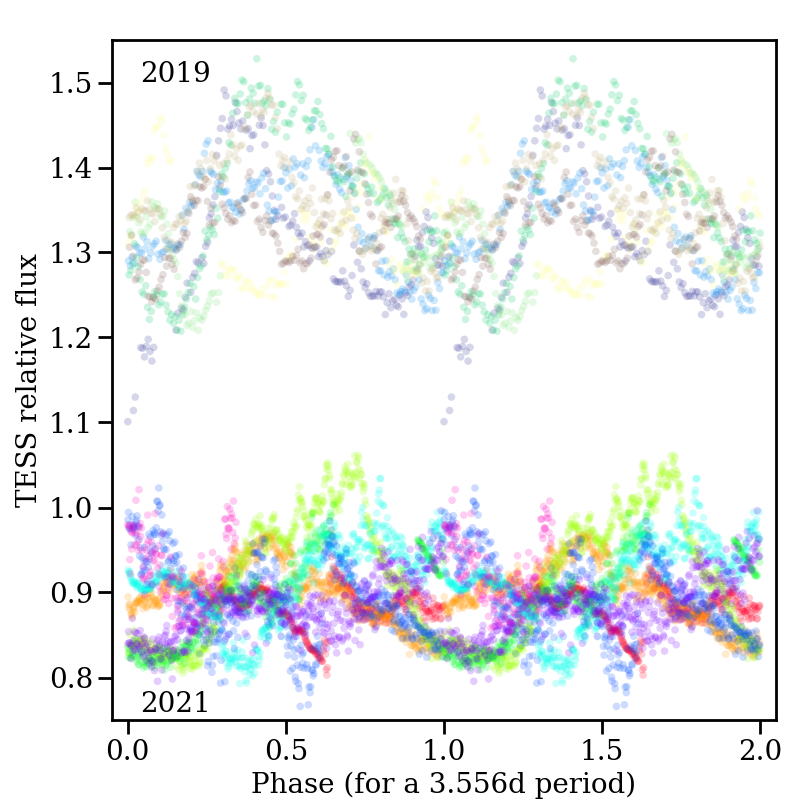}\\
    \end{tabular}
    \begin{tabular}{c}
    {\bf AAVSO}\\
    \includegraphics[width=15cm]{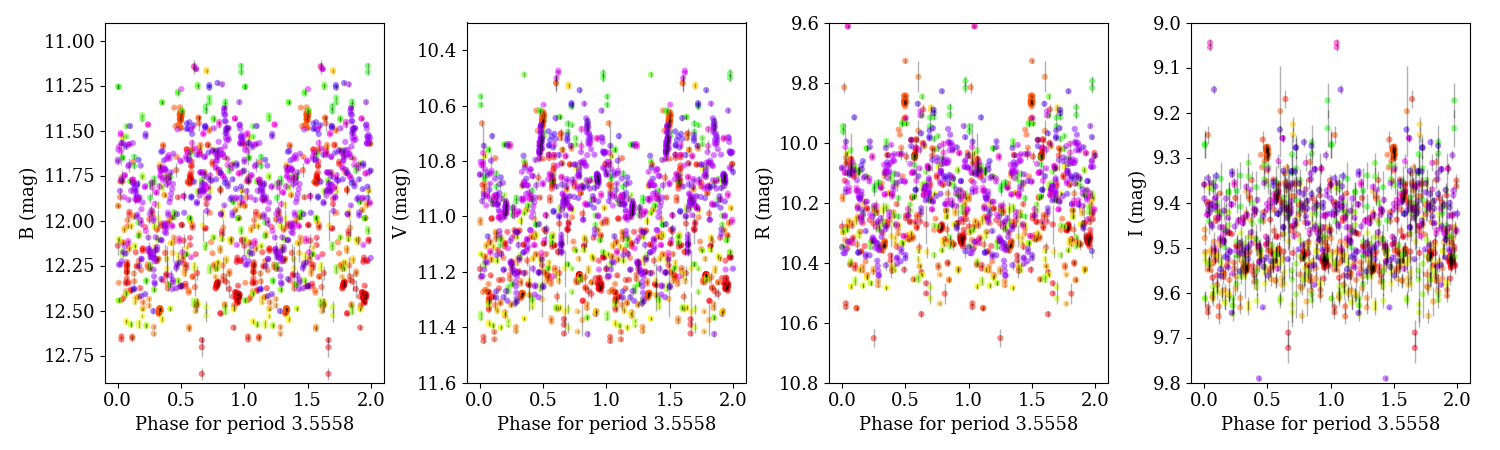} \\
    \includegraphics[width=15cm]{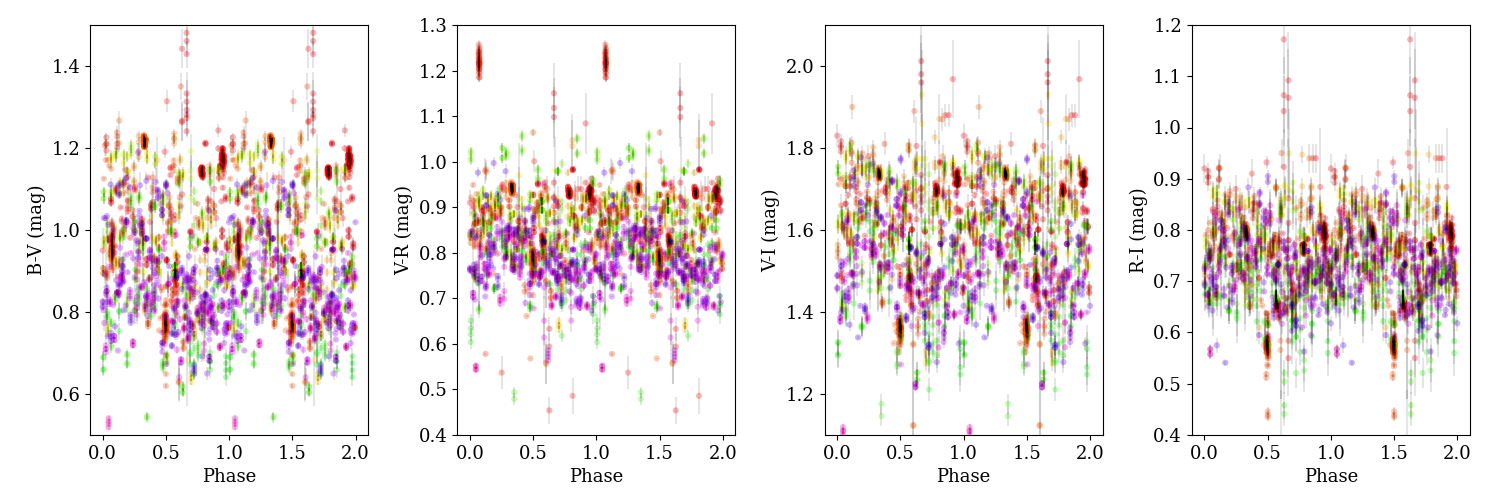}\\
    {\bf AAVSO for JD=2459640-2459680}\\
    \includegraphics[width=15cm]{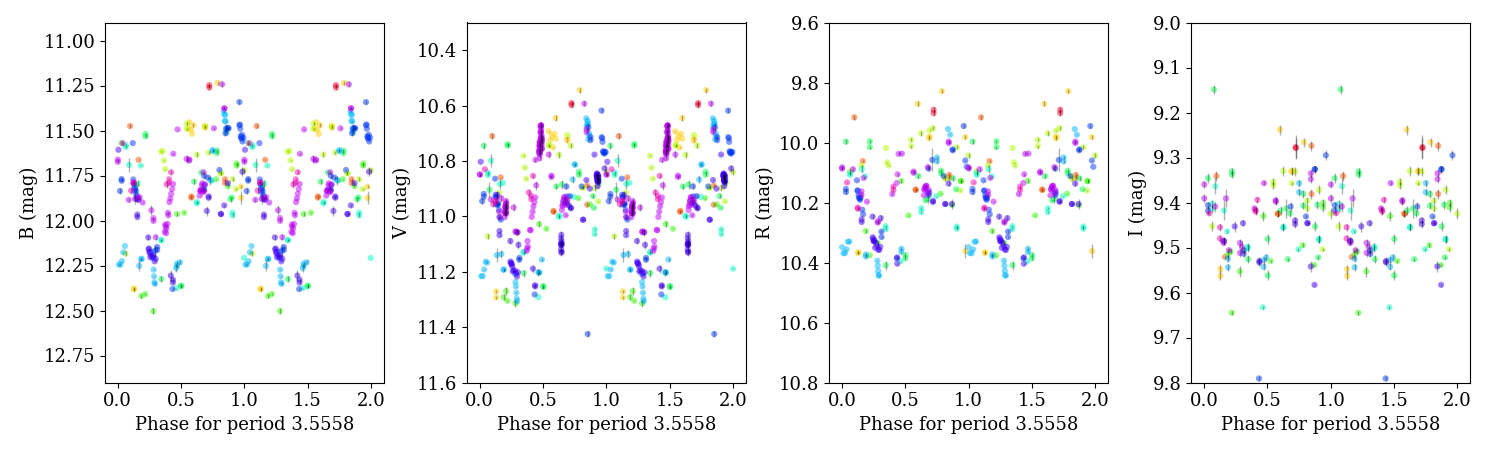}\\
    \end{tabular}
    \caption{Same as Fig.  \ref{exlup-phase} but for TW Hya, using the period of the He I 5016\AA\ line (3.5558d) and a starting point JD=2458543.827644~d. }
    \label{twhya-phase}
\end{figure*}

In this section, we first analyse the NC emission lines, which provide the best information about stellar rotation in both objects as first step to derive a coherent picture, adding later the multi-wavelength, multi-cadence photometry data. We start with the processes on the stellar surface or close to it (e.g. spots), later moving onto the BC emission lines, which are related to extended and more complex structures. Our aim is to compare all the data, so we begin by displaying the phase-folded data for both objects in Fig. \ref{exlup-phase} and \ref{twhya-phase}.  They are the reference for the analysis and discussion that follow.

\subsection{STAR-MELT line extraction and basic analysis}

The line identification and fitting was done with STAR-MELT \citep{campbellwhite21}. The STAR-MELT extraction and analysis of the NC for EX Lupi was presented in \citet{campbellwhite21}, revealing a very stable (or very slowly-changing, with timescales of years), single accretion hot spot, its location not affected by variations in the accretion rate, dominating the NC emission. This work thus concentrates on the BC, discussed in Section \ref{exlupBC-sect}. 

\begin{figure*}
\begin{tabular}{ccc} 
{\bf Fe~II 5018\AA} & {\bf He~I 5016 \AA} &{\bf Ca~II 8498\AA} \\
\includegraphics[width=5.cm]{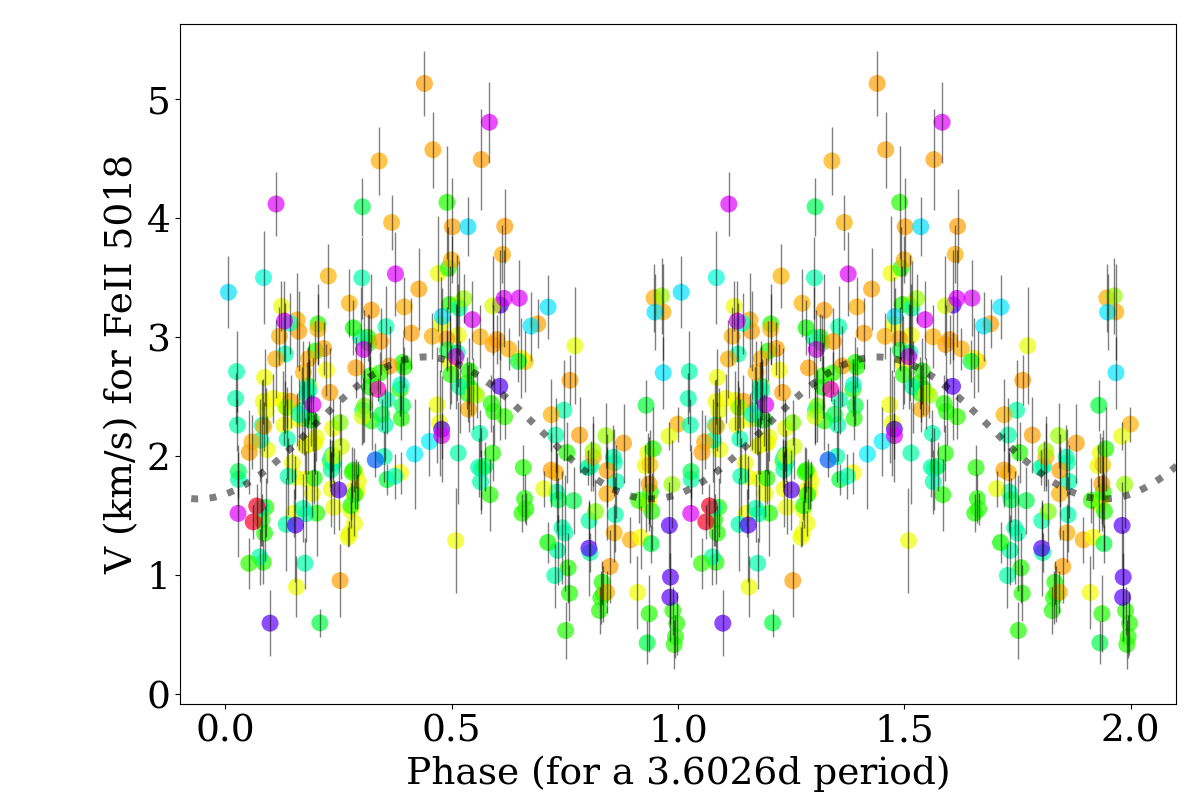} &
\includegraphics[width=5.cm]{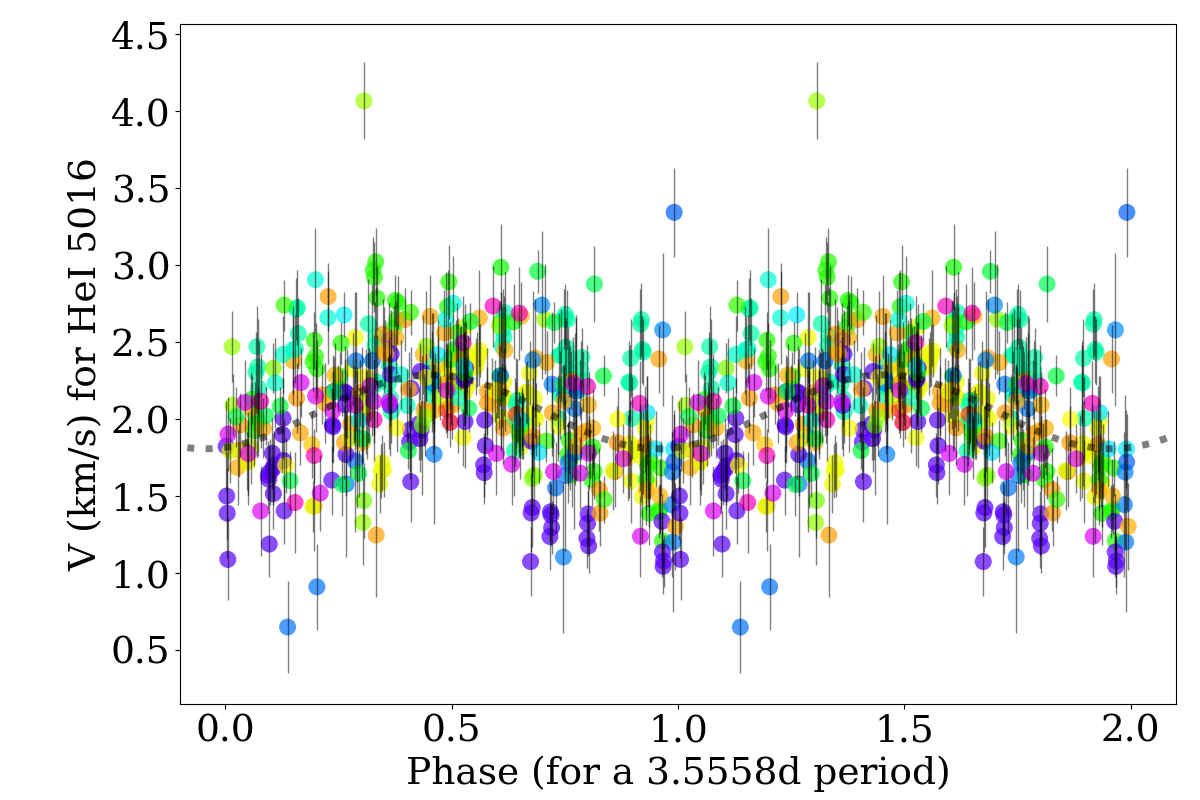}  &
\includegraphics[width=5.cm]{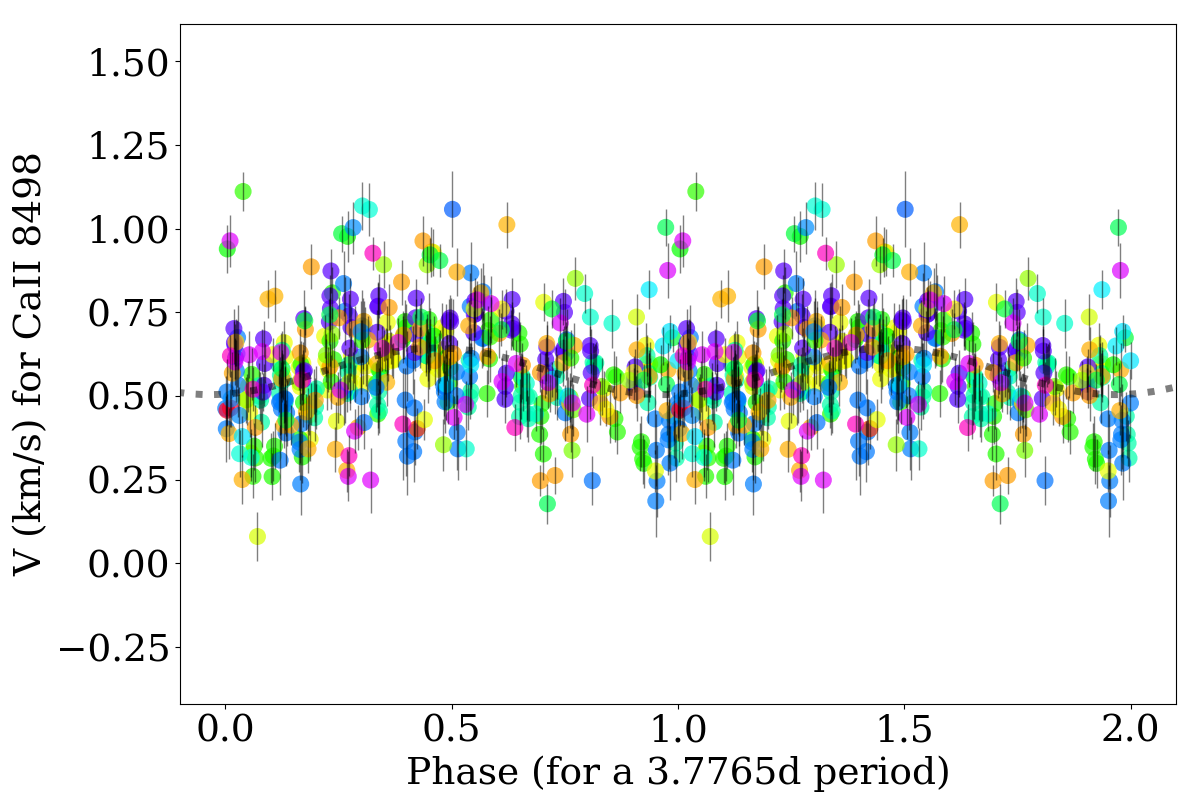}
\\
	\includegraphics[width=5.cm]{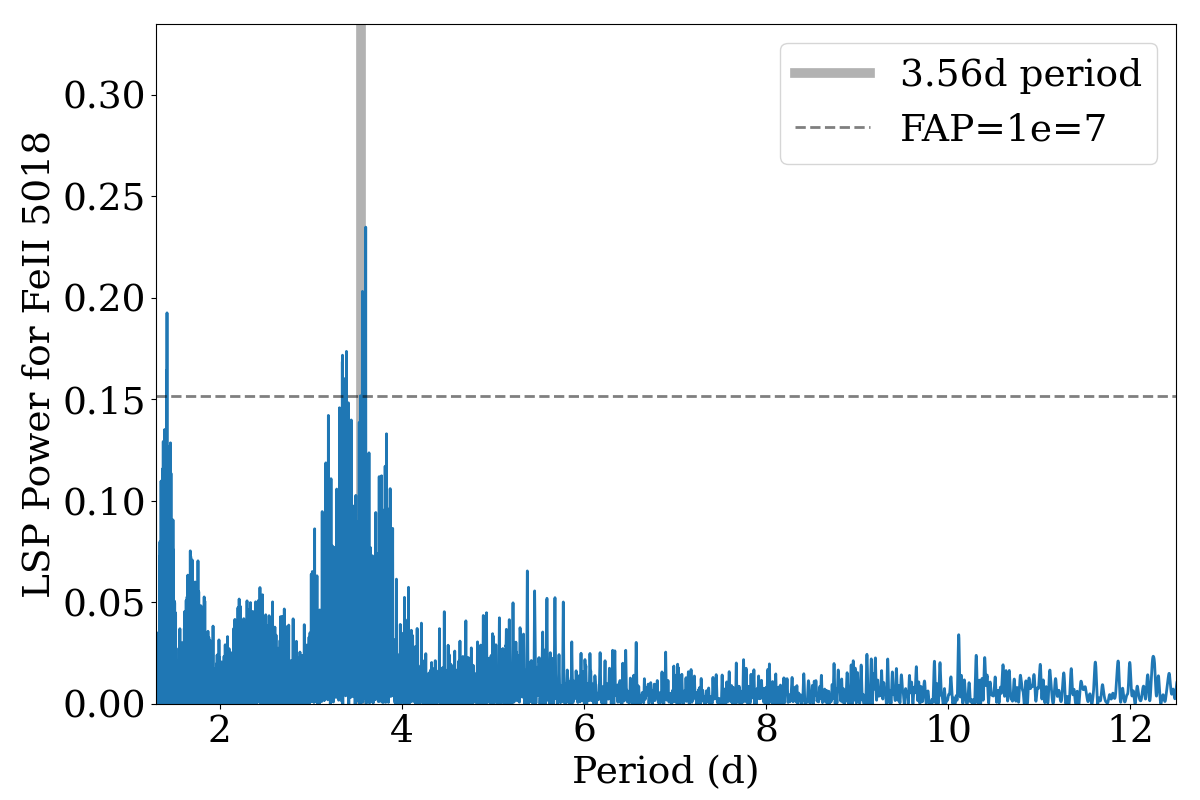} &
	\includegraphics[width=5cm]{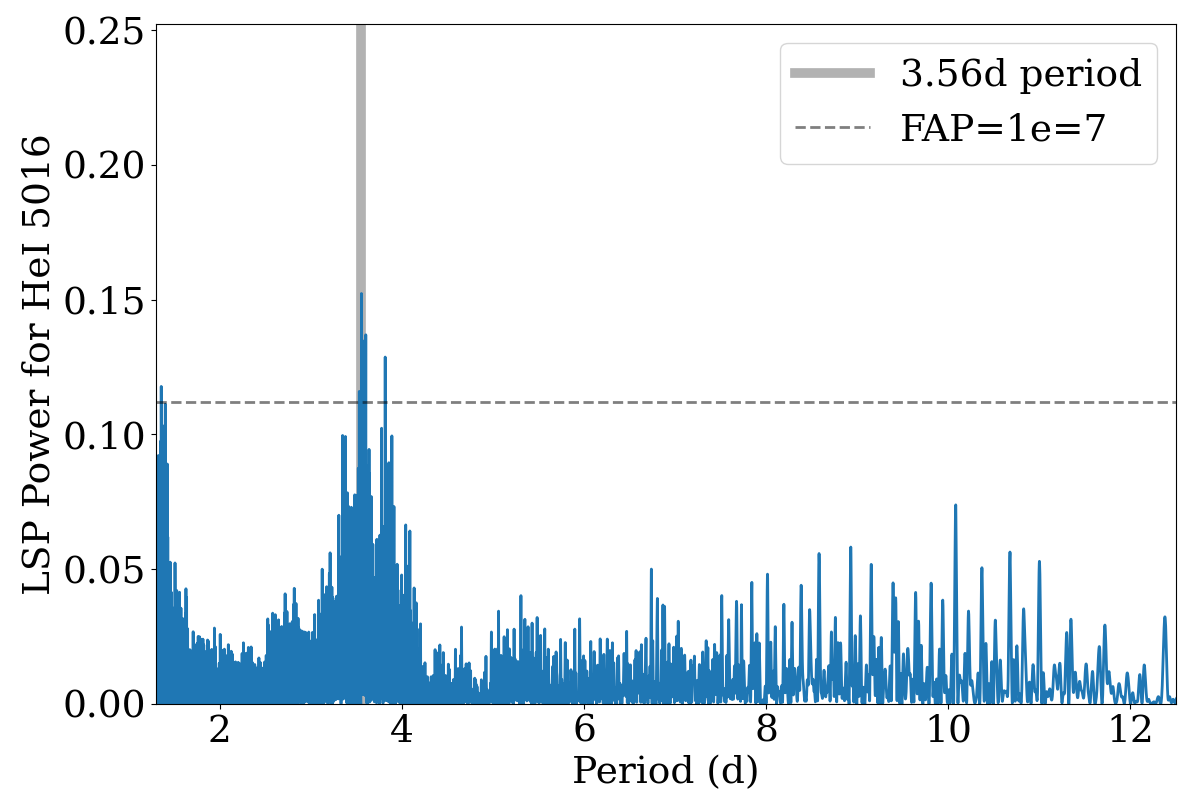} &
	\includegraphics[width=5cm]{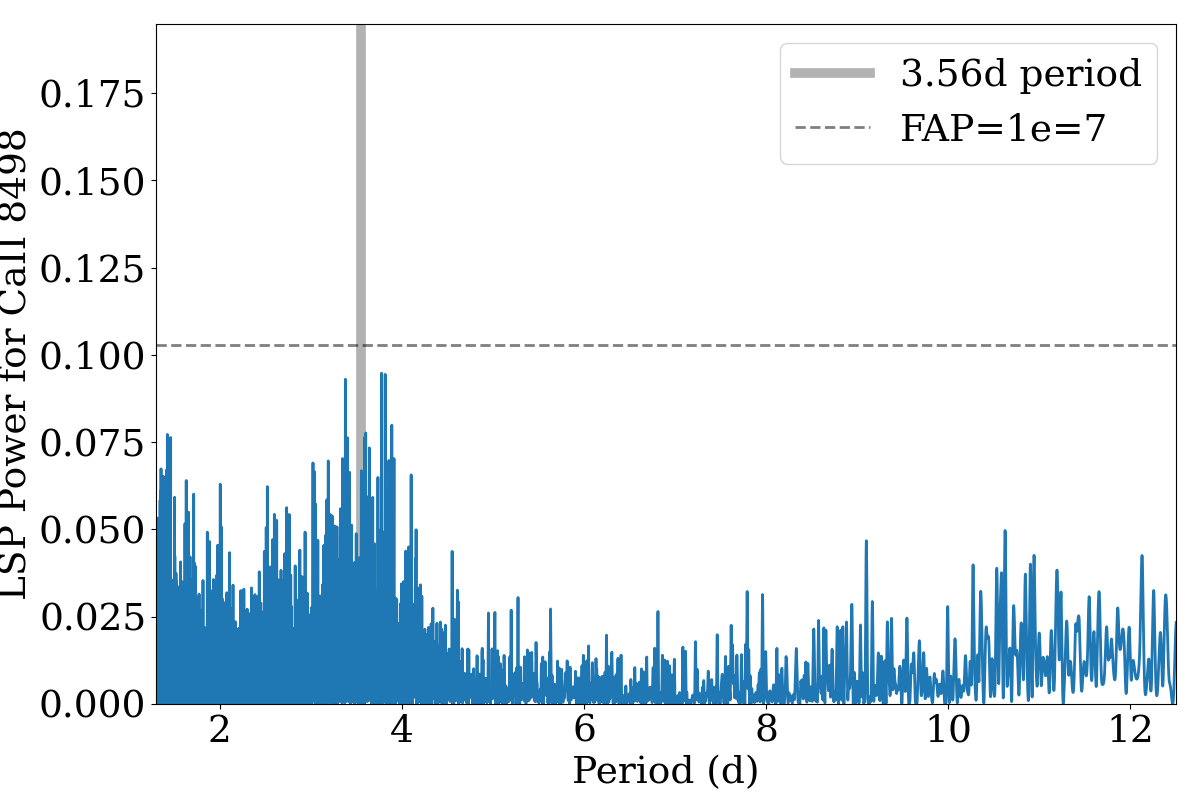} \\
\end{tabular}
\caption{Examples of phase-folded line velocity data (top) and LSP (bottom) for several of the lines observed for TWHya. We include lines for which we detect a highly-significant peak around 3.56d (e.g. Fe II 5018, He I 5016) and
Ca~II 8498\AA, with lower significance.  The colours in the phase-folded data reflect the date (see Fig. \ref{timeline-fig}). The data for each line has been phase-folded according to its individual best period, which explains the differences and small phase offsets. 
}
\label{fig:TWHya-GLS}
\end{figure*}

For TW Hya, we performed the same process as in \citet{campbellwhite21}, finding the emission lines in the average spectrum, creating a list of lines, and then fitting the NC with one positive Gaussian to determine the line velocity. Not all the lines were detected in all spectra due to differences in spectral coverage between instruments, variations in line strength (e.g. related to accretion variability), and signal-to-noise ratio (S/N). Only lines detected well over 3$\sigma$ were analysed. The analysis relies on the line velocity, not its intensity. The fit line velocity was corrected by the stellar radial velocity, estimated by STAR-MELT via cross-correlating regions of the spectra without emission lines with a template with a similar spectral type. For TW Hya, we measured radial velocity $v_r$=12.26$\pm$0.11 km/s and projected rotational velocity v$sini$=4.6$\pm$0.7 km/s.
Both were derived for each high-resolution instrument (FEROS, ESPaDOnS, HARPS) independently, 
and the final result is the weighted average. All the results are consistent, with HARPS giving a slightly lower value (v$sini$=3.3$\pm$0.5 km/s), while the rest of measurements range from 4.3 to 5.3 km/s with a typical uncertainty $\sim$0.3--0.5 km/s.

The NC lines were fitted using a 20--70 km/s window, chosen to cover the line width and to minimize the goodness-of-fit parameter (GOF, similar to a $\chi^2$) over the largest number of lines. The Ca~II~IR lines are broad and best fit with a 70km/s window, while for He~I and H~I lines 30--50 km/s is more appropriate, and narrow Fe~II lines are best done with a 20 km/s window. The results of using different windows are essentially identical.
We rejected any line for which the Gaussian fit failed or GOF$\geq$0.3--1 (depending on the line width, since the GOF is worse for broader lines), those with uncertain velocities (compared with the typical line velocity and velocity width) and/or deviating more than 3$\sigma$ with respect to the rest of the measurements.
As a brief comment on line width, the typical FWHM of the lines is 10--20~km/s for the metallic NC, too large to be of thermal origin. This suggests a highly turbulent region as expected in the accretion post-shock region \citep{hartmann_accretion_2016}. 
The lines observed in TW Hya are listed in Table \ref{tab:twhya-linelist}.

\begin{table}
    \centering
    \begin{tabular}{lcccl}
    \hline
Species & Wavelength & A$_{ki}$ & E$_{i}$-E$_{k}$ & Notes\\
& (\AA) & (s$^{-1}$) & (eV) & \\
\hline  
\hline
Ca II & 3933.66 & 1.47e8 & 0.000 - 3.151 & ZCMa/EXLup  \\
Ca II & 3968.47 & 1.40e8 & 0.000 - 3.123 & ZCMa/EXLup   \\
H I & 3970.04 & 4.39e5 & 10.199 - 13.321 & EXLup  \\
He I & 4026.19 & 1.16e7 & 20.964 - 24.043 & EXLup  \\
H I & 4101.71 & 2.86e6 & 10.199 - 13.221 & ZCMa/EXLup   \\
H I & 4340.47 & 2.53e6 & 10.199 - 13.055 & ZCMa/EXLup    \\
He I & 4471.48 & 2.46e7 & 20.964 - 23.736 & EXLup  \\
He II & 4685.8 & 2.21e8 & 48.372 - 51.017 &EXLup  \\
\hline
\end{tabular}
\caption{Lines detected by STAR-MELT in the TW Hya spectra. Information on the laboratory wavelength (in air), transition probability and energies of the lower and upper level \citep[from the NIST database;][]{a_kramida_nist_2020}, as well as on observations of the line in other stars \citep[EX Lupi, ZCMa, and ASASSN-13db, see][]{sicilia-aguilar_optical_2012,sicilia-aguilar_accretion_2015,sicilia-aguilar_2014-2017_2017,sicilia20zcma,campbellwhite21} is also provided. Only 8 lines are shown here, with the rest being available online via CDS. }
\label{tab:twhya-linelist}
\end{table}

The line velocity data were examined using a generalised Lomb-Scargle Periodogram \citep[LSP;][]{scargle82,horne86,zechmeister09,vanderplas18} as available via Python $astropy$, to identify the most likely periodic signals in the data. Lines that have a LSP peak around the rotational period are considered rotationally modulated. 
The strength of a periodic signal is evaluated via the false-alarm probability (FAP), which we calculated using the Baluev approximation that assumes that the noise is white and uncorrelated. This provides a lower limit to the FAP, since time-resolved observations are affected to some extent by red noise and cadence-related issues. Our aim is not to independently derive a rotational period for every line, but to check which lines are consistent with rotational modulation, so this approach is suitable. Further verification of the periodicity is done via phase-folded lightcurves.

If an emission line is produced on the stellar surface and rotationally modulated, the radial velocity of the line (RV$_{line}$) as the star rotates describes a sinusoidal curve as given by
\begin{equation}
    RV_{line} = V_0' + vsini\, cos \theta_s\, sin \phi, \label{rota-eq}
\end{equation}
where $V_0'$ is the offset velocity after correcting for the stellar radial velocity,  v$sini$ is the projected rotational velocity,  $\theta_s$ is the latitude of the spot, and $\phi$ the phase angle \citep{sicilia-aguilar_accretion_2015, mcginnis_magnetic_2020,campbellwhite21}. The offset velocity can be due to additional motions of the line-producing material, such as infall, and changes from line to line depending on its optical depth \citep{sicilia-aguilar_accretion_2015}. It may also reflect systematic errors in the laboratory wavelength of the line. The maximum velocity amplitude due to rotational modulation is v$sini$, for an equatorial spot. Nevertheless, lines produced at some height over the stellar surface may have a higher amplitude, assuming that solid-body rotation dominates on that spatial scale.

\subsection{Spectroscopy analysis \label{twhya-sect}}

\subsubsection{Line velocity periodicity for TW Hya}

\begin{table}
    \centering
    \begin{tabular}{lcccl}
    \hline
Species & Period & FAP & \#  & Comments\\
     &  (d)   &     &  Points & \\ 
\hline  
\hline
CaII 3934 & 3.44408 & 4e-4  & 303 & Quasiperiod \\
HeI 4026  & 3.34022  & 8e-3 & 271 &  \\
HI 4102  & 3.65501 & 3e-3 & 312 & Also shows a 6.2~d period\\
HeI 4471  & 3.56944  & 2e-6 & 352 & \\
HeII 4686  & 3.55568 & 0.05 & 145 & Very weak line\\
HeI 4713  & 3.34873  & 4e-5 & 356 & \\
HeI 4922  &  3.61393  & 6e-7  & 397 & \\
FeII 4924 & 3.56778 & 4e-11 & 289 & \\
HeI 5016  & 3.55583 & 8e-12 & 417 & \\
FeII 5018  & 3.60264 & 3e-14 & 296  & Strongest line \\
HeI 5875  & 3.56879 & 4e-6  & 459 & \\
HeI 6678  & 3.57531 & 1e-6 & 408 & \\
HeI 7065  & 3.08375 & 1e-3 & 84  &Modulated despite tellurics\\
CaII 8498  & 3.56927 & 3e-11  & 458 &  \\
CaII 8542  & 3.13435  & 9e-11 & 284 & Quasiperiod 3.1-3.8d\\
CaII 8662  & 3.56837 & 1e-4 & 457 & Also shows a 1.4~d period \\
\hline
\end{tabular}
\caption{List of periods observed for TW Hya for all the emission lines with
significant S/N. The FAP is determined using the Baluev method, being a lower limit. Note that the Ca~II lines display
signatures of quasi-periodicity, probably due to
extra variability in the line wings (e.g. from accretion, winds). }
\label{twhya-periods}
\end{table}

\begin{figure*}
    \centering
    \includegraphics[width=17cm]{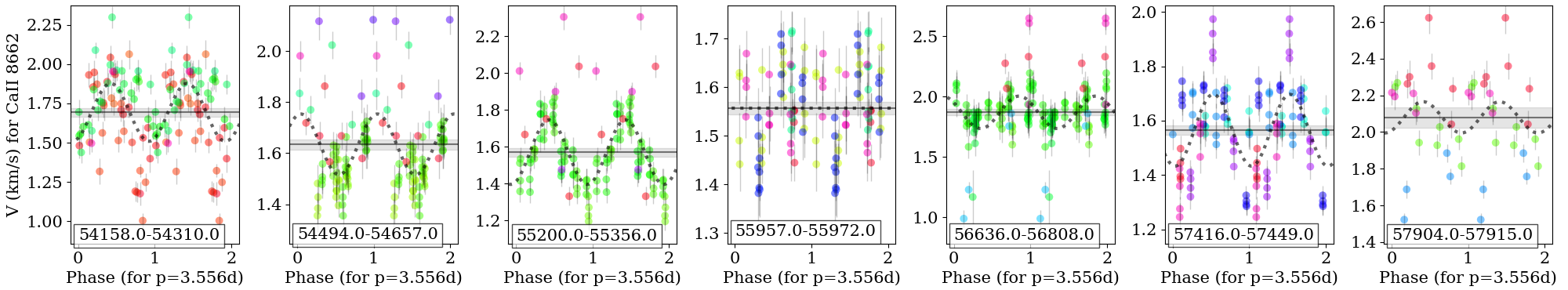} 
    \includegraphics[width=17cm]{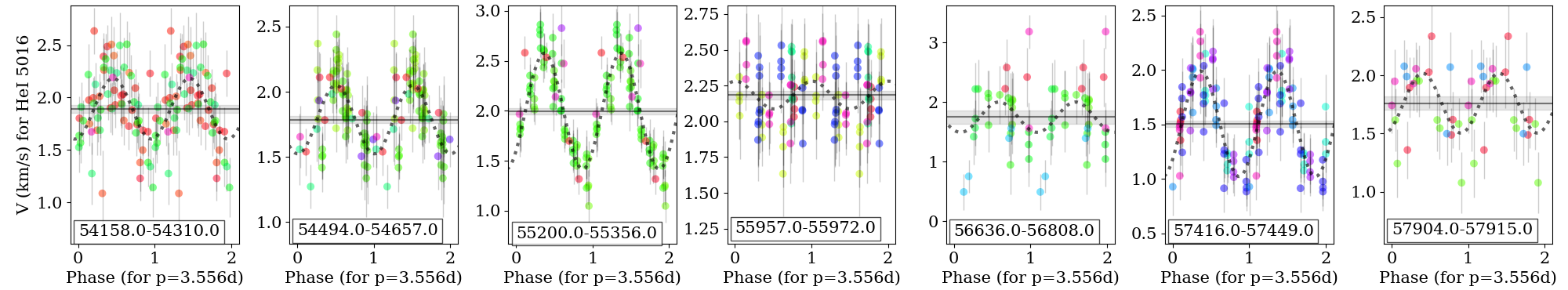}
    \caption{Rotational modulation in the Ca~II 8662\AA\ and He~I 5016\AA\ lines in TW Hya as a function of date, considering the 3.556d period derived from the He~I 5016\AA\ line. We only consider time intervals for which there are enough consecutive (no more than a few days between observations) data to try to fit a modulation, and the colour scheme here is set to vary from period to period within each epoch, for clarity. A similar behaviour is observed in other lines.}
    \label{twhya_intervals}
\end{figure*}

TW Hya has significantly less NC emission lines than EX Lupi, but they provide a strikingly clear 3.56 d period, fully consistent with the expected rotational period \citep{huelamo08}. The rotational period is detected in essentially every line that has enough S/N and a sufficient number of observations, with typical differences in period of less than $\sim$1~h (see Fig. \ref{fig:TWHya-GLS} and Table \ref{twhya-periods}). Although the photometry data reveal an erratic burster behaviour (Fig. \ref{twhya-phase}), the rotational modulation of the emission lines is clean, stable and robust during the 13 years covered by the spectra. The velocity amplitudes observed are always below v$sini$ and thus consistent with an origin on the stellar surface. As in EX Lupi \citep{kospal_radial_2014}, stable accretion spots may cause the spurious radial velocity variations initially interpreted as a companion \citep{setiawan08} and latter explained as a cold spot \citep{huelamo08}. The amplitudes are smaller than in EX Lupi, likely due to the nearly-pole-on position.

As for EX Lupi, higher-energy lines such as He~I 5016~\AA\ and the Fe~II multiplet at 4924 and 5018~\AA\ show very clean modulations through the entire dataset. They are also the strongest lines without blueshifted absorption from winds, which may contribute to their cleanliness. Stronger He~I and Ca~II lines are also rotationally modulated, albeit with higher FAP. 
The rotational modulation is evident even in lines such as the He~I 7065\AA, strongly affected by tellurics. 

The only lines with dominant LSP peak substantially different from 3.6~d are H$\delta$ and Ca~II 8662\AA. H$\delta$ is the only one of the Balmer lines without self-absorption or blends. It has a low-significance period of 6.2~d, close to twice the rotational period and only slightly stronger than the second peak, at 3.66~d. Considering that, even if the line seems symmetric and narrow, it may still be affected by absorption, the result cannot be considered as discrepant.
The short period of the Ca~II 8662\AA\ line\footnote{ESPaDOnS and FEROS have gaps near the 8498 and 8542\AA\ lines, respectively, so Ca~II 8662\AA\ has more observations. } is harder to interpret, 
although the line also shows a 3.57~d period.  Ca~II 8542\AA\ has a second peak at 1.46~d, which could mean that the Ca~II~IR lines have a different signature from the rest.  The $\sim$1.4~d peak is particularly evident when the line is fitted with a broader window, which suggests it is caused by uncertainties and a residual BC in very variable lines. TW Hya has very weak BC's, but the complex shape of EX Lupi's BC also displays a shorter period (see Section \ref{exlupBC-sect}), and TW Hya may be similar. Phase-folding confirms that the Ca~II~IR lines are rotationally modulated, especially considering data over short time intervals.

Significant differences in the accretion rate (as inferred from the strength of the H$\alpha$ line) do not affect the presence or absence of rotational modulation. For the metallic lines, the modulation is cleaner when the accretion rate is higher, but this seems a consequence of S/N, since the NC lines are stronger when accretion is stronger. For stronger lines such as the Ca~II~IR triplet or the Ca~II~H and K lines, and He~I lines, the modulation is messier when accretion rate is higher, which could be due to wind- or accretion-related absorption or BC distorting the profile. The ubiquitous modulation, independent of the accretion rate, is an additional similarity between TW Hya and EX Lupi, where the NC velocity modulations are not even affected by outbursts \citep{sicilia-aguilar_accretion_2015}. Due to its fewer lines, and since the strongest lines are too narrow and affected by collisional de-excitation, we cannot further estimate the structure of the line region using e.g. the Sobolev approximation.

\subsubsection{Determining the location of the accretion column footprints \label{columnlocation}}

\begin{figure*}
    \centering
    \begin{tabular}{ccc}
    \includegraphics[height=5.0cm]{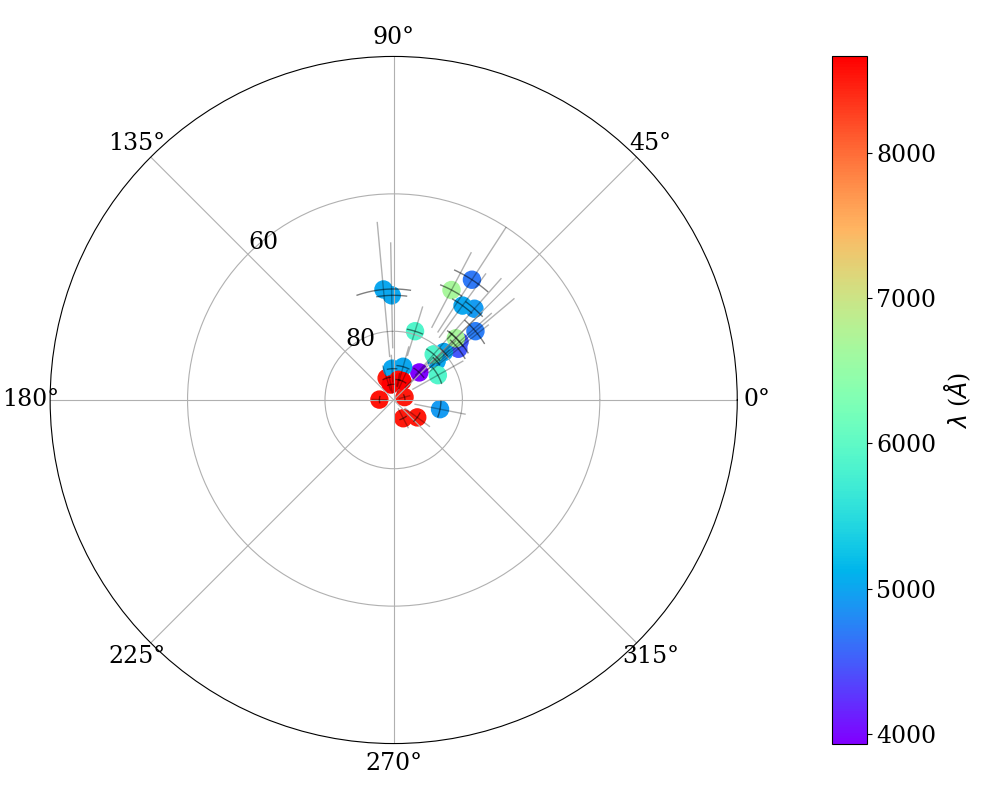} &
     \includegraphics[height=5.0cm]{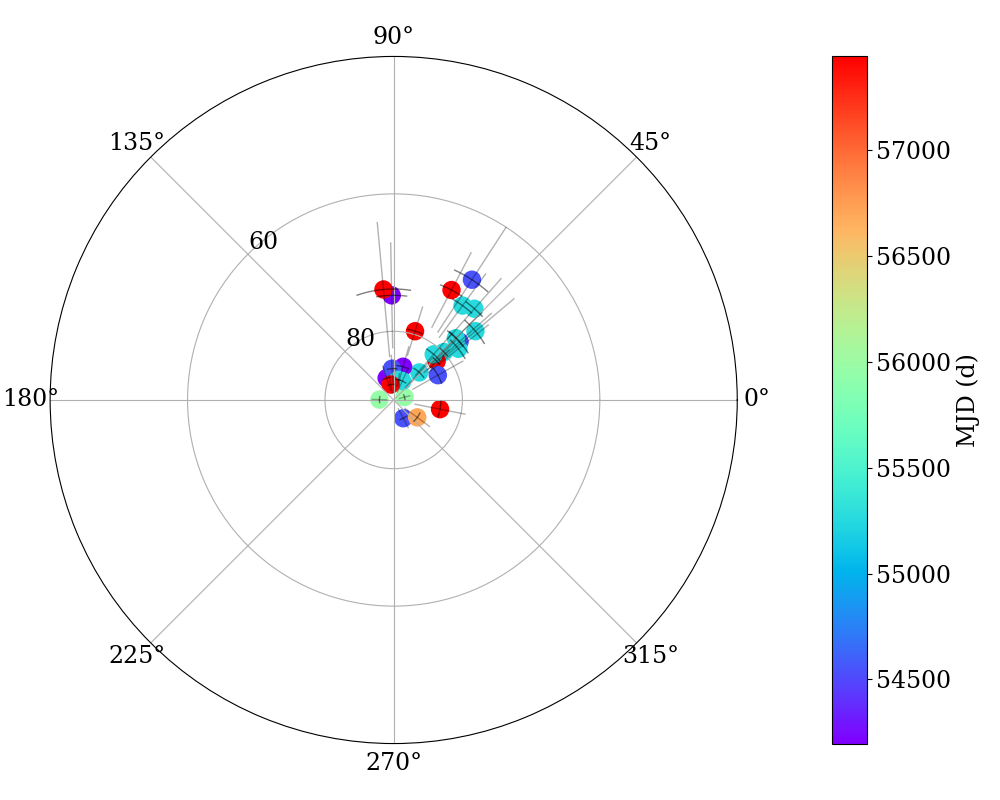} &
      \includegraphics[height=5.0cm]{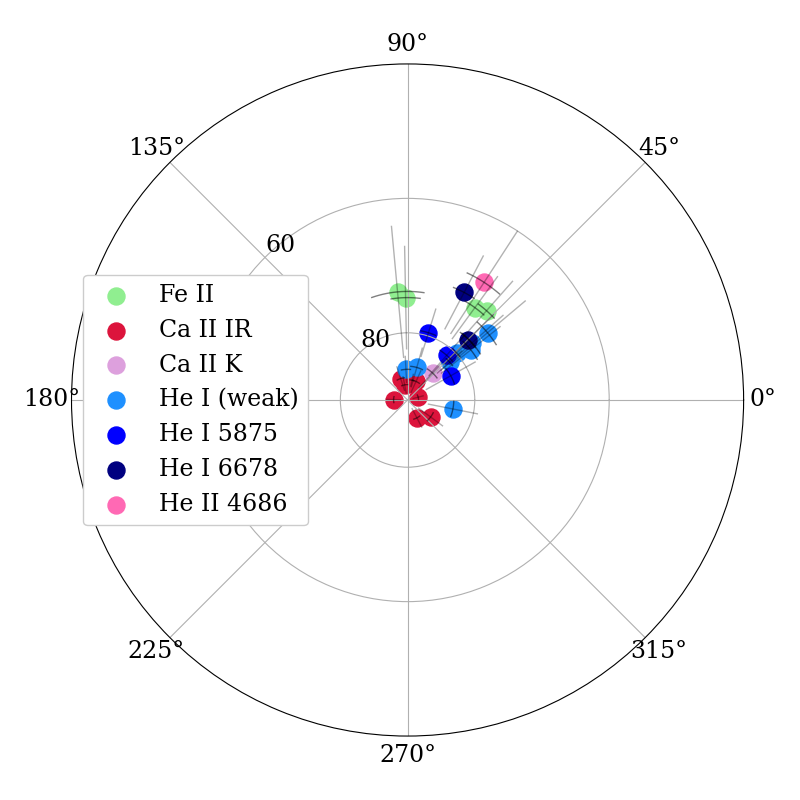}\\
    \end{tabular}
    \caption{Polar plots for TW Hya, showing the positions of the line-emitting region derived from line fitting for various species over different time intervals, as shown in Fig. \ref{twhya_intervals} (see the list of lines and dates in Table \ref{twhya-mjdfit}). The left plot shows the dependency with the wavelength, the middle plot shows the variation over time, and the right plot shows the distribution of different species. Note that, since the spots are very close to the pole, the figures are zoomed so the outer ring corresponds to 40 degrees latitude and not to the equator. The latitudes have been estimated considering the v$sini$=3.3 km/s, as derived from the HARPS data, and the rotational period of  3.5558d obtained for the He I 5016\AA\ line. The phase is given relative to the arbitrary date MJD=53460.22582d. Note that, although the data for certain lines had significantly higher uncertainties and noise, there is a very good agreement for all lines, especially when considering a given date.}
    \label{TWHya-polar}
\end{figure*}

Following  Eq. \ref{rota-eq}, we can determine the location of the NC-emitting region for each species. Since we observe changes in the amplitude over time that do not simply respond to noise, we followed the same procedure for TW Hya as we did for EX Lupi in \citet{campbellwhite21}, and divided the data in various epochs, taking intervals that are as short as possible while still containing a large-enough number of points for the fit to be meaningful. The number of time intervals that produce significant results varies from line to line due to differences in line S/N. We explored the location as a function of wavelength, date, and species, which can reveal information on the temperature and density structure in the accretion column \citep{sicilia-aguilar_accretion_2015,campbellwhite21}.  

Figure \ref{twhya_intervals} shows examples of the fits (listed in Table \ref{twhya-mjdfit}), which can be transformed into locations on the stellar surface. and Fig.\ref{TWHya-polar} shows the transformation of the fitted values per line, species and epoch into a polar plot with longitude and latitude location for different lines and epochs: from Eq. \ref{rota-eq}, a change in the amplitude of the modulation corresponds to a change in latitude for the NC-emitting region. The sinusoidal modulation is cleaner over shorter time spans, and on certain dates, and we also observe significant changes in amplitude that evidence changes in location. Observational noise is responsible for the lack of modulation in some epochs, which are not used in the polar plot. Due to the nearly pole-on inclination, detecting the modulation would be impossible if the lines were arising from a mildly extended region, confirming their origin in a very compact region.

Vertical changes or stratification in the location of the post-shock region, resulting for instance from a variation of the accretion column density, could also induce differences, and not all NC need to originate at the same height. In EX Lupi, the phase offsets between the lines suggested that the structure, albeit small, is extended and contains trailing parts with slightly different phases, which is consistent with a vertical scale over the accretion footprint \citep{sicilia-aguilar_accretion_2015,campbellwhite21}. Changes to this scale, rotating at the same pace as the star, could allow for larger velocities if the NC-emitting region is further away from the stellar surface. 

To test this hypothesis, we examined the NC velocity amplitudes for both objects. For EX Lupi, the Ca~II~8662\AA\ line has an amplitude of 0.37$\pm$0.06 km/s, while He~II~4686\AA\ has 2.5$\pm$0.3 km/s \citep{campbellwhite21}. If we assume that the Ca~II line forms at the stellar surface, and that He II forms at the same latitude but higher up, it would require a very large and unphysical difference in radius, nearly a factor of 7. For TW Hya, we took the data centred around MJD=55200 d, which have the largest and cleanest number of measurements (see Table \ref{twhya-mjdfit} and Fig.  \ref{twhya_intervals}). There is a clear difference between the amplitude (and phase) of the Ca~II lines and the rest, and comparing it with the high amplitude He~I 5016 and Fe~II 5018~\AA\ lines, the difference in radius would be a factor of $\sim$5-6, which is again too large. Even the differences between strong lines (with amplitudes $\sim$1 km/s) and others (with amplitudes $\sim$0.6 km/s) produce a very large factor, $\sim$1-2. 
In contrast, the latitude changes are moderate, as seen in Fig.  \ref{TWHya-polar}. 
Therefore, although some vertical structure is unavoidable considering the energies and transition probabilities of different lines, vertical stratification alone cannot explain the variations in the radial velocity amplitude.

Despite the variations in amplitude and phase, the location of the hot and dense region that emits the NC in TW Hya (in particular, for the energetic lines) is very well-defined, very compact and close to the stellar pole (see Fig. \ref{TWHya-polar}). A close-to-polar dominant accretion spot is in agreement with the single spot revealed by magnetic field mapping at a different epoch \citep{donati11}, so the NC behaviour suggests this configuration may be usual in TW Hya. EX Lupi shows no significant change in the phase of the line modulation over the 13 years of observations despite some phase (and amplitude) differences between species \citep{campbellwhite21}. TW Hya lines show stronger changes in the phase (or longitude) of the lines:
Although the modulation is always evident and the NC-emitting regions are concentrated at latitudes over 60$^{\circ}$, the longitude does change in time as if the line-emitting region shifted around the star, so the spot location is not as stable as in EX Lupi, varying by about 120$^{\circ}$ and showing a larger spread within a single epoch, while in EX Lupi they are always concentrated within the same quadrant.

There is also a difference between the location of the Ca~II~IR lines and the rest, with the Ca~II~IR lines having lower amplitudes and thus being closer to the pole. They also have the lowest FAP among strong lines, the largest deviations from the 3.56d period, and are broader than the rest of metallic lines. 
The Ca~II~IR modulation in EX Lupi also has lower amplitude than for the rest of lines \citep{sicilia-aguilar_accretion_2015, campbellwhite21}, but the sinusoidal modulation is still clear in both stars. This points towards a physical reason for the difference between the spatial origin of different lines, such as observed in the structured spot found by \citet{espaillat21}, and is a signature of stratification (in temperature/density) rather than simply dilution due to an origin in extended regions.  

As a final note, we explored whether the slight LSP period variations from line to line could be due to differential rotation, but we did not find any correlation with the excitation energy or transition probability of the line, nor the modulation amplitude, and the period, for any of the two objects. Since all lines are consistent with the same rotational period in both objects, we conclude that, although differences in phase and amplitude are significant, the differences in period are not.
Section \ref{discussion} discusses all the possibilities in more detail.

\subsubsection{Unveiling underlying radial velocity variations}

In addition to the variations in amplitude (or latitude), we also observe some changes in the zero-point offset of the NC from epoch to epoch for both TW Hya and EX Lupi. This suggests that there could be additional residual radial velocity variations.  Note that, although the global offset for a given line is affected by the uncertainty in the wavelengths (our sub-km/s uncertainties are often better than the precision in the NIST database wavelengths), the differences between epochs are robust and agree across observations done with different instruments \citep[see also][who confirmed the agreement between instruments]{campbellwhite23}, ruling out relative calibration issues. The differences in offset velocity are clear when shorter time intervals are taken into account (Figs. \ref{fig:TWHya-GLS} and \ref{twhya_intervals}).
A change in the offset velocity produced by a massive companion would affect all lines in the same way. Nevertheless, if the line-emitting region is not so simple or not locked to the stellar surface, things such as small changes in the infall velocity within a stratified post-shock region, or line-dependent variations in optical depth (so that the observed emission changes its location along the stratified region) could cause similar but line-dependent radial velocity variations.

We used the offset calculated over different periods of time to extract any further potential signature of radial velocity variation. For both EX Lupi and TW Hya, any remnant periodic signature has low-significance and is similar to the rotational period of the star, thus suggesting that it is a relic of the rotational modulation itself. This, plus the fact that not all the lines have the same offsets at any given time (see Fig.  \ref{twhya_intervals}), makes it unlikely that the variations are caused by a global change in stellar velocity (e.g. as induced by a companion), at least, down to the typical $\sim$0.1 km/s uncertainties (see Table \ref{twhya-mjdfit}). Therefore, the leading causes remain changes in optical depth and/or in the infall velocity within the stratified post-shock region, or changes in the stellar photosphere or spot structure(s), all of which can be rotationally-modulated. Changes due to optical depth effects that cause lines to arise from slightly different places have been documented in EX Lupi \citep{sicilia-aguilar_accretion_2015}.
Although one reasonable cause for these variations in the post-shock region could be a change in the accretion rate, we do not observe a clear correlation between the strength of accretion \citep[or line intensity,][]{fang_star_2009} and the changes in amplitude, phase, or offset velocity. Further variations on the global size of the line-emitting region, or near-but-distinct secondary spots with a variety of temperatures and densities, may be also behind this.

\subsection{TESS photometry analysis \label{tess-photo}}

\begin{figure}
    \centering
    \begin{tabular}{cc}
     {\large{\bf TW Hya}} & {\large{\bf EX Lupi}} \\
    \includegraphics[height=3.2cm]{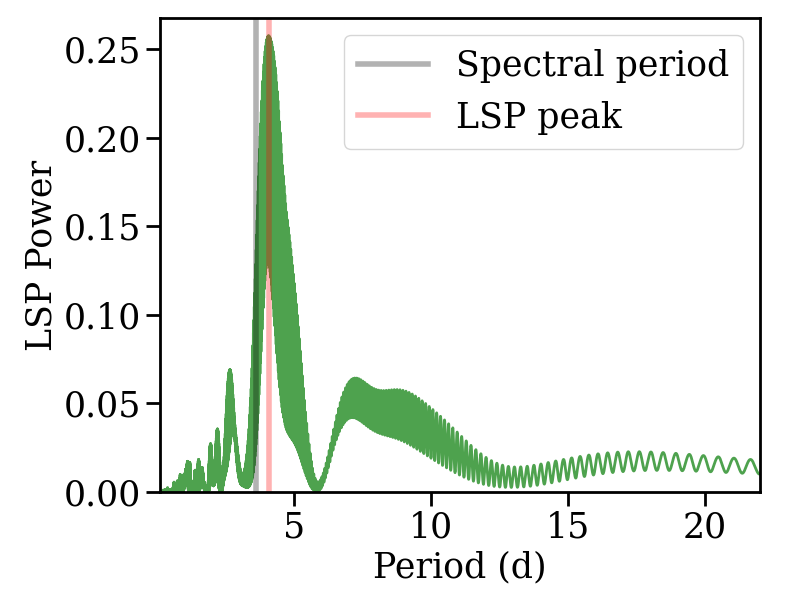} 
    &
     \includegraphics[height=3.2cm]{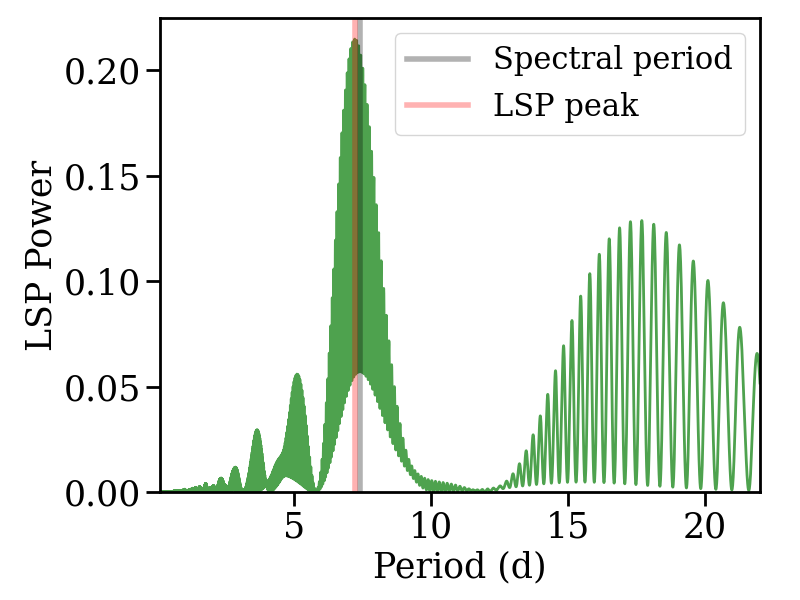} \\
     \end{tabular}
    \caption{LSP for the TESS data for TW Hya (left) and EX Lupi (right). The LSP peak and the period derived from the emission lines are both marked. The diagrams include the data from both sectors together. }
    \label{twhyaexlup-tess}
\end{figure}

Phase-folding of the TESS data confirms the rotational periods detected by the NC.
The lightcurve for TW Hya is very irregular. The best LSP period is 4.066~d (Fig.  \ref{twhyaexlup-tess}),  but phase-folding favours a 3.6~d period. If the two datasets are fitted separately, we find that sector 36 data has roughly the same  4.060~d period, while sector 9 has a 4.595~d period. The rotational period is more evident in the 2019 data, when the star was brighter by about 0.4 mag. The periodicity in Sector 36 is less clear and the maxima of sector 9 and sector 36 are off by nearly 180 degrees (for a 3.56~d period), as if the hot spot causing the excess luminosity moved to the other side of the star (see also Section \ref{discussion}).
These $\sim$4~d periods are reminiscent of those reported by \citet[][]{siwak14} and \citet{dupree_tw_2012} and likely linked to the same phenomena.

The TESS data for EX Lupi show the sinusoidal modulation expected for a source dominated by a single, stable spot, with a period essentially identical to the spectroscopic period \citep{sicilia-aguilar_accretion_2015,campbellwhite21}.
Sector 12 was discussed in \citet{campbellwhite21}; sector 39 data reinforces these results, although it also shows that, unlike the NC-related spot, the continuum-emitting region gets fully out of sight on certain epochs, producing a flattened sinusoidal. The LSP has a broad peak at 7.206d, and phase-folding the data is consistent with the spectroscopic period, 7.417d (Fig.  \ref{twhyaexlup-tess}). As observed by TESS, EX Lupi was on average a factor of 1.38 brighter ($\sim$0.35 mag) in 2021 than in 2019, likely related to an increase in the accretion rate that also seems to trigger more frequent, rapid small-scale bursting variations such as from accretion variability. 
Unlike in TW Hya, EX Lupi displays a cleaner modulation during the faint stage. EX Lupi gets brighter when the accretion rate increases \citep[e.g.][]{herbig08,juhasz12}, so its lightcurve is more irregular during higher accretion phases, maybe due to additional accretion-related spots or extended parts of the accretion structures becoming brighter. 

Subtraction of the sinusoidal modulation and further LSP analysis of the 2021 data produces a peak at 3.77d, which is essentially half of the rotational period and suggest that most of these changes are indeed related to a weaker secondary spot.
Spots with variable intensity on opposite sides of a star are common, and their relative intensity may be linked to increases in the accretion rate \citep[e.g.][]{sicilia20j1604}. A second spot on the same visible hemisphere could also explain the changes observed in TW Hya, where the phase change between the two TESS epochs may result from the dominant spot in the first epoch switching to be the secondary one later on.

Finally, we detect several flares in the TESS data for EX Lupi, three of them in 2019, and at least one (and potentially a second) in 2021. Identifying flares in TW Hya is harder due to its irregular lightcurve, although we also see peaks consistent with flares. 
Analysis of the bursting behaviour reveals no significant periodicity for the bursts, but the timescales of the small brightness variations in EX Lupi are of the order of 3h, while the same small-scale spikes in TW Hya appear to last rather 5-6h. The bursts may be related to other spots from minor accretion events, which suggests that the lifetime of these events in TW Hya (the fastest rotator and the oldest star) is longer. 

\subsection{Analysis of the broad-band photometry \label{bbphoto-sect}}

\subsubsection{EX Lupi: stable hot spots that grow and shrink}

\begin{figure}
    \centering
    \includegraphics[width=9cm]{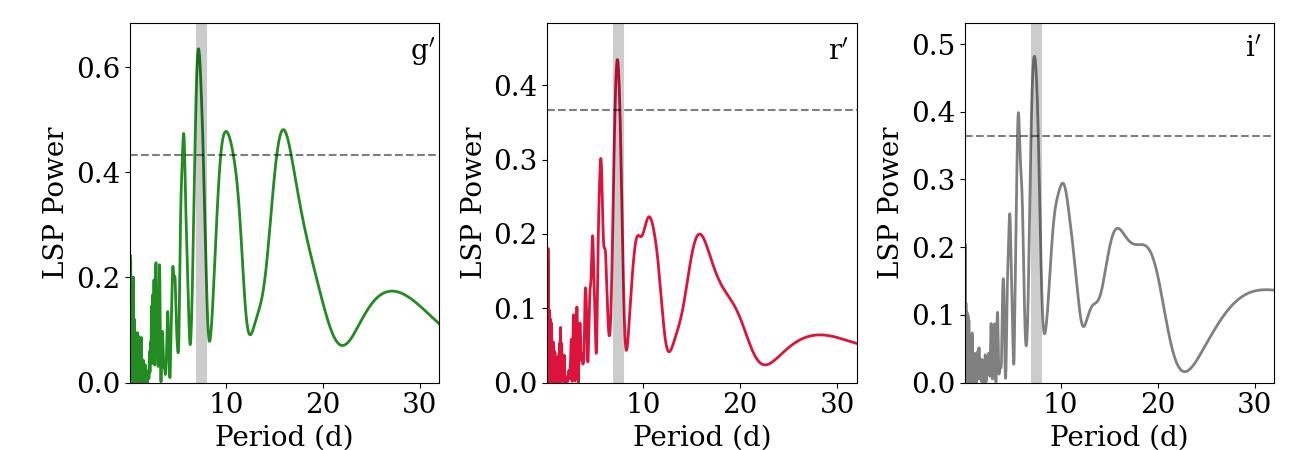}
    \caption{Periodograms for the broad-band photometry (g', r', and i' filters) in EX Lupi, showing the 1 $10^{-7}$ FAP level for white noise (see text). }
    \label{exlup-photoper}
\end{figure}

EX Lupi photometry is rotationally modulated on virtually every epoch observed, including all AAVSO bands, obtained during the rise and fade-off phase of an intermediate-magnitude outburst. The LCOGT data are consistent with the spectroscopy period of 7.417d. The LSP for the filters for which there is a reasonable number of datapoints have peaks at 7.167~d (g'), 7.324~d (r') and 7.281~d (i') with a virtually nil FAP. Phase-folding confirms the results, which are also similar to \citet{cruz23}.

The photometry data were used to explore the colour variations associated with the modulation.
The change in colour during quiescence (LCOGT data) is very small (Fig.  \ref{lcogt-curve}), but there is some evidence of rotational modulation of the colour for the 7.417d period, especially in g'-r'. The rotational modulation is much weaker in the r'-i' data, so that the structure causing it rather affects the bluer colours (see Fig. \ref{exlup-phase}). The AAVSO data also shows that the star is at its bluest (and brightest) during the peak of the outburst, suggesting it is related to hot spots.

To quantify the modulation, we used a model with a single spot. As in \citet{sicilia-aguilar_2014-2017_2017}, the simple model assumes the spot irradiates like a black body and compares the variations expected when the spot at maximum and minimum visibility in each band (BVRI for the AAVSO data, g'r'i' for the LCOGT data), deriving the best-fit in terms of spot temperature and coverage. This is also similar to \citet{froebrich22}. We use a 3800K effective temperature from \citet[][]{sipos09} to represent the photosphere. For the quiescence data (LCOGT and the last part of the AAVSO data) we checked for both hot and cold spots, while for the high luminosity state only hot spots are feasible. Spot temperatures between 2700 K and 10000 K were considered, as well as spot coverage from 1\% to 90\%. 
Since the model only focuses on the colour change, what we measure is just the part of the spot that comes in and out of sight and, if part of the spot is visible all the time, this would not be taken into account. For EX Lupi moderate inclination, and if we consider the NC-emitting region, it is likely that part of the continuum-emitting region is visible most of the time and thus the spot coverage obtained by our model is a lower limit. A spot coverage over 50\% means that most of the visible surface of the star is covered by \textit{spot} rather than by photosphere, but we wanted to explore this possibility.

For the model, the data were binned in 15 equal intervals within a full turn (large enough to be representative of the typical magnitude and noise, and small enough to minimize the sinusoidal variation as the star rotates) to estimate the average magnitudes in each filter. This allows to reduce noise and rapid photometric changes not related to the spot. The standard deviation within each bin is taken as the uncertainty. The maximum and minimum magnitudes were taken to be the maximum and minimum of the averages over the 15 bins, and considered as caused by the changes in spot visibility. The lowest $\chi^2$ provides the best-fit model, and the models within 3$\chi^2_{min}$ are used to derive the uncertainty.  The results of the simple spot model fit are shown in Table \ref{exlupitwhya-simplespot}. The minimum $\chi^2$ is well-defined, with the best models concentrating around the same part of the parameter space (see Fig. \ref{spot-params}). The minimum $\chi^2$ is also lower than expected even for a very good fit, which indicates that the standard deviation overestimates what is noise, likely due to other causes of variability beyond the spot.

\begin{table}
    \centering
    \begin{tabular}{ccccl}
    \hline
       <MJD> & $\chi^2_{best}$ & T$_{spot}$ [3$\chi^2$] & f$_{spot}$ [3$\chi^2$] & Notes\\
       (d) & & (K) & & \\
       \hline
       \hline
       {\bf EX Lupi} &&&\\
       \hline
     2457219 & 1.21 & 2700 [2700-2910] & 0.20 [0.19-0.25] & 15 Q (cold) \\
     2457219 & 0.25  & 4000 [4000-4225] & 0.58 [0.44-0.63] & 15 Q (hot) \\
     \hline
     2459661 & 1.45 & 4700 [4700-4776] & 0.88 [0.80-0.88] & 22 O\\
     2459675 & 0.31 & 4400 [4400-4477] & 0.81 [0.70-0.83] & 22 O \\
     2459688 & 0.05 & 4500 [4420-4600] & 0.35 [0.30-0.42] & 22 O \\
     2459704 & 0.17 & 5200 [5120-5630] & 0.20 [0.18-0.23] & 22 O \\
     2459716 & 0.91 & 5200 [5070-5200] & 0.11 [0.11-0.15] & 22 O\\
     \hline
     \hline
      {\bf TW Hya} &&&\\
       \hline
       2459668 & 0.02 & 6350 [6240-6550] & 0.04 [0.03-0.05] & Full dataset \\
       2459657 & 0.10 & 6400 [6190-6630] & 0.06 [0.05-0.07] & First half \\
       2459674 & 0.15 & 5800 [5620-5990] & 0.07 [0.06-0.08] & Second half \\
\hline
    \end{tabular}
    \caption{Temperature and spot coverage vs time for the simple spot model for EX Lupi and TW Hya. The temperature and coverage range for fits with a $\chi^2$ up to 3$\times$ that of the best fit is also indicated. The notes indicate the year (2015 or 2022), the quiescence/outburst status (Q/O), and whether the spot is cold or hot (only relevant for the quiescence status, see text). The cold spot model gives a worse fit for EX Lupi even during quiescence, which suggests that the hot spot is always dominating the photometric variation. For TW Hya, we only consider the epochs with sinusoidal photometry variation and hot spot models, since cold spot ones give much larger $\chi^2$. }
    \label{exlupitwhya-simplespot}
\end{table}

We find that all the data are consistent with a hot spot with a moderate temperature. The spot temperature is only slightly higher than the photosphere ($\sim$4000-5600 K), but we find that an extraordinary large coverage is needed to reproduce the luminosity, especially during outburst, suggesting that the spots that produce the photometric variability extend beyond the stellar surface.
This situation is similar to the 2008 large outburst \citep{juhasz12},  where the luminosity variation was attributed to a single-temperature spot at 6500 K dominating 80-100\%  of the emission \citep{juhasz12}.  The 2022 EX Lupi outburst temperature is significantly lower, which is reasonable for a weaker burst. It also agrees with the ASASSN-13db 2014 outburst, dominated by a 5800 K large spot not constrained to the stellar surface \citep{sicilia-aguilar_2014-2017_2017}.

The AAVSO data is particularly interesting to explore the evolution of a burst.  The object was decreasing in luminosity very rapidly, so we divided the available data in three epochs, each one corresponding roughly to two rotational periods. Surprisingly, although the temperature varies slowly, the spot coverage goes from 88\% to 11\%, meaning that a very large spot rapidly decreases in size within a few rotations, as the outburst fades. The size decreases throughout the burst, but the temperature increases towards the end. This suggests that, during outbursts, we see a much larger (but still rotationally-modulated) luminous part around the star, which is later replaced by a small, compact spot. Although it could seem contradictory, the increase in temperature when the accretion rate is lower suggests that a region hotter than what dominates the continuum excess during burst, but smaller, becomes relatively more prominent as the burst component rapidly shrinks in size. This fast evolution is consistent with the observed fading of spectroscopic signatures of accretion outbursts, which can disappear within days \citep[e.g.][]{holoien14}.

Repeating the analysis for the LCOGT data, which cover several periods in quiescence, the results are less evident. The magnitude variations also favour a hot spot, but the difference in significance is lower. The fits are also worse, probably because they are derived from worse sampled data covering a longer time span. The results are also listed in Table \ref{exlupitwhya-simplespot}, with the parameter space shown in  Fig. \ref{spot-params}, for comparison.

The single temperature of the model is a limitation, but the relative variation in different bands is very sensitive to temperature and the fit is reasonable. Although there are much hotter parts in the post-shock region, such as those producing the He II and Fe II NC emission, these are not photometrically observed due to their size or lack of continuum emission. Note that even the hottest spot temperatures that we measure are too low compared with the temperatures required for line emission, indicating that the region that emits the NC is significantly smaller and/or lacks continuum emission. The filling factors derived, especially in outburst, are quite large, and the temperature differences between the spot and the photosphere are not very high, which is consistent with what has been observed in other variable objects \citep[e.g., ][]{bozhinova16,froebrich22,herbert23}.

\subsubsection{TW Hya: The shape-shifter}

\begin{figure*}
    \centering
    \includegraphics[width=15truecm]{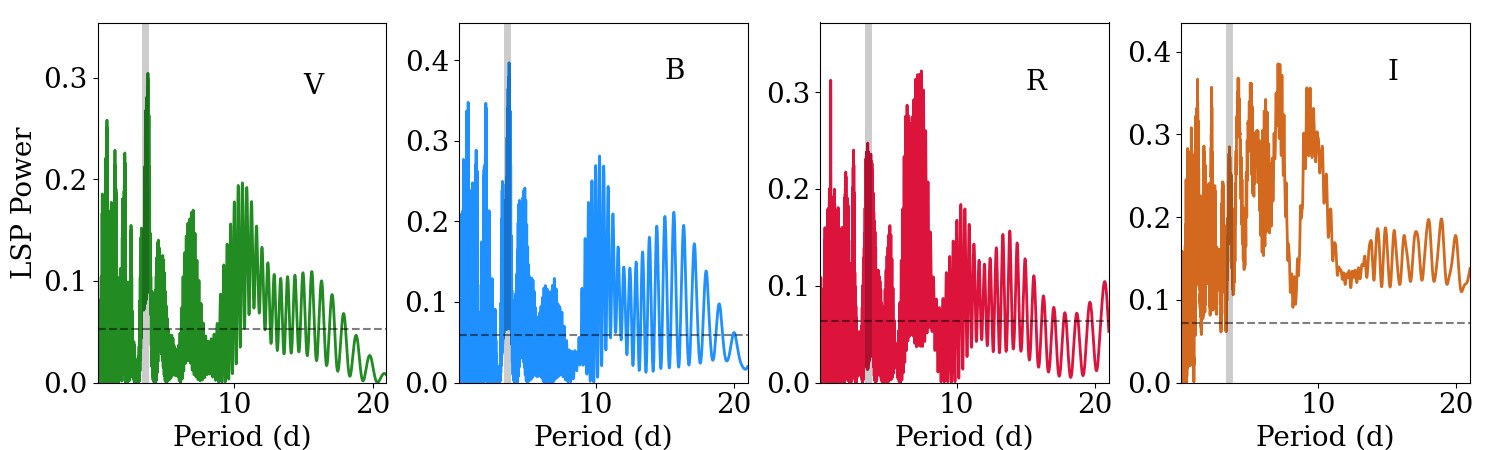}
    \caption{LSP for TW Hya in the various Johnsons filters as observed by the AAVSO. The grey line marks the region consistent with the rotational period derived from the lines, 3.556 d. The horizontal line marks a FAP of 10e-7 as derived by the Baluev method. }
    \label{aavso-periods}
\end{figure*}

For TW Hya, we used the LSP to derive photometric periods from the AAVSO data. This was done for each band separately, and the results are shown in Fig. \ref{aavso-periods}. Due to red noise, the periods have low significance, but we detect a peak around 3.7d in B and V. Both the phase-folded photometry and colour are also consistent with the spectral line period of 3.56d. 
As mentioned in Section \ref{sourceprops}, the rotational signatures of TW Hya are particularly hard to infer from photometry due to the bursting nature of the curve. Interestingly, the photometric periods derived with the TESS and 
AAVSO photometry and those in the literature \citep{dupree_tw_2012,siwak14} are all slightly longer (3.7-4.1d) than the period derived from the lines, which is the opposite to what we observe in EX Lupi. This suggests that the structure producing the photometric variations in TW Hya tends to lag behind the location of the emission-line spot and is more irregular, to the point that the rotational modulation is often absent.

The colour variations (Figs. \ref{twhyaexlupi-colors} and \ref{twhya-phase}) could be consistent with extinction, but are more likely related to spots since the ligthcurve and phase-folded colours do not show dipper-like events and simple spot models can reproduce the variations. When the sinusoidal modulation is visible, the amplitude of the variation is stronger in the bluer bands, albeit not as extreme as in EX Lupi. As observed with TESS, the modulation is cleaner when the star is brighter, from which we can infer that a single hot spot dominates, causing stronger modulations, during enhanced accretion phases. 

Modelling the photometry of TW Hya with a simple spot model is in general not possible, except for the epoch around JD=2459640-2459680 d, when the object shows the expected sinusoidal curve. We fitted a spot model for the 2459641-2459666 d epoch, another for 245967-245980 d, as well as another for the entire dataset, to test for uncertainties caused by the structure variability in a few-periods timescale. The data were split simply to have a similar number of observations in each set. The results are given in Table \ref{exlupitwhya-simplespot}. Although the additional, non-spot variability is quite large (which also causes the estimated $\chi^2$ to be very small), one important difference with EX Lupi is that the temperature is significantly higher than the stellar photosphere, and the spot coverage is always very small. As for the case of EX Lupi, the results for a cold spot model are significantly worse, with minimum $\chi^2$ over an order of magnitude larger than for the hot spot models, thus ruling them out. Given the low inclination angle, the spot sizes estimated are also lower limits since they do not include parts of the spot that remain always visible.

\subsection{EX Lupi: BC spectroscopy analysis \label{exlupBC-sect}}

The strongest lines display BC during epochs of increased accretion, to the extent that essentially all lines have BC during large accretion outbursts \citep{lehmann95,herbig07,kospal08,aspin10,sicilia-aguilar_optical_2012}. Accretion rates a factor of few to one order of magnitude over the quiescence rate \citep[\.{M}$\sim 10^{-9}$M$_\odot$/yr;][]{sicilia-aguilar_accretion_2015,alcala17} are enough for the BC to appear around strong lines.  Day-to-day velocity variations in the BC are observed both during outburst and quiescence \citep{sicilia-aguilar_optical_2012,sicilia-aguilar_accretion_2015}.

\begin{figure}
    \centering
    \includegraphics[height=6.5cm]{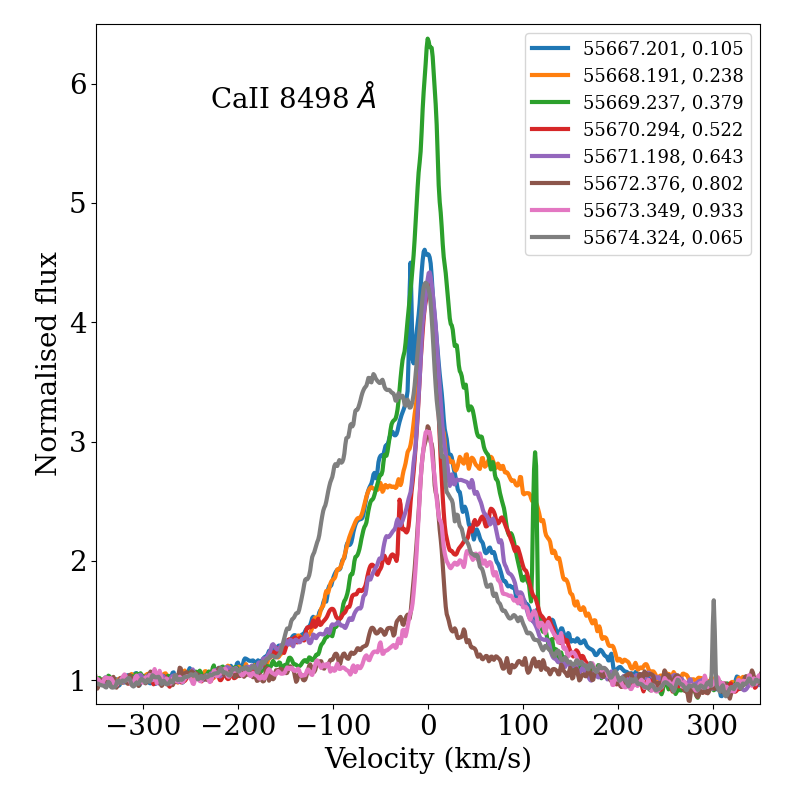} \\
     \includegraphics[height=6.5cm]{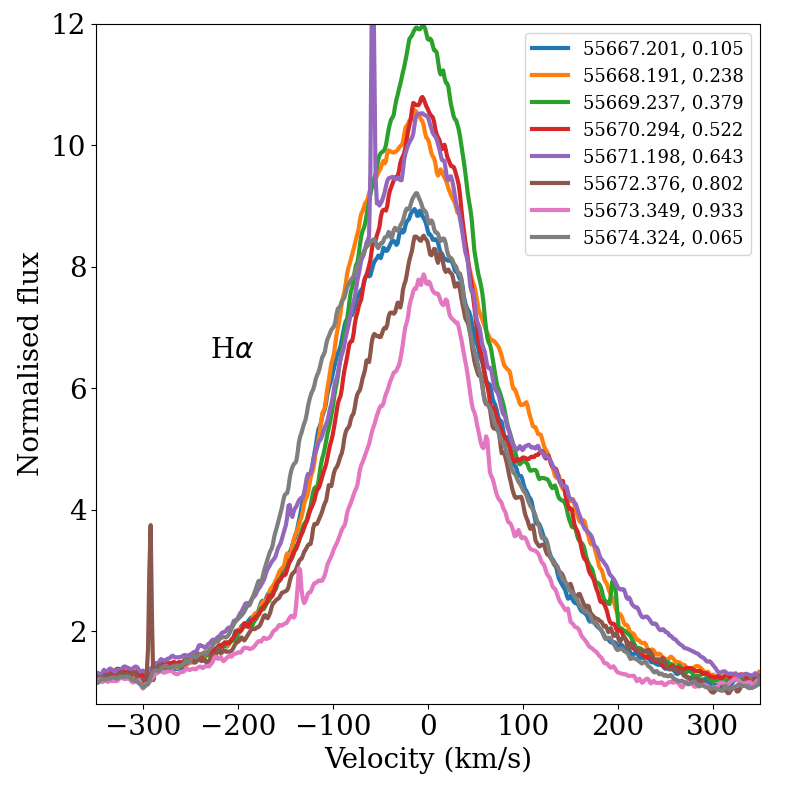} \\
    \caption{Example of the day-to-day variation in the Ca~II 8498 \AA, and H$\alpha$ 
    lines  for EX Lupi  during one of the increased accretion episodes during the observed epochs. The phase is given relative to the arbitrary date MJD=54309.1147d to match the modulation of the photometry and NC. }
    \label{CaIIvary-together}
\end{figure}

The variation of the BC line profile in quiescence is reminiscent of what was observed for the neutral metallic lines during outburst \citep[][see also Fig.  \ref{CaIIvary-together}]{sicilia-aguilar_optical_2012}. 
The change from a blueshifted component to a centred one (with red and blue sides) and then to a redshifted component and back hints towards something non-axisymmetric rotating around the star. 
The BC varies in shape and strength more than the NC, getting stronger when the accretion rate increases and sinking into the photosphere at lower rates, so that connecting data distant in time can be difficult. 
We concentrated the analysis on the Ca~II~IR triplet lines, which have the most prominent BC and are detected in a larger number of observations. Other strong lines with BC display a similar behaviour (e.g. Fe~II 5018/4923\AA), but their BC are weaker and often lack enough S/N.

\begin{figure}
    \centering
    \includegraphics[width=7.5cm]{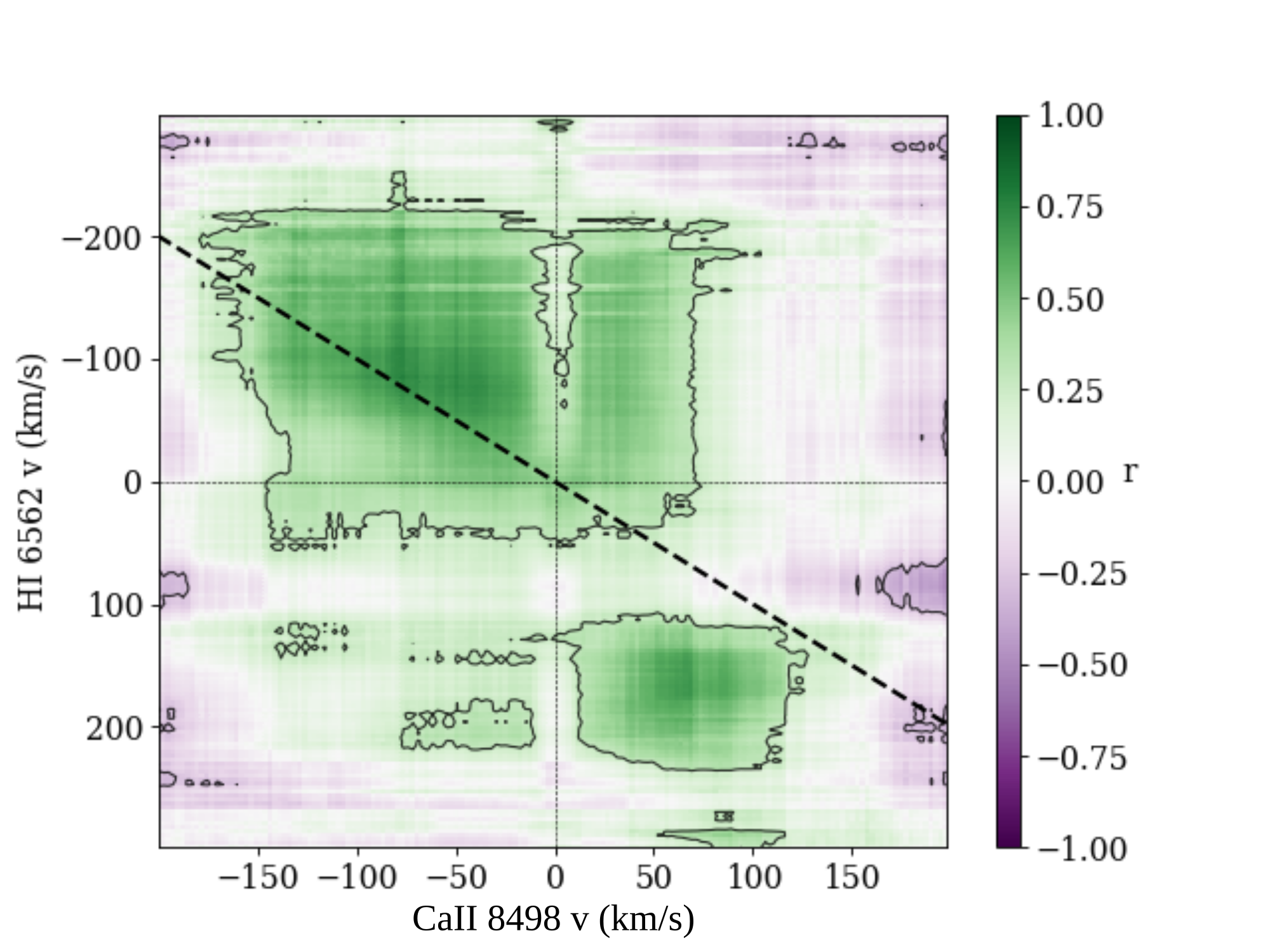}
    \includegraphics[width=7.5cm]{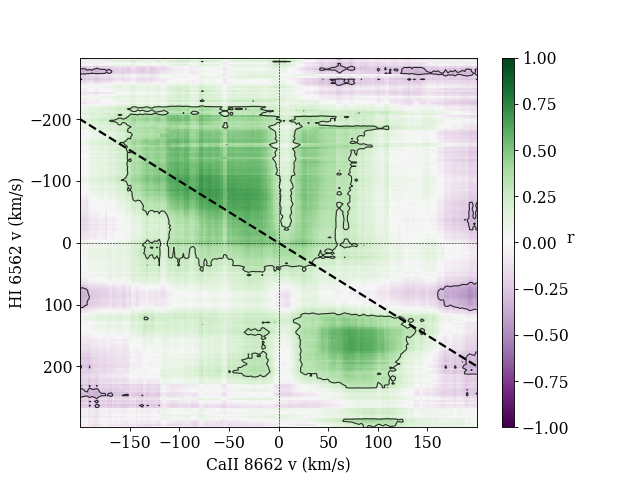}
    \caption{Cross-correlation between H$\alpha$ and two of the Ca II IR lines for EX Lupi during periods of increased accretion with visible broad components. Regions where the correlation is significant are surrounded by black contours, and the colour scale on the left indicates the positive or negative correlation coefficient.The data are smoothed by a 7-pixel both to reduce the noise.}
    \label{EXLupi-xcorr}
\end{figure}

We started by checking if the BC velocity variation was due to moving emission components, or to moving absorption components changing over a relatively stable emission.
For instance, accretion columns seen along the line-of-sight can show inverse P Cygni profiles \citep[e.g.][]{batalha01}, and blueshifted absorptions from variable, non-axisymmetric winds have been observed in young stars \citep[e.g.][]{sicilia-aguilar_2014-2017_2017,sicilia20zcma}.
Comparing with the metallic lines, we noted that the H~I lines show the inverse behaviour: when the BC of the metallic lines is redshifted, we observe a redshifted absorption in the lower Balmer lines at the same velocity. This is as expected from a gas parcel moving away (redshifted) along the line-of-sight, that is dense (seen in absorption) in H I but appears in emission in less abundant species. Although this effect is not as evident when the BC is blueshifted (probably due to geometry and being on the far side), we can identify simultaneous blueshifted absorption in the H~I lines, so that the variations are similar to those observed in non-axisymmetric rotating accretion columns \citep[e.g.][]{alencar_accretion_2012,kurosawa_spectral_2013}. There is also a positive correlation between the width and strength of the Ca~II and H~I lines (see Fig. \ref{EXLupi-xcorr}), although this is dominated by the lines getting broader with stronger accretion. Therefore, we conclude that what we see as redshifted or blueshifted BC in the metallic lines is in emission and not the result of a variable absorption, so we will explore scenarios characterised by emission from moving, dense matter parcels.

\subsubsection{STAR-MELT analysis of EX Lupi BC}

\begin{figure}
\begin{center}
{\bf \large{Line Velocity Analysis:}}\\
\begin{tabular}{cc}
{\bf Ca~II 8498\AA\ Full Period} &{\bf Ca~II 8498\AA\ Half Period} \\
\includegraphics[width=4.2cm]{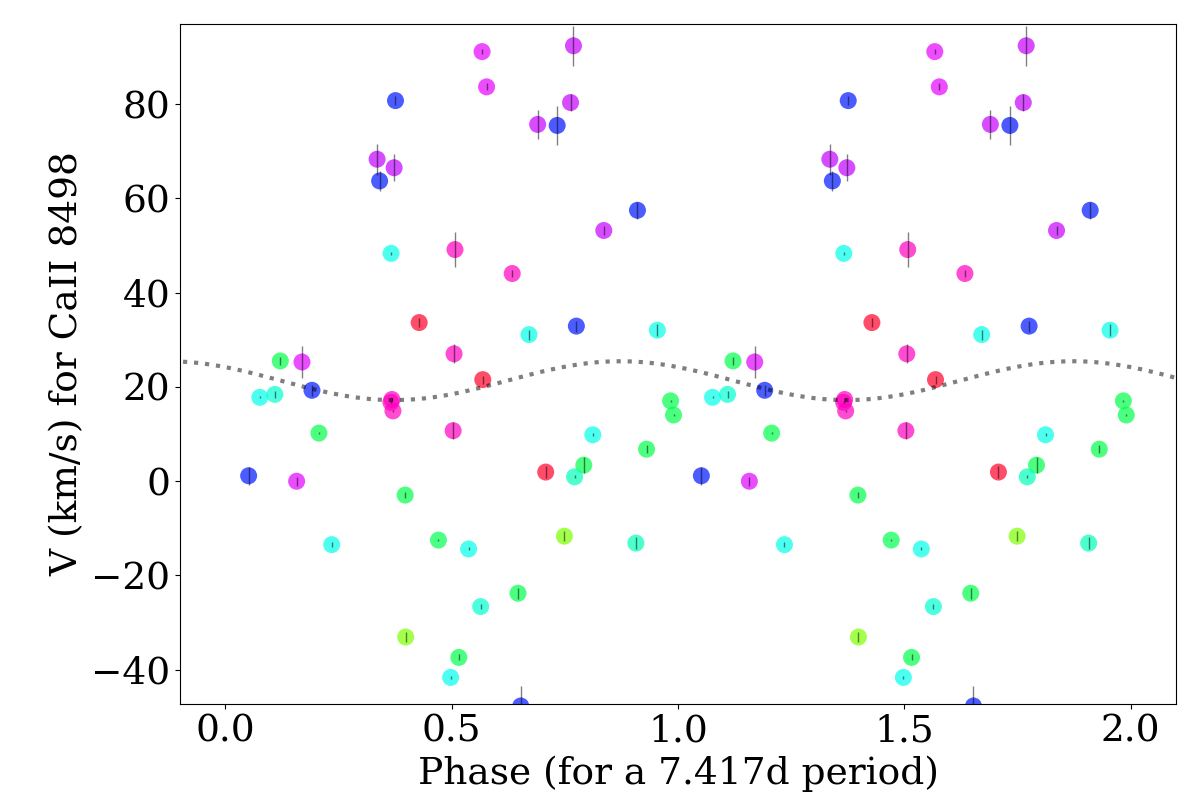} &
\includegraphics[width=4.2cm]{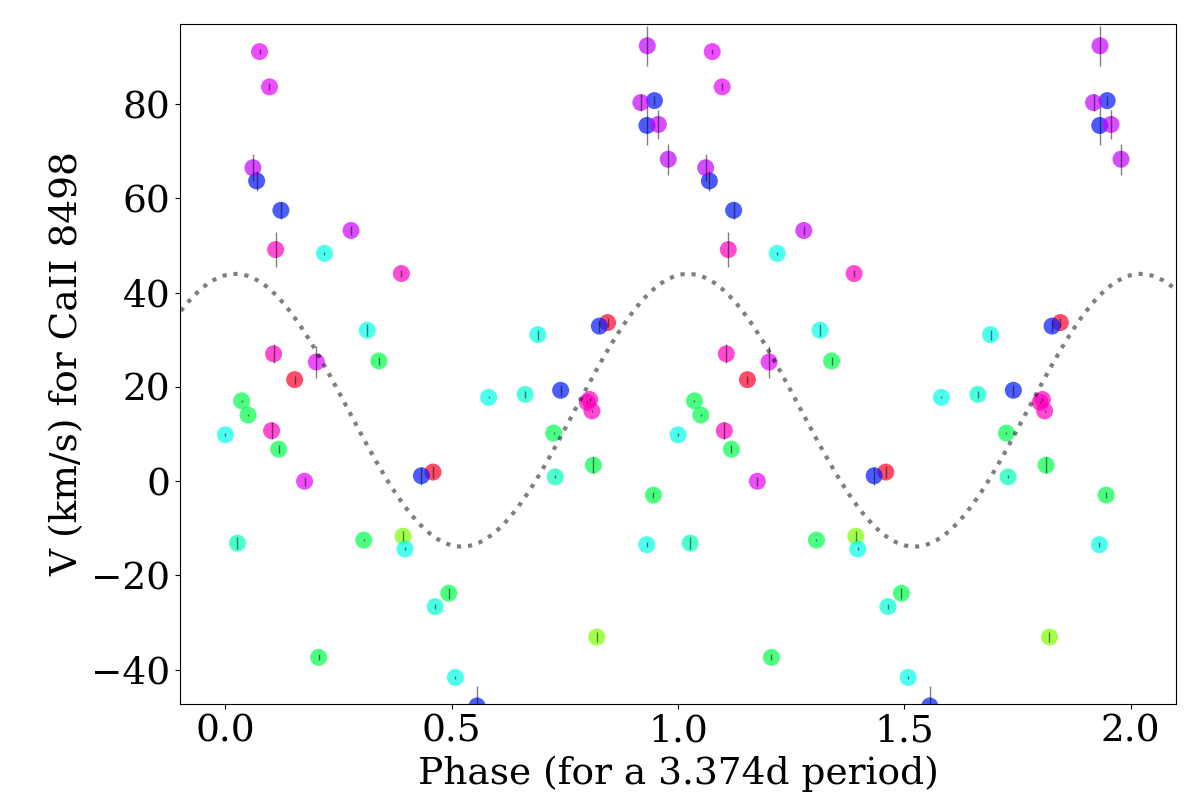} \\
 {\bf Ca~II 8662\AA\ Full Period} &{\bf Ca~II 8662\AA\ Half Period}\\
\includegraphics[width=4.2cm]{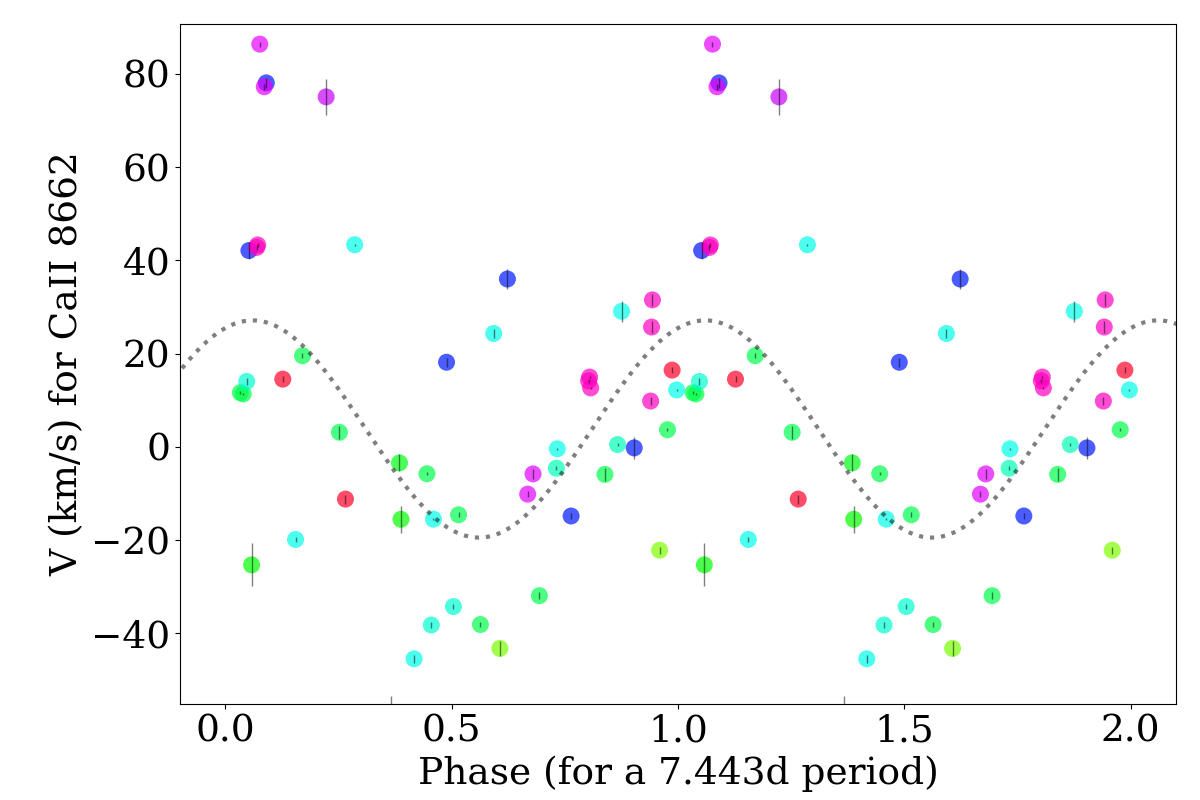} &
\includegraphics[width=4.2cm]{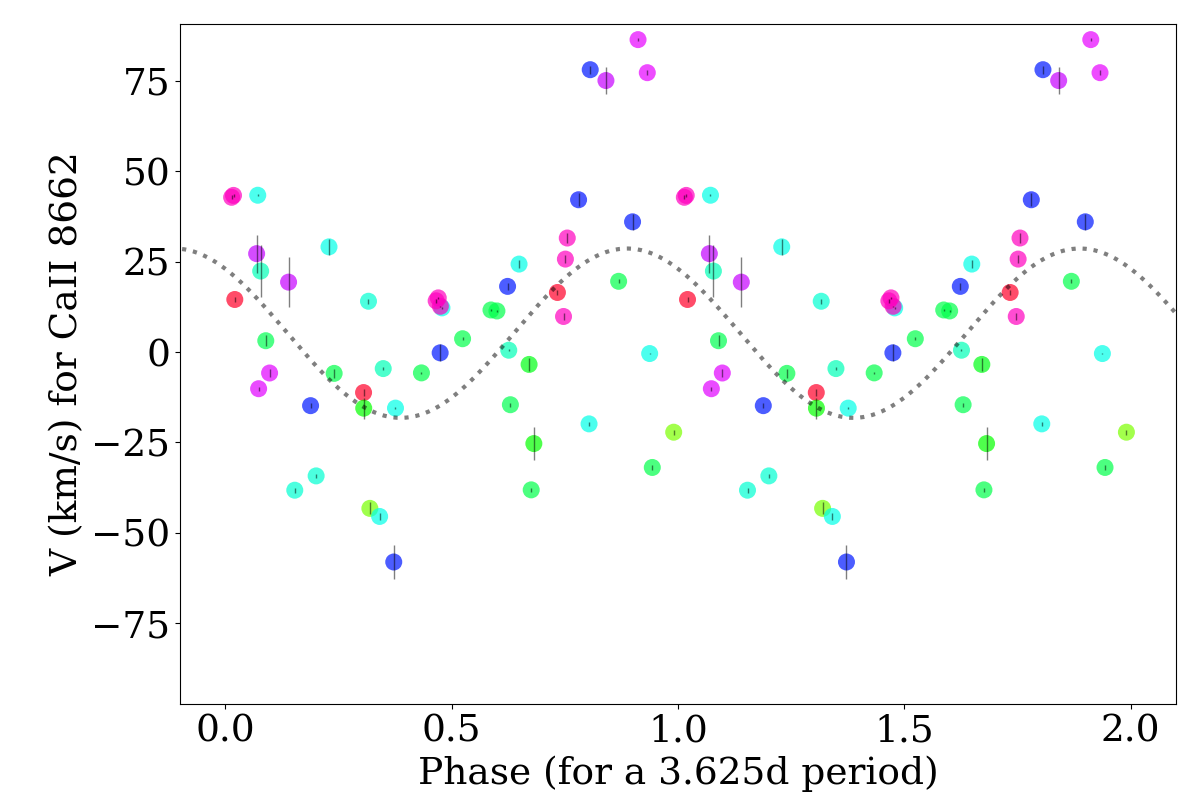} \\
\end{tabular}
\vskip 0.5truecm
{\bf \large{ Line Asymmetry Analysis:} } \\
\begin{tabular}{cc}
{\bf Ca~II 8498\AA} &  {\bf Ca~II 8662\AA} \\
\includegraphics[width=4.2cm]{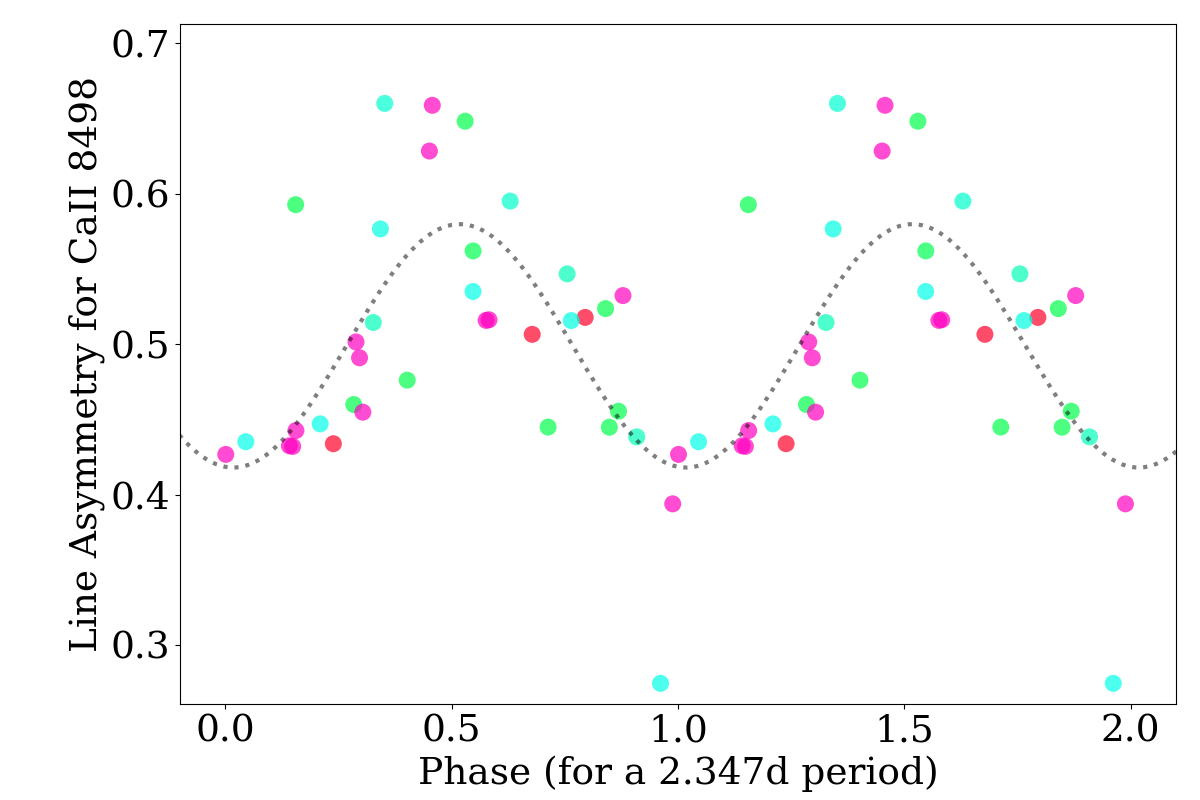} &
\includegraphics[width=4.2cm]{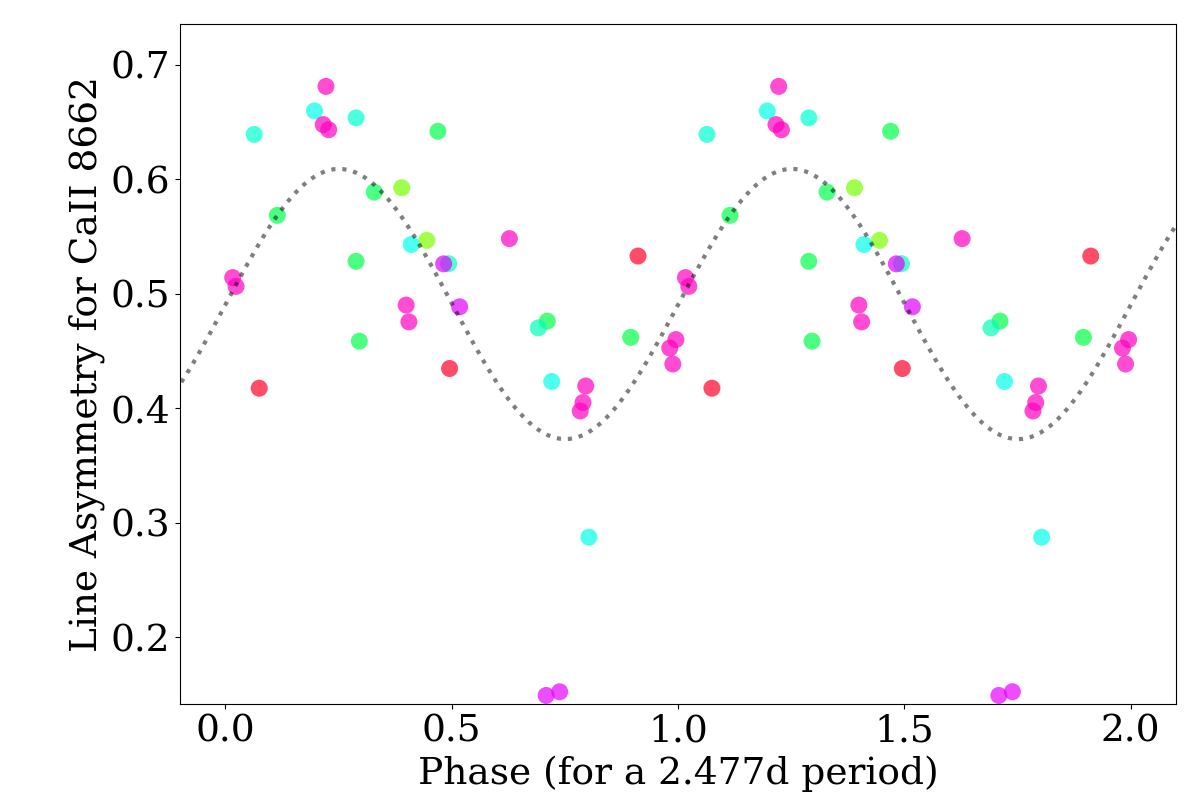} \\
\end{tabular}
\end{center}
\caption{Top panels: EX Lupi broad component phase-folded velocities for the BC of the Ca~II lines (top: 8498\AA\, line; middle: 8662\AA\, line). Note that both sets of BC are consistent with the rotational period, or half of the rotational period.
Bottom panels: Phase-folded line asymmetry.  Although there are some signatures of periodicity around the 7.4~d rotation, especially for Ca~II 8498\AA\, the main period observed is in the range of 2.3-2.4~d, especially, for the line asymmetry. The phase is given with respect to an arbitrary date, MJD=54309.1147d.
The uncertainties are in most cases smaller than the symbols. The best sinusoidal fit to the data is plotted as a dotted line, although the modulations are not sinusoidal.}
\label{EXLup-CaIIBC} 
\end{figure}

We first used STAR-MELT to examine the long-term behaviour of the BC in EX Lupi.
The data were fitted with STAR-MELT, including the BC and NC, using a multi-Gausian fit. Multi-Gaussian fits tend to be degenerated, so we added conditions on the NC \citep[restricting it to radial velocities of $\pm$ 10 km/s and FWHM$<$20 km/s, as done in][]{sicilia-aguilar_2014-2017_2017}. For the BC, velocities were allowed to be within $\pm$120 km/s, and the line width was constrained so that FWHM$>$50 km/s. Further requirements included a minimum GOF of 1 and amplitude over the continuum of at least 3$\sigma$. All fits were also examined by eye. 
The typical velocity for the Ca~II~IR BC is higher than what was observed for metallic lines in outburst \citep{sicilia-aguilar_optical_2012}, reaching up to $\pm$100 km/s, but the fit is hard to use at face value, since the BC shape differs from a combination of Gaussians.  Following \citet{sicilia-aguilar_2014-2017_2017}, we also used STAR-MELT to measure the asymmetry of the line  (ratio of the blue flux vs total line flux, measured from the centre of the line according to the radial velocity), a much more robust and non-degenerated way to characterise the lines. Figure \ref{EXLup-CaIIBC} shows the phase-folded plots.

A LSP reveals that both the velocity and line asymmetry have, if any, periods shorter than the stellar rotation. While using the velocities of each single redshifted or blueshifted Gaussian is weakly consistent with a half-period modulation, the line asymmetry has a stronger periodic modulation around 2.3-2.5 days (see Fig. \ref{EXLup-CaIIBC}). Visual inspection also reveals this faster periodicity in observations taken during the same rotational period (to minimise the effect of variable accretion), although there are only two relatively complete sequences. The lines get weaker (and mostly redshifted) at two phase ranges, 0.8-1.0 and 0.3-0.6 (Fig. \ref{CaIIvary-together}). This suggests either the presence of more than one emitting structure, a structure not locked to the star (different period), or additional effects, such distinct matter infall episodes. Viewing more than one structure, such as two accretion columns, is unlikely because the blue- and red-shifted components are very rarely seen at the same time. If the shorter period is caused by a gas parcel in Keplerian rotation, it would be located at around 4-5 R$_*$ (inside the corotation radius of 8.43R$_*$). If caused by distinct infall episodes, there would need to be some underlying periodicity for the events. We explore these possibilities in the next section.

\subsubsection{Models and origin of the BC and its modulation}

We used different models to find the simplest structure that can account for the BC modulation.
The BC must be produced in a non-axisymmetric region that is moving or rotating relatively fast, to produce the high velocities. A non-axisymmetric structure is also the natural match of the stable accretion footprint traced by the NC. We thus explored two main scenarios: a non-axisymmetric magnetospheric accretion column rotating as a solid body together with the star \citep[modifying the model of][to limit the azimuthal extension of the column]{hartmann94}, or a hot, non-axisymmetric part of the disk \citep[somewhat similar to the keplerian disk observed by][around ZCMa NW in outburst, but with a limited azimuthal angle]{sicilia20zcma}. A similar disk sector has also been proposed for FU Orionis \citep{siwak18fuor}.
For the first model, we assume that the column starts at the corotation radius, while for the second we assume a disk sector extending slightly inwards or outwards of the corotation radius. We also experimented with a combination accretion column and disk segment. In all cases, we probed azimuthal extensions between 60 and 120 degrees, since larger ones dilute the observed 'wobbling'. Note that our models aim to reproduce the velocity modulation, not the flux, which would require a much more complex treatment beyond the scope of this work.

For all models, we consider the stellar radius (R$_*$=1.6~R$_\odot$), mass (M$_*$=0.6~M$_\odot$), and start with a stellar inclination angle $\theta \sim 20^o$ (from inner disk modelling)  from \citet{sipos09}, as well as the corotation radius inferred from the 7.417d period (8.43~R$_*$). Our NC analysis also agrees with a low angle.
Nevertheless, an inclination of 20-35$^{\circ}$ is not enough to explain the line wing velocities, even if we include infall and rotation. In fact, the best match for the models was obtained for an angle around 45$^{\circ}$, suggesting that the innermost structures (or the star) are slightly inclined with respect to the outer disk.
Since the lines are too broad to be accounted for by thermal broadening alone, we also add an extra turbulent broadening factor of 5 times the thermal velocity.

\begin{figure}
    \centering
    \includegraphics[width=6.80cm]{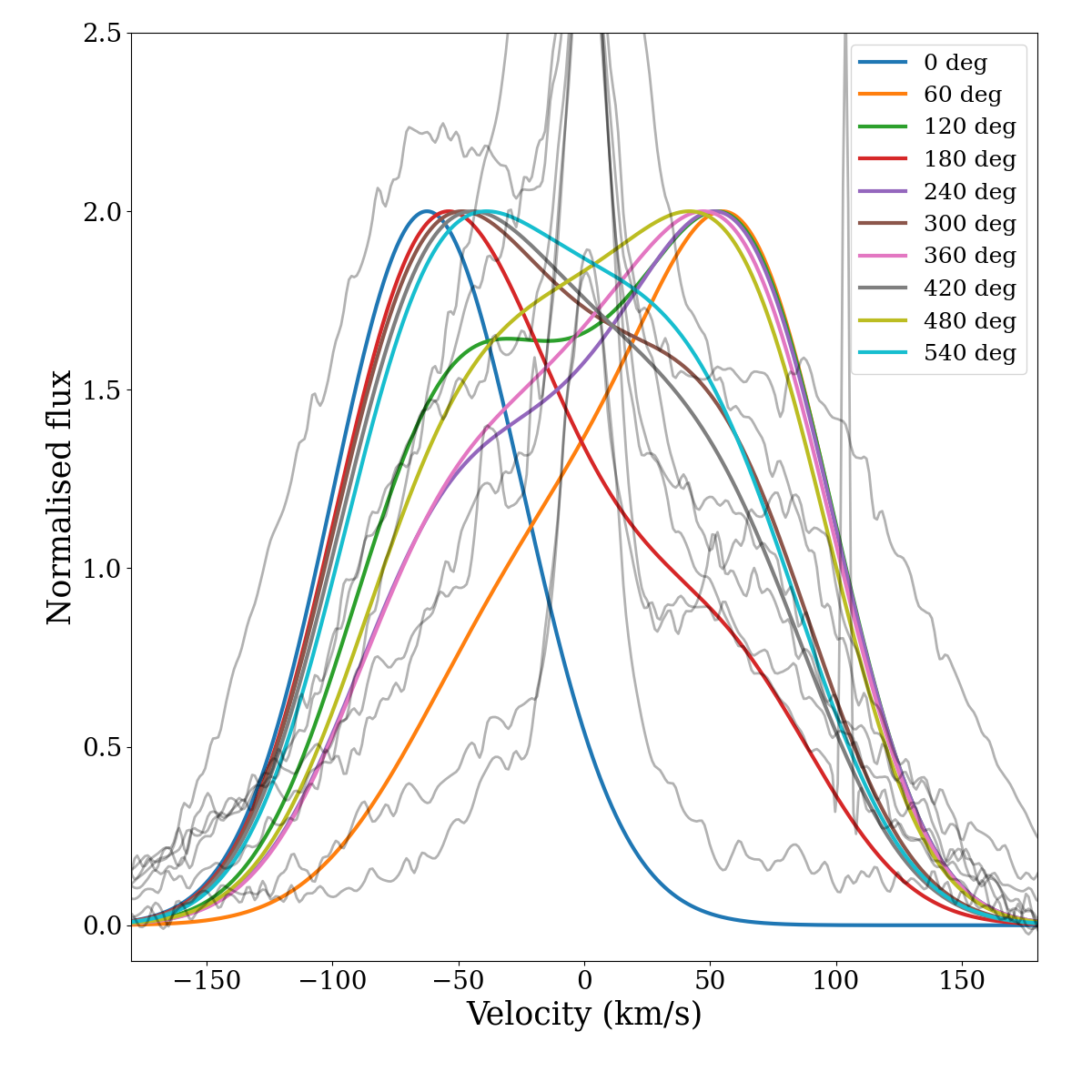}
    \includegraphics[width=6.8cm]{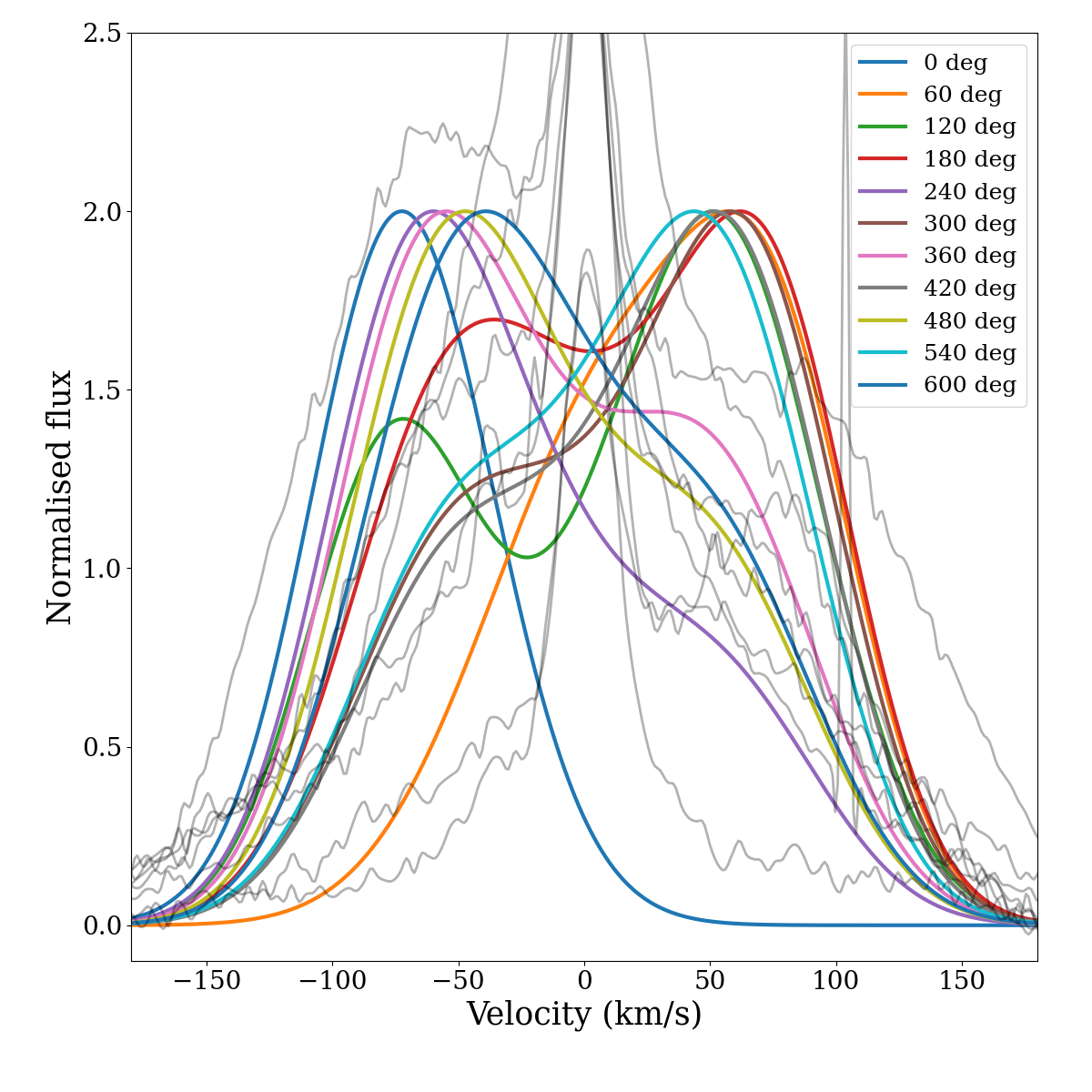}
    \caption{Models for a 120$^{\circ}$ (top) and 60$^{\circ}$ (bottom) sector of a Keplerian disk, extending from the corotation radius (R$_{cor}$) down to 0.6R$_{cor}$, compared with the Ca II 8662\AA\ line as observed during an episode of increased accretion between MJD=55667-55674 d. The rotation angles are given, and it can be observed how the shape of the disk changes in time as it spirals in faster in the inside, thus producing a blueshifted and redshifted maximum that changes sides faster than the rotational period.}
    \label{diskmodel2}
\end{figure}

For the accretion column model, we followed the magnetospheric structures and accretion velocities from \citet{hartmann94}, assuming that the accretion column starts at the corotation radius, R$_{cor}$, and limiting the azimuthal extension. In the magnetospheric accretion column, the material flows along the magnetic field lines, which are defined by
\begin{equation}
   r = r_{in} sin^2 \theta, 
\end{equation}
where r is the radial distance to the centre of the star, $r_{in}$ is the location in the disk where the material comes from (R$_{cor}$) and $\theta$ is the angle with the z-axis. The accretion velocities come from a simple magnetospheric model where the material leaving the disk follows the field lines essentially in free-fall, landing in a position that depends on the radius where the magnetic line originated \citep{ghosh77,hartmann94}.
 
We consider a single magnetically-channelled column locked at the corotation radius, which lands at $\sim$20$^{\circ}$ from the pole, in agreement with the typical ranges observed for any NC line, between 10-50$^{\circ}$ \citep{sicilia-aguilar_accretion_2015,campbellwhite21}. 
We consider a single temperature of 10,000 K for the magnetospheric column, a typical value for accretion models, noting that this has only limited effect on the observed velocity variation. The accretion column rotates with the star as a solid, while the disk part would be in Keplerian rotation.

For the non-axisymmetric disk part of the model, we started with the simple irradiated disk model from \citet{sicilia20zcma}, limiting the azimuthal extension of the disk.
Since the Keplerian period decreases towards the star (only the part at the corotation radius moves together with the star), a disk sector winds up and evolves in shape with each rotation, which makes the line profile very variable. The temperature in the rotating Keplerian disk is achieved by irradiation, decreasing from this maximum value down as $R^{-3/4}$ \citep{dalessio99} but, if the emitting structure is not very extended, the temperature variation is minimal and the choice of temperature structure is quite irrelevant.

The first result from these models is that the BC emission cannot originate in a magnetospheric accretion column. The main problem is that, given the inclination of the system, it produces a systematically blueshifted component due to infall, which would be visible at any time during a rotational period but is not detected. On the other hand, a Keplerian disk winding up towards the inside of the corotation radius can reproduce the basic features of the BC, as well as the messy and faster-than-rotational-period modulation.
Figure \ref{diskmodel2} shows the results for the disk models. A non-axisymmetric, winding, slightly extended ($\sim$5-8.4 R$_*$) disk is thus the simplest model that can explain the observations. This does not exclude by any means magnetospheric accretion onto the source, but suggests that the BC emission observed is produced in a hot, non-axisymmetric sector of a  disk. 

It is worth noting that a similar non-axisymmetric or azimuthally limited disk component was also identified in IR CO lines during the 2008 outburst \citep{goto11}. The CO observations revealed a rapidly-changing component at around 0.04-0.06 au, which is consistent with a structure inwards of the corotation radius (5.4-8 R$_*$).
The non-axisymmetric disk model can also explain what is observed in the optical emission lines during the outburst, as long as the inner disk is non-axisymmetric far beyond the corotation radius. The outburst spectra revealed a variable BC in most of the lines, shifting between the blue and the red, but with lower velocities than observed here for the quiescence Ca~II BC \citep[][]{sicilia-aguilar_optical_2012}. Assuming that the maximum shift was due to Keplerian rotation, the estimated distance of this spiraling-in structure was $\sim$0.1-0.2au \citep[$\sim$13-26 R$*$;][]{sicilia-aguilar_optical_2012}, extending well beyond R$_{cor}$. 
Disk emission, albeit from a complete disk, has been seen in around ZCMa NW, when the disk is heated by an accretion burst \citep{sicilia20zcma}.  A disk scenario also naturally explains why the velocity offsets are smaller in outburst than in quiescence: In outburst, more distant parts of the disk, with lower Keplerian velocities, can be heated and produce BC emission. As the accretion rate drops, the disk cools down and the BC-emitting region of the disk moves inwards. The available data are also consistent with more energetic lines (e.g. Fe~II) having larger BC velocities, again in agreement with a disk heated by irradiation, although due to the SN and the changes from period to period this is not easy to quantify.

To conclude, although there are not so evident BC in the spectra of TW Hya, inner disk rotating material is also thought to affect the photometry \citep{siwak11} and thus the global structure may not be so different in both objects.

\section{Discussion: Unveiling the structure of accretion columns}\label{discussion}

\begin{table*}
    \centering
%    \begin{footnotesize}
    \resizebox{\textwidth}{!}{%
    \begin{tabular}{lccccccc}
    \hline
Object & TESS  & TESS  & LCOGT   & LCOGT colour & AAVSO  & AAVSO colour  & Lines \\
 & (2019) & (2021) & (g'/r'/i') & (g'-r'/g'-i'/r'-i') & (B/V/R/I) & (B-V/V-R/V-I/R-I) & \\
\hline
\hline
EX Lupi &  0.14	& 0.17	& 0.64/0.61/0.6: & 0.7/0.65:/0.2: & 0.83-0.74/0.82-0.74/0.88-0.77/0.85-0.89	& 0.67-0.75/0.73-0.75/0.69-0.71/0.77-0.7:	&   0.60 \\
TW Hya &  0.5:	& --	& --	& --	&  0.79/0.74/0.78/0.74	& 0.8/0.8/0.7:/--	& 0.96$^a$/0.94$^b$	  \\
\hline
\end{tabular}
}
%\end{footnotesize}
\caption{Summary of the phases observed in the photometry and lines. The phases are given considering the maximum photometric brightness, the bluest colour, and the maximum line blueshift, respectively, as discussed in the text. They are all given considering the maximum blueshifted velocity (for the lines) and the minimum magnitude (or colour) for the photometry and phase-folded with respect to the arbitrary JD 2454309.6147 (for EX Lupi) and 2458543.827644~d (for TW Hya). The values are measured from the plots, and the uncertainties are of the order of the last digit given. When the modulation is observed to change over time, we give the entire range. Particularly uncertain values are marked with ":". $^a$ Phase for Fe II 5018\AA\ line. $^b$ Phase for He I 5016\AA\ line.}
\label{tab:phases}
\end{table*}

We now use the information obtained from the phases and periods from the NC line velocity, the photometric magnitudes, and colour variations, to reconstruct a picture of the environment around both stars. Several structures are needed to account for the range of temperatures and variability patterns. Lines such as He II and Fe II require very hot parts of the post-shock region, while the photometric modulations can be explained by regions typically 500-2000 K hotter than the photosphere.
Therefore, we have a \textit{continuum spot} (or rather, feature) that affects the photometry, and a \textit{line spot} where NC lines originate, which may or may not overlap.
The NC-emitting regions need to be much smaller in size and/or do not produce a significant continuum, or they would rapidly increase the temperature estimated by photometry.

The photometry could also be affected by other structures, such as circumstellar extinction events, which may be also correlated with the passage of an accretion column \citep[e.g. as in][]{sicilia20j1604}. Examining the colour-magnitude diagrams and lightcurves for both objects (Fig. \ref{twhyaexlupi-colors}), the behaviour of EX Lupi is different from extinction and in agreement with an evolving hot spot that changes in area and temperature, producing a smooth, sinusoidal lightcurve, unlike typical extinction/dipper variability. TW Hya data lay closer to the slope expected from extinction, but is not fully consistent, especially when all colours are taken into account. We also observe some parallel trends, which are redder (and brighter) in TW Hya, and bluer and fainter in EX Lupi, compared to the most populated trends. They are more evident in the redder colours (V-I and R-I) and (for EX Lupi) towards the middle of the outburst.  A bluer and fainter parallel trend, observed mostly in R vs R-I, could also be produced by scattering and extinction (as in UXor variables), but 
since both objects switch very rapidly between these two parallel trends, so it is more likely an spot visibility effect rather than a physical change in temperature/area.  Since the NC emission is visible at all times, drastic changes in spot visibility further points toward photometric modulations and NC lines arising from different locations.
 
In the general case of a simple, single spot, the NC lines achieve their maximum blueshifted velocity when the NC-emitting region moves into the hemisphere facing the observer, and the maximum redshifted velocity is attained as it turns into the other hemisphere.
In contrast, a star is photometrically brighter when a continuum hot spot is facing us, and a cold spot would have the opposite effect. In either case, the star is bluer when it is brighter. Therefore, if the \textit{line spot} and \textit{continuum feature} belong to the same simple, symmetric structure (e.g. a spot formed by a hotter core surrounded by extended warm regions), we would expect the star becoming brighter approximately at a phase 0.25 after the maximum blueshift of the line velocity is observed. Occultations of the NC-emitting region by an extended column could produce a shift as well, but the lines are always observed. For systems at low inclination, it is unlikely that the spot(s) in the visible hemisphere go totally out of sight. 

If, instead, the accretion structure has a hotter footprint (where the NC emission occurs) but the  continuum feature is trailing this footprint or is produced a different part of the system, phase offsets are expected. 
If we have a combination of spots with different temperatures, or a single spot with a complex shape and/or temperature distribution, offsets between the phase at which the star is brightest vs the time when it is bluest can also occur. Finally, note that line emission can significantly affect some photometric bands if the accretion rate is large \citep[e.g. R, see][]{juhasz12}. Table \ref{tab:phases} summarises the observed phase offsets that are used to track the location of the different structures. Our intention in this section is to discuss all these possible scenarios, keeping in mind their complexity.

\subsection{EX Lupi: stable structures, but very variable accretion}

We start with EX Lupi since it has a cleaner modulation. We observe small phase offsets between lines from different species, which were used to pinpoint the location of the line-emitting region for each species \citep{sicilia-aguilar_accretion_2015, campbellwhite21}, but individual species do not vary in phase significantly. Nevertheless, the photometry phase is not as stable, and the colour is phase-dependent, although the rotational period is always evident.

Figure \ref{exlup-phase} contains the phase-folded variations in magnitude and colour, as well as in He II 4686\AA\ velocity, taking as reference MJD=54309.1147d. Phase-folding over long timescales risks introducing spurious phase offsets if the period has a small uncertainty, but the phase of the NC velocity changes by very little over the nearly 12 years of data \citep{campbellwhite21}. The maximum NC blueshift occurs at $\phi \sim$0.6. In contrast, there is a clear change in phase over time for the magnitudes and colours, sometimes even for data spanning only a few rotational periods. The magnitude brightness peak varies greatly, being at $\phi\sim$0.14-0.17 for both TESS epochs, at $\phi\sim$0.6 for the LCOGT data (which is bluest at about $\phi\sim$0.2-0.7, mildly consistent with a simple hot spot), and $\phi\sim$0.8 for the AAVSO data for both brightest and bluest magnitude. This means that, although the photometry alone behaves as expected from a simple, hot \textit{spot}, compared to the stable velocity of the NC, the TESS data peaks about $\Delta\phi\sim$0.55 later, the LCOGT is essentially simultaneous, and the AAVSO data have a phase offset $\Delta\phi\sim$0.2-0.25. Therefore, only the NC and the AAVSO outburst data (which corresponds to a very large spot) are consistent with the simple, symmetric, single spot model ($\Delta\phi\sim$0.25). Therefore, we conclude that the photometric maximum is not related to the hot region that produces the NC, probably because there are less hot but more extended structures that contribute a bigger change to the photometry. 

The photometry phase shift is irregular, so the relative location of the hot continuum feature and the NC region is not symmetric - nor stable over time. Since the NC-emitting region changes very little, the \textit{continuum feature} is sometimes advanced, and sometimes trailing it. There is nevertheless a tendency for photometric periods to be shorter, particularly in the TESS and LCOGT photometry, which suggests that the continuum, photometry feature is typically catching up with and passing the line-emitting part and could be located slightly inwards of the corotation radius (e.g. at 8.27~R$_*$ or 0.98 R$_{cor}$, for a period of 7.2d instead of 7.417d, for instance).
The phase changes on consecutive rotations in the AAVSO data. This can be explained if a trailing structure rapidly shrinks as the burst fades: its phase would shift towards more advanced values, as observed. In contrast, the phase is stable during the quiescence LCOGT observations, which roughly cover the same number of periods. 

We thus conclude that the variations in the shape and location of the hot continuum feature that produces the photometric variability depend on the accretion rate, while the NC radial velocity signatures do not. The photometry feature may follow the accretion column at different radii, as we observe for the non-axisymmetric disk traced by the BC. The drastic phase changes, plus the visibility changes, together with its rapid changes and its large size in outburst compared to the stellar surface, also mean that the feature is not just sitting on top of the star and may extend over a hot and dense accretion column or part of the corotating inner disk, especially during outburst. Together with the location of the BC emitting material, this suggests that the region inwards of the corotation radius is more complex than anticipated. 
The small difference between maximum brightness and bluest colour also hints that the feature is not uniform in temperature although, as seen in Fig.  \ref{exlup-phase}, the colour variations are not very large due to the small temperature difference with the photosphere. 

\begin{figure}
    \centering
    \includegraphics[width=9cm]{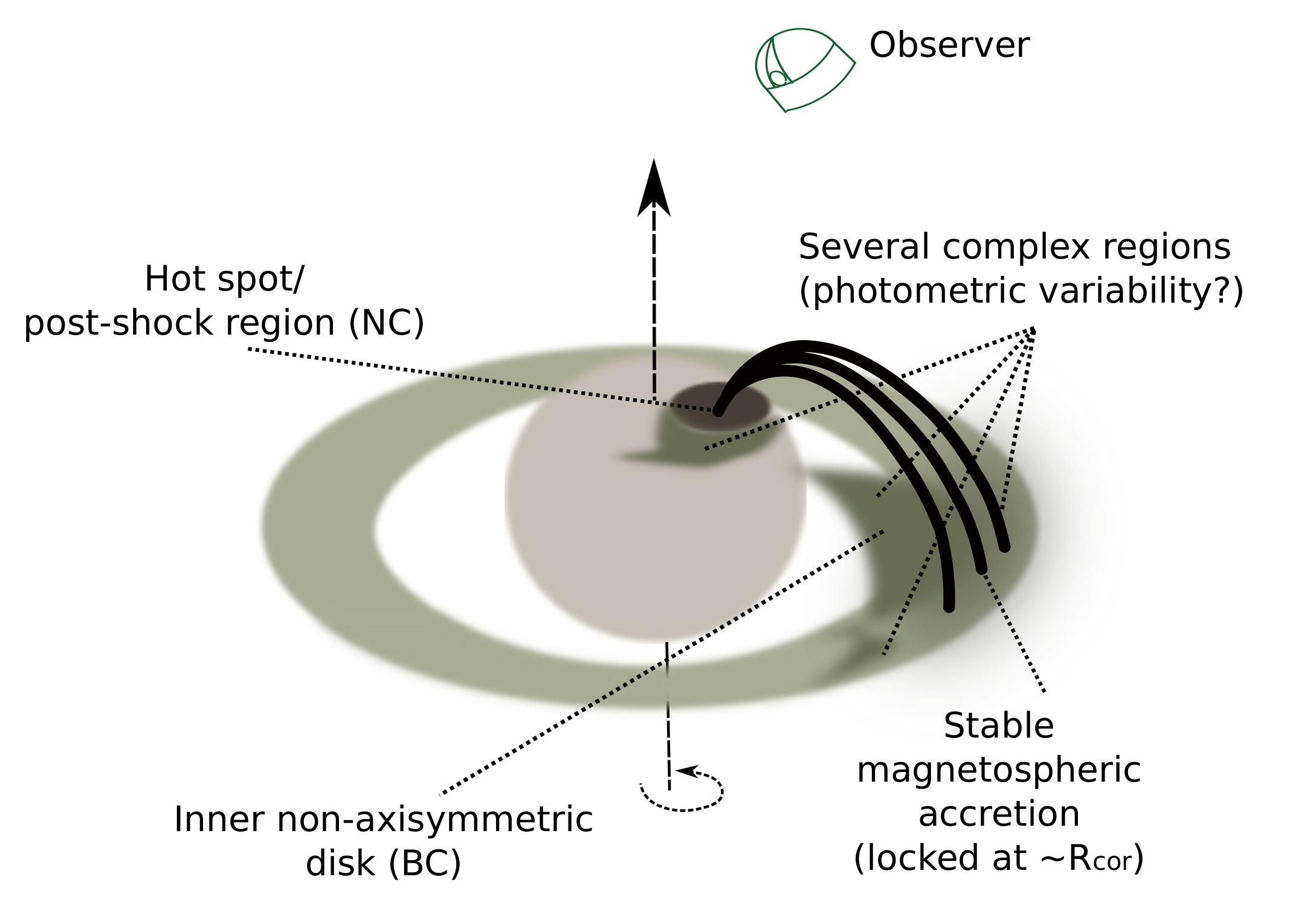}
    \caption{A sketch of the structures in the visible hemisphere of EX Lupi, which are probably similar, albeit less variable, for TW Hya. The NC of the lines is produced at the footprint of the accretion column or post-shock region, being very stable in time. The BC corresponds to non-axisymmetric material inwards of the corotation radius. The photometric variability is harder to interpret, corresponding to variable hot spots on the stellar surface and beyond, which do not necessarily overlap with the NC region and do not necessarily have the same latitude/inclination (to account for rapid visibility changes). Lower-scale effects from extinction and scattering by material in the inner disk or accretion columns may also need to be included.}
    \label{fig:sketch}
\end{figure}

The non-axisymmetric disk inside the corotation radius, as inferred from the BC analysis, is a location where periods can be shorter than the rotational one, and a slightly advanced photometry feature may extend in this direction, even if not as far as the BC emitting region, since the photometry period is still very close to the rotational one. For clarity and reference, we have sketched the different structures resulting from this analysis in Figure \ref{fig:sketch}.

\subsection{TW Hya: unstable columns for a quite stable accretion rate}

For TW Hya, the simple spot analysis can be done only during the epochs when the photometric modulation is sinusoidal. In other epochs, the data reveals a much more complex structure that is not suitable for the model. Figure \ref{twhya-phase} shows the phase-folded line radial velocities, magnitudes, and colours from the various datasets. Although we can rule out that the NC-emitting region is extended or has multiple components with similar strength, this is not true for the photometry. 
The photometric variability for TW Hya is not only messy period-wise, but it also varies in phase: assuming a 3.556~d period, the TESS photometry maxima for 2019 and 2021 display large period-to-period variabity and some $\sim$180$^{\circ}$ offsets.
The average NC phase for the maximum blueshift for TW Hya in strong lines, using JD=2458543.827644~d as the reference point, is around $\phi \sim$0.95.
The AAVSO photometry in the most stable epoch shows peak brightness at $\phi \sim$0.75, which is strongly offset compared with the NC velocity for a simple spot model. The bluest color is achieved at  $\phi \sim$0.8, so, despite the mismatch with the NC velocities, the photometry feature is consistent with a simple structure dominated by a single temperature. 
There is also evidence of variable extinction in TW Hya on certain epochs \citep{siwak14}, which could make the colour interpretation harder, although the effect does not seem very large in the data we analyse here. Finally, it is worth mentioning the cold spot structure inferred from radial velocity variations and magnetic mapping \citep{huelamo08,donati11}, which adds to the complexity of the accretion structures around TW Hya.

In any case, the analysis of TW Hya reveals that its accretion structures are shape-shifting and more variable than in the case of EX Lupi and on sub-period timescales, especially regarding those responsible for photometric changes. The structure for TW Hya at any given time may not be too different from what is shown in Fig. \ref{fig:sketch}, but the entire inner disk and accretion columns will be in this case by far more unstable. Since the photometric periods tend to be larger than those of the NC, the continuum features may have a tendency to extend beyond the corotation radius, rather than inwards.

\subsection{Understanding the inner regions around young stars}

Combining the results for TW Hya and EX Lupi, we conclude that both have remarkably stable accretion columns footprints, tracked via the NC emission lines, which remain locked to a particular location on the star for years.  
Although the temperature and density of the NC spot and, probably, its area, may change slightly (as suggested by the line strength), the location is essentially constant, with some variations in latitude, rather than longitude, evidenced by changes in the amplitude of the modulation. 

While the region where the NC lines are produced is very stable, the continuum-producing region, albeit consistent with a hot feature, is not stable and has a complex and variable temperature, size, and shape. 
In EX Lupi, the rotational modulation is evident from the photometry, revealing a stable structure overall from more extended regions down to the footprint.
In TW Hya, the lightcurve is very complex, while the NC line velocities are surprisingly clean for an object that is nearly pole-on. The hot continuum-emitting region is highly variable, but less sensitive to rotation,  explaining why efforts to determine any rotational periodic behaviour are often problematic. This may be in part caused by the low inclination, although the variability range in TW Hya is not so different from that of EX Lupi in quiescence.

The geometry and stability of accretion structures is linked to the properties of the magnetic field \citep{gregory06,jardine11, ireland22}.
Comparing our results with existing models, the observations are, for both objects, highly consistent with the situation expected for young stars with an inclined dipole magnetic field. The magnetic field of both objects is similar in strength: 3.5 kG for TW Hya \citep{lavail19} vs 3 kG for EX Lupi \citep{white20}, but there is a lack of contemporary information on the structure and the ratio of dipolar vs octupolar components, which, combined with the known variability of magnetic fields in young stars \citep{morin10,donati17} precludes detailed modeling.  What causes the difference in two stars so similar in mass remains thus unknown, but it may be related to the inner disk structure, and/or the magnetic field topology, which evolves with stellar age.

In general, the dipole field controls the location at which the disk is truncated, its tilt sets up the phase, and the structure of the accretion columns is regulated by the ratio between the octupolar and dipolar magnetic field components \citep{gregory06,gregory08}. The octupolar component can break the axisymmetry of the accretion column, so that a strong octupolar component results in a small, well-localised spot \citep{gregory11}, which is similar to what is observed here. The tilt of the magnetic field with respect to the rotation axis is well-known in the case of TW Hya \citep{donati11}, so we expect EX Lupi to be also tilted.

The photometric rotational period tends towards shorter values for EX Lupi and longer values for TW Hya, compared to the NC period. This suggests that at least part of the continuum feature(s) are located inwards vs outwards of the corotation radius, respectively. The BC in EX Lupi originates inwards of R$_{cor}$, a further confirmation of hot material present in that region. The accretion mode and whether the accreted disk material feeds from inside or outside of the corotation radius may affect accretion stability \citep{dangelo12,ireland22}, with accretion from inwards (outwards) of the corotation radius resulting in stable (unstable) accretion. For the case of EX Lupi, accreting from inwards of the corotation radius can explain the stability of the accretion column and footprint independently of the accretion rate, while the strong accretion bursts in timescales of years may result from accumulation of material in the inner disk that is not directly accreted due to this over-regulated accretion and, eventually, triggers an instability. Accumulation of material in the inner disk that is emptied after outburst has indeed been observed \citep{banzatti15}, and it is worth exploring in the future. Unstable accretion columns can explain the rapid, messy, photometric variations observed in TW Hya \citep[see also][for rapid changes in unstable accretion columns]{kurosawa_spectral_2013}. If the changes concern only the extended accretion structures, the column footprint on the stellar surface does not need to be affected, as we expect from the NC observations.

Further examining the photometric burster behaviour present in both objects, we find that the typical timescales for the irregular, minor bursts are shorter in EX Lupi than in TW Hya, despite TW Hya being a faster rotator. The timescales of these irregular bursts are a few  hours, which may suggest a connection to flares or to rapid changes in the magnetic field structure. In case of TW Hya, the variations may betray the weakness of the large-scale magnetic field and changes in the octupolar component, which has been observed to produce flare-like bursts \citep{donati11}. Measuring the magnetic fields at the same time as the NC and photometry signatures could thus help to disentangle the many substructures in accretion columns going beyond the scales of spatially-resolved data.

\section{Summary and conclusions \label{conclu}}
\label{conclu}

In this work, we examine how time-resolved spectroscopy and multi-band photometry unveil the structure of the accretion process in young stars, using EX Lupi and TW Hya as test benches. 

We find that the velocity of NC lines is a better tracer of rotation in young stars than the photometry. This happens because the NC-emitting region is very stable and small, even when the photometric variability is very complex. Individual line velocities are stable and rotationally modulated even when the accretion rate changes up to outburst levels, while photometric changes are much more complex and accretion rate variations change the size, location, and temperature of continuum spot(s). The NC and photometry spot are physically different, pointing out the diversity and complexity of accretion structures around young stars.

As it had been found for EX Lupi \citep{sicilia-aguilar_accretion_2015,campbellwhite21}, the 3.56d rotational period of TW Hya comes out clearly in many NC line velocities, and phase-folding reveals modulation in any NC line with enough S/N, over more than a decade. 
The NC line analysis for TW Hya confirms that the footprints of its accretion structures are non-axisymmetric, non-polar, and very stable.  Since TW Hya is viewed close to pole-on, the detection of the rotational modulation also demonstrates the capabilities of STAR-MELT to accurately measure very small radial velocities.

For both objects, the differences in the peak-to-peak velocity of the NC modulations are consistent with latitudinal distribution. The footprints of lines with different energies are different, indicating that the NC line spot, while covering a very small part of the stellar photosphere, has a complex structure with a range of temperatures and densities. Some lines such as Ca~II IR have low velocity amplitude (high latitude location). Comparing lines with different energies (e.g. Ca II IR lines vs He I or Fe II lines), we find the same location trends in EX Lupi and TW Hya, suggesting a common physical cause (such as stratified structures) behind this.

The hot spots that produce the photometric changes are in contrast highly variable and spatially independent of the NC-emitting regions. The photometry is in agreement with simple hot spots, but the  multi-band data reveals changes in their temperature, shape and size over time, which explain the lack of periodicity during certain epochs. In EX Lupi, we even witness how the size and temperature of the photometry/continuum spot change as the accretion rate decreases, while the NC velocity signatures remain stable. Photospheric spot sizes grow when the accretion rate increases, and the large filling factors that they achieve at high accretion rates suggest that they are not constrained to the stellar surface. This had been observed in EXor outbursts \citep[e.g.][]{juhasz12,sicilia-aguilar_2014-2017_2017}, but we now infer it at lower rates as well.

The hot spot(s) needed to explain the photometry are not much hotter than the stellar photosphere, but much colder than the temperatures needed to raise emission in energetic lines (e.g. He I, He II, Fe II). For the continuum spot, hotter temperatures may dominate at lower accretion. Since the structure producing the photometric modulation decreases in size when the accretion rate goes down, this may also reflect the relative contribution of other regions (e.g. NC line-emitting ones) that are very small but significantly hotter, or the continuum-emitting region moving closer to the stellar surface.

The analysis of the BC of the emission lines in EX Lupi reveals that its velocity modulations can be explained as emission from a non-axisymmetric disk in Keplerian rotation, winding up around the star inwards of the corotation radius. Outer parts of this asymmetric disk (with lower orbital velocities) become visible when the accretion rate increases, as it has been also observed in the intermediate-mass star ZCMa NW \citep{sicilia20zcma}. The shift towards lower velocities (larger Keplerian radii) during outburst is also in agreement with this picture.

The differences in the accretion column stability between TW Hya (where accretion is more dynamic, but the accretion rate is less variable) and EX Lupi (where the accretion rate is very variable, but the accretion structures are remarkably stable) bring up the question of whether their variability is related to properties of the star or to the inner disk. The magnetic field strengths for both objects are similar \citep[][but note that the topologies at the time of our investigation are unknown]{lavail19,white20}. The stability of accretion structures does not seem correlated with the accretion rate variability, although the differences between NC velocity and photometry periods suggest that EX Lupi's magnetospheric accretion may feed from inwards of the corotation radius, leading to very stable accretion \citep{ireland22} even if the accretion rate changes, while TW Hya would accrete from outward of the corotation radius, resulting in very dynamic extended accretion structures but with relatively stable footprints.  EX Lupi has thus more stable accretion structures than TW Hya, raising questions about the type of instability that triggers the accretion outbursts. It is worth to investigate in the future whether the over-stable accretion structures in EX Lupi are indirectly responsible for the outbursts, for instance, by leading to matter accumulation in the inner disk and, eventually, outbursts. This scenario would suggest the outbursts are rather triggered by the star, and not by the disk.

Since accretion footprints are so stable and can be well-tracked by the NC line velocity allows us to investigate underlying radial velocity variations, removing the accretion and spot signatures. For the two objects examined here, there is no evidence of residual radial velocity variations caused by companions, at least at the 0.1 km/s typical uncertainty level. Nevertheless, targeted high signal-to-noise time-resolved spectra could be used to explore the presence of companions in these and other objects with similarly stable accretion. We thus conclude that determining whether stable accretion is common among young stars is a key to identify young planets in the future.

\section*{Acknowledgements}

We thank the anonymous referee for their careful revision and suggestions, which contributed to clarify this paper.
ASA and JCW were supported by the STFC grant number ST/S000399/1 ("The Planet-Disk Connection: Accretion, Disk Structure, and Planet Formation").
JCW is funded by the European Union under the European Union’s Horizon Europe Research \& Innovation Programme 101039452 (WANDA). VR acknowledges the support of the Italian National Institute of Astrophysics (INAF) through the INAF GTO Grant  ERIS \& SHARK GTO data exploitation". AK and FCSM received funding from the European Research Council (ERC) under the European Union's Horizon 2020 research and innovation programme under grant agreement No 716155 (SACCRED).
This paper includes data collected by the TESS mission. Funding for the TESS mission is provided by the NASA's Science Mission Directorate.
This paper includes data collected with the TESS mission, obtained from the MAST data archive at the Space Telescope Science Institute (STScI). Funding for the TESS mission is provided by the NASA Explorer Program. STScI is operated by the Association of Universities for Research in Astronomy, Inc., under NASA contract NAS 5-26555. 
This work makes use of observations from the Las Cumbres Observatory global telescope network.
We acknowledge with thanks the variable star observations from the \textit{AAVSO International Database} contributed by observers worldwide and used in this research and, in particular, observers F.-J. Hambsch, A. Pearce, J. Pierce, L. Monard, M. Millward,  and K. Menzies. We are also indebted to J. Setiawan, who took many of the available archival spectra.
JD was supported by a Summer Undergraduate Bursary Fellowship by the Royal Astronomical Society in summer 2020. 
Based on data obtained from the ESO Science Archive Facility under requests by $asicilia$ and $justyncw$.
Based on observations obtained at the Canada-France-Hawaii Telescope (CFHT) which is operated by the National Research Council of Canada, the Institut National des Sciences de l´Univers of the Centre National de la Recherche Scientique of France, and the University of Hawaii. 
This work made use of Astropy:\footnote{http://www.astropy.org} a community-developed core Python package and an ecosystem of tools and resources for astronomy \citep{astropy_collaboration_astropy_2013,astropy_collaboration_astropy_2018}. 

%%%%%%%%%%%%%%%%%%%%%%%%%%%%%%%%%%%%%%%%%%%%%%%%%%
\section*{Data Availability}

All the spectral data used in this work is publicly available in the ESO Science Archive (for FEROS and HARPS data\footnote{http://archive.eso.org/scienceportal/home}) and the CFHT Archive (for ESPaDOnS data\footnote{https://www.cadc-ccda.hia-iha.nrc-cnrc.gc.ca/en/cfht/}). The LCOGT photometry data is already public and available from the archive \footnote{https://lco.global/documentation/data/archive/}, and the AAVSO photometry is available via their website\footnote{https://www.aavso.org}. The Python program STAR-MELT \citep{campbellwhite21} is also publicly available via GitHub\footnote{https://github.com/justyncw/STAR\_MELT}.



\bibliographystyle{mnras}
\bibliography{exluptwhya.bib} % if your bibtex file is called example.bib

%\begin{thebibliography}{99}
%\end{thebibliography}

%%%%%%%%%%%%%%%%%%%%%%%%%%%%%%%%%%%%%%%%%%%%%%%%%%

%%%%%%%%%%%%%%%%% APPENDICES %%%%%%%%%%%%%%%%%%%%%

\appendix

\section{Complete data tables}
This appendix contains the tables listing the data and the corresponding fits used in our work. Table \ref{tab:Spectra} contains a summary of the public observations used in the spectroscopy analysis. Table \ref{tab:LCOGTphoto} list the dates and magnitudes for the LCOGT observations. Finally, Table \ref{twhya-mjdfit} includes the result of fitting the different lines observed in TW Hya over different time intervals, which were used to trace the location of the accretion column footprint over time (see Section \ref{columnlocation}).

\begin{table}
    \centering
    \begin{tabular}{lccc}
    \hline
Object & MJD & Instrument & Wavelength range\\
     &  (d)   &  & (nm)     \\ 
\hline  
\hline
EX Lupi & 54309.11469827 & FEROS & 353 - 922\\ 
EX Lupi & 54310.15650255 & FEROS & 353 - 922\\ 
EX Lupi & 54311.18745653 & FEROS & 353 - 922\\ 
EX Lupi & 55053.17519784 & FEROS & 353 - 922\\ 
EX Lupi & 55057.99258892 & FEROS & 353 - 922\\ 
EX Lupi & 55309.39403788 & FEROS & 353 - 922\\ 
\hline 
TW Hya & 53460.22581782 & FEROS & 353 - 922\\ 
TW Hya & 53460.25440568 & FEROS & 353 - 922\\ 
TW Hya & 53461.26353177 & HARPS & 378 - 691\\ 
TW Hya & 53466.07513617 & HARPS & 378 - 691\\ 
TW Hya & 53495.09996845 & HARPS & 378 - 691\\ 
TW Hya & 53500.06376298 & HARPS & 378 - 691\\ 
\hline
     \end{tabular}
    \caption{Spectra used in this paper. All the data are available via the ESO and CFHT archives. The table displays a small subset of data, the complete table is available from CDS.}
    \label{tab:Spectra}
\end{table}

\begin{table}
    \centering
    \begin{tabular}{llc}
    \hline
Filter & MJD & Magnitude \\
     &  (d)   &  (mag)     \\ 
\hline  
\hline
g' & 57185.987284 & 0.056$\pm$0.045  \\
g' & 57185.987896 & 0.048$\pm$0.009  \\
g' & 57187.710939 & 0.049$\pm$0.052  \\
g' & 57188.037580 & 0.044$\pm$0.045  \\
\hline
r' & 57230.825144 & -0.094$\pm$0.017  \\
r' & 57230.825602 & -0.077$\pm$0.006  \\
r' & 57230.920987 & -0.096$\pm$0.019  \\
r' & 57230.921538 & -0.111$\pm$0.007  \\
\hline
i'& 57230.827079 & -0.098$\pm$0.015  \\
i'& 57230.827569 & -0.047$\pm$0.005  \\
i'& 57230.923069 & -0.095$\pm$0.016  \\
i'& 57230.923617 & -0.089$\pm$0.006  \\
\hline
u' & 57205.872875 & -0.173$\pm$0.088  \\
u' & 57207.189643 & -0.110$\pm$0.080  \\
u' & 57230.068783 & -0.061$\pm$0.103  \\
u' & 57231.365465 & 0.092$\pm$0.178  \\
\hline
     \end{tabular}
    \caption{LCOGT relative magnitudes and MJD. Only a small subset of datapoints is shown here. The complete table is available from CDS.}
    \label{tab:LCOGTphoto}
\end{table}

\begin{table*}
    \centering
    \begin{tabular}{lcccccc}
    \hline
Species &   $<$MJD$>$ & MJD$_{fin}-$MJD$_{ini}$ & Offset & Amplitude & Phase & \# \\
     &    (d)   & (d)   &  (km/s) & (km/s) & (deg) & Points  \\ 
\hline       
\hline
FeII4924 & 55259 & 55200 - 55356 & 1.35$\pm$0.07 & 1.01$\pm$0.10 & 49$\pm$5 & 50 \\
FeII4924 & 57440 & 57416 - 57449 & 1.40$\pm$0.14 & 0.73$\pm$0.19 & 85$\pm$15 & 41 \\
\hline
FeII5018 & 54198 & 54158 - 54310 & 2.51$\pm$0.09 & 0.87$\pm$0.13 & 91$\pm$8 & 59 \\
FeII5018 & 55259 & 55200 - 55356 & 1.71$\pm$0.07 & 0.97$\pm$0.10 & 54$\pm$6 & 60 \\
FeII5018 & 55965 & 55957 - 55972 & 1.89$\pm$0.07 & 0.44$\pm$0.10 & -113$\pm$12 & 54 \\
FeII5018 & 57443 & 57435 - 57448 & 1.66$\pm$0.13 & 0.92$\pm$0.17 & 95$\pm$14 & 10 \\
\hline
HeI4026 & 54195 & 54158 - 54233 & -7.65$\pm$0.11 & 0.47$\pm$0.15 & 38$\pm$19 & 49 \\
HeI4026 & 55260 & 55200 - 55356 & -7.70$\pm$0.11 & 0.58$\pm$0.15 & 69$\pm$15 & 46 \\
\hline
HeI4471 & 54550 & 54494 - 54657 & -1.14$\pm$0.08 & 0.73$\pm$0.11 & 42$\pm$9 & 62 \\
HeI4471 & 55259 & 55200 - 55356 & -1.00$\pm$0.07 & 0.69$\pm$0.10 & 38$\pm$8 & 60 \\
HeI4471 & 55965 & 55957 - 55972 & -0.77$\pm$0.11 & 0.46$\pm$0.17 & -77$\pm$16 & 54 \\
HeI4471 & 57441 & 57417 - 57448 & -1.19$\pm$0.08 & 0.39$\pm$0.13 & 65$\pm$16 & 39 \\
HeI4471 & 57909 & 57904 - 57915 & -0.23$\pm$0.10 & 0.43$\pm$0.14 & -54$\pm$18 & 20 \\
\hline
HeI4713 & 54197 & 54158 - 54310 & 5.06$\pm$0.10 & 0.55$\pm$0.14 & 65$\pm$14 & 61 \\
HeI4713 & 55259 & 55200 - 55356 & 3.80$\pm$0.09 & 0.88$\pm$0.12 & 40$\pm$8 & 60 \\
HeI4713 & 55965 & 55957 - 55972 & 4.58$\pm$0.12 & 0.41$\pm$0.19 & -67$\pm$22 & 52 \\
HeI4713 & 57440 & 57416 - 57449 & 3.73$\pm$0.12 & 0.55$\pm$0.17 & 19$\pm$17 & 45 \\
\hline
HeI4922 & 54197 & 54158 - 54310 & 5.08$\pm$0.04 & 0.16$\pm$0.06 & 115$\pm$21 & 65 \\
HeI4922 & 54551 & 54494 - 54657 & 4.95$\pm$0.04 & 0.24$\pm$0.05 & -64$\pm$13 & 67 \\
HeI4922 & 55259 & 55200 - 55356 & 4.72$\pm$0.04 & 0.23$\pm$0.06 & 74$\pm$14 & 59 \\
HeI4922 & 57441 & 57416 - 57449 & 4.72$\pm$0.05 & 0.40$\pm$0.07 & -11$\pm$10 & 59 \\
\hline
HeI5016 & 54197 & 54158 - 54310 & 1.89$\pm$0.04 & 0.29$\pm$0.05 & 74$\pm$10 & 64 \\
HeI5016 & 54551 & 54494 - 54657 & 1.79$\pm$0.03 & 0.26$\pm$0.04 & 93$\pm$8 & 67 \\
HeI5016 & 55259 & 55200 - 55356 & 2.00$\pm$0.03 & 0.58$\pm$0.04 & 44$\pm$4 & 60 \\
HeI5016 & 57441 & 57416 - 57449 & 1.51$\pm$0.03 & 0.48$\pm$0.04 & 42$\pm$5 & 60 \\
\hline
HeI5875 & 54551 & 54494 - 54657 & 7.06$\pm$0.05 & 0.42$\pm$0.07 & 29$\pm$10 & 67 \\
HeI5875 & 55259 & 55200 - 55356 & 7.42$\pm$0.05 & 0.51$\pm$0.07 & 49$\pm$8 & 60 \\
HeI5875 & 55965 & 55957 - 55972 & 7.35$\pm$0.04 & 0.25$\pm$0.06 & -812$\pm$11 & 56 \\
HeI5875 & 57441 & 57416 - 57449 & 6.65$\pm$0.05 & 0.60$\pm$0.06 & 73$\pm$6 & 60 \\
\hline
HeI6678 & 55259 & 55200 - 55356 & 6.70$\pm$0.06 & 0.73$\pm$0.09 & 45$\pm$7 & 60 \\
HeI6678 & 57441 & 57416 - 57449 & 7.14$\pm$0.07 & 1.02$\pm$0.10 & 62$\pm$5 & 60 \\
\hline
HeII4686 & 54548 & 54494 - 54657 & 5.97$\pm$0.12 & 1.18$\pm$0.15 & 57$\pm$8 & 50 \\
HeII4686 & 55253 & 55200 - 55354 & 7.74$\pm$0.18 & 0.64$\pm$0.24 & 28$\pm$21 & 22 \\
HeII4686 & 57910 & 57904 - 57915 & 5.08$\pm$0.18 & 0.70$\pm$0.22 & -17$\pm$25 & 18 \\
\hline
CaII3934 & 55257 & 55250 - 55263 & 1.15$\pm$0.04 & 0.31$\pm$0.05 & 47$\pm$10 & 51 \\
CaII3934 & 55965 & 55957 - 55972 & 1.03$\pm$0.03 & 0.15$\pm$0.04 & -163$\pm$20 & 56 \\
CaII3934 & 57441 & 57416 - 57449 & 0.86$\pm$0.05 & 0.31$\pm$0.07 & 83$\pm$13 & 54 \\
\hline
CaII8498 & 54197 & 54158 - 54310 & 0.38$\pm$0.02 & 0.19$\pm$0.03 & 108$\pm$9 & 64 \\
CaII8498 & 54551 & 54494 - 54657 & 0.27$\pm$0.02 & 0.17$\pm$0.03 & -62$\pm$9 & 67 \\
CaII8498 & 55259 & 55200 - 55356 & 0.20$\pm$0.03 & 0.17$\pm$0.04 & 109$\pm$14 & 60 \\
CaII8498 & 55965 & 55957 - 55972 & 0.13$\pm$0.01 & 0.09$\pm$0.02 & 15$\pm$12 & 56 \\
CaII8498 & 56705 & 56636 - 56808 & 0.12$\pm$0.03 & 0.25$\pm$0.04 & -36$\pm$10 & 70 \\
CaII8498 & 57441 & 57416 - 57449 & 0.17$\pm$0.02 & 0.13$\pm$0.03 & -12$\pm$11 & 60 \\
\hline
CaII8542 & 54548 & 54540 - 54553 & 1.21$\pm$0.02 & 0.08$\pm$0.03 & 205$\pm$20 & 52 \\
CaII8542 & 55257 & 55250 - 55263 & 1.13$\pm$0.01 & 0.17$\pm$0.02 & 77$\pm$6 & 52 \\
CaII8542 & 55965 & 55957 - 55972 & 1.30$\pm$0.02 & 0.12$\pm$0.02 & 178$\pm$12 & 56 \\
CaII8542 & 56702 & 56693 - 56710 & 1.20$\pm$0.02 & 0.10$\pm$0.03 & 222$\pm$14 & 64 \\
\hline
CaII8662 & 54196 & 54158 - 54310 & 1.69$\pm$0.03 & 0.18$\pm$0.04 & 62$\pm$12 & 63 \\
CaII8662 & 54551 & 54494 - 54657 & 1.63$\pm$0.02 & 0.12$\pm$0.03 & -72$\pm$13 & 66 \\
CaII8662 & 55259 & 55200 - 55356 & 1.57$\pm$0.02 & 0.17$\pm$0.03 & 65$\pm$11 & 60 \\
CaII8662 & 57441 & 57416 - 57449 & 1.56$\pm$0.02 & 0.13$\pm$0.02 & 101$\pm$11 & 60 \\
\hline
\end{tabular}
\caption{Results of fitting the modulation over several time intervals for TW Hya, using the period 3.5558d measured for the
HeI 5016\AA\ line. Only intervals and lines with a reasonable number of low-noise datapoints are included. The phase is given using as zero-point the arbitrary date MJD=58543.327644~d. }
\label{twhya-mjdfit}
\end{table*}

\section{Supplementary figures for the spot models}

\begin{figure*}
\begin{center}
{\bf \large{EX Lupi}}\\
\begin{tabular}{cccc}
{\bf Quiescence (hot)} & {\bf Quiescence (cold)} & {\bf Outburst, JD 2459661} & {\bf Outburst, JD 2459675}\\ 
\includegraphics[width=4.3cm]{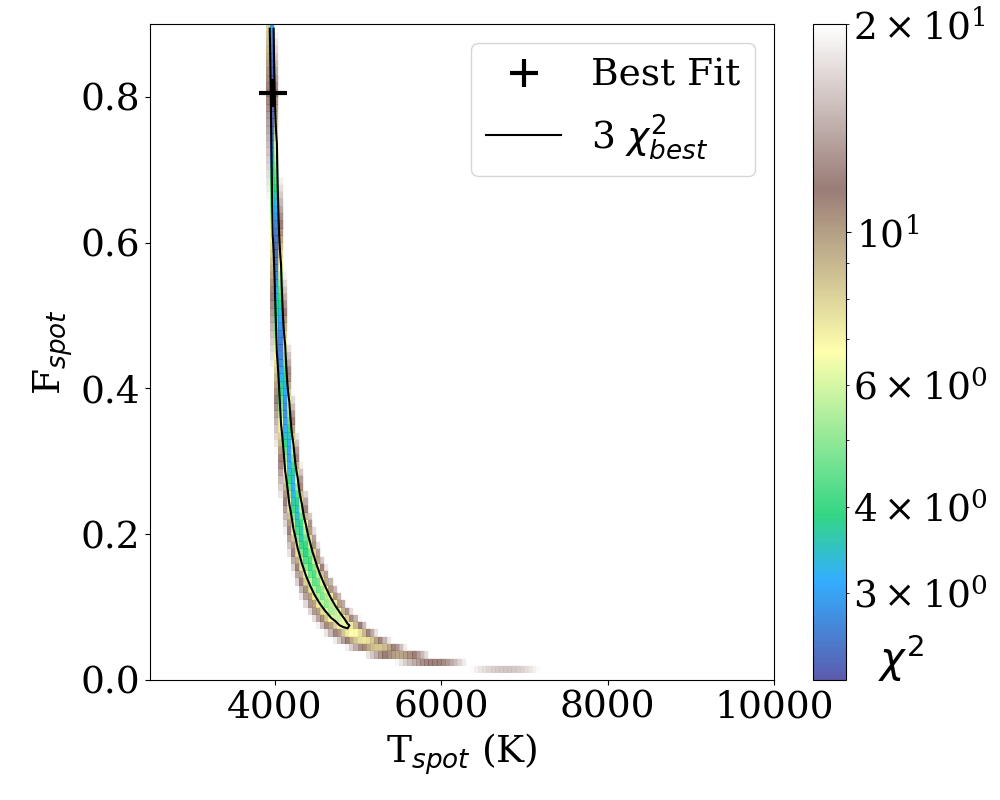} & 
\includegraphics[width=4.3cm]{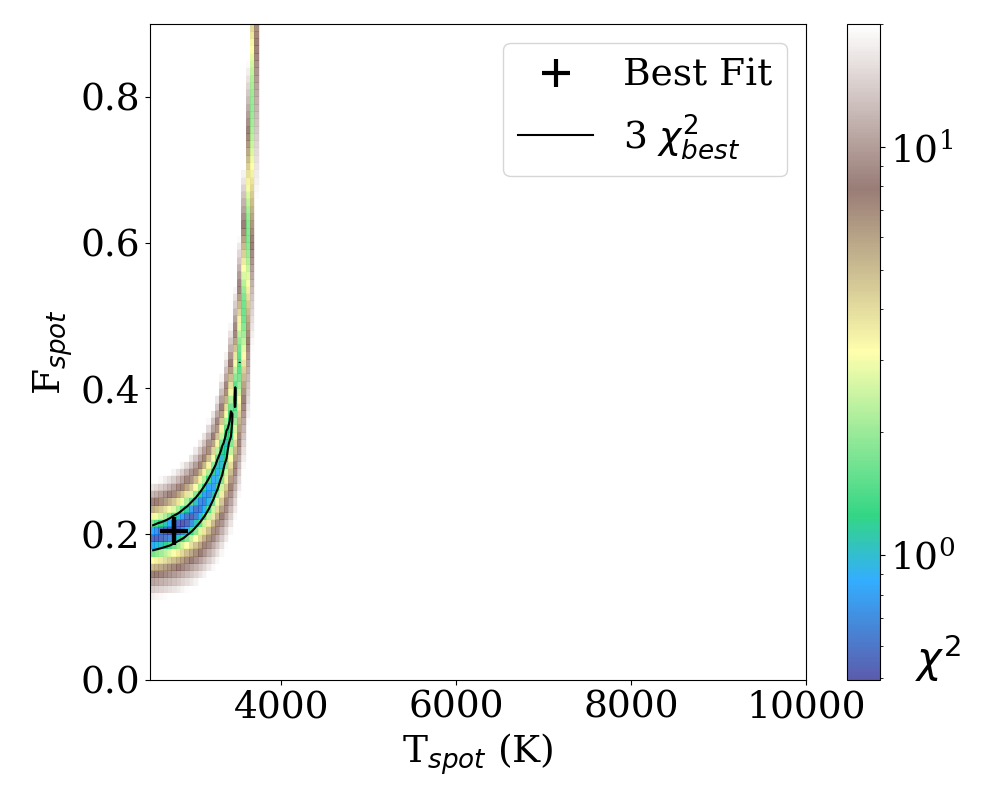} &
\includegraphics[width=4.3cm]{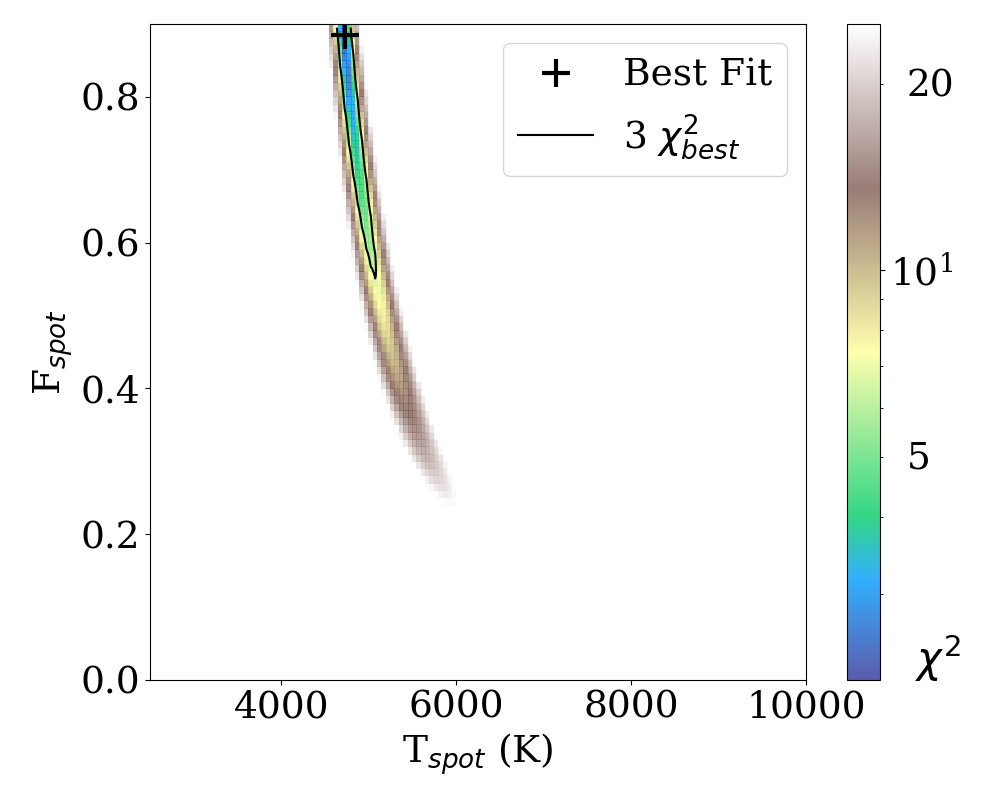} &
\includegraphics[width=4.3cm]{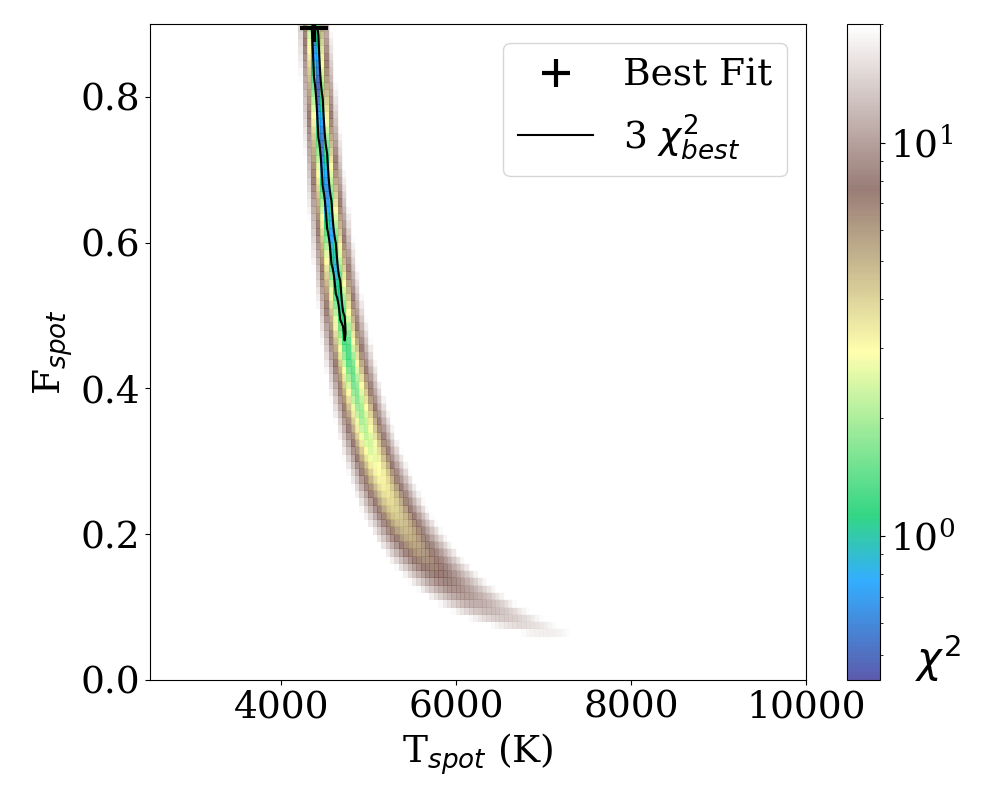}\\
\end{tabular}
\begin{tabular}{ccc}
{\bf Outburst, JD 2459688} &  {\bf Outburst, JD 24597041} & {\bf Outburst, JD 2459716} \\ 
\includegraphics[width=4.3cm]{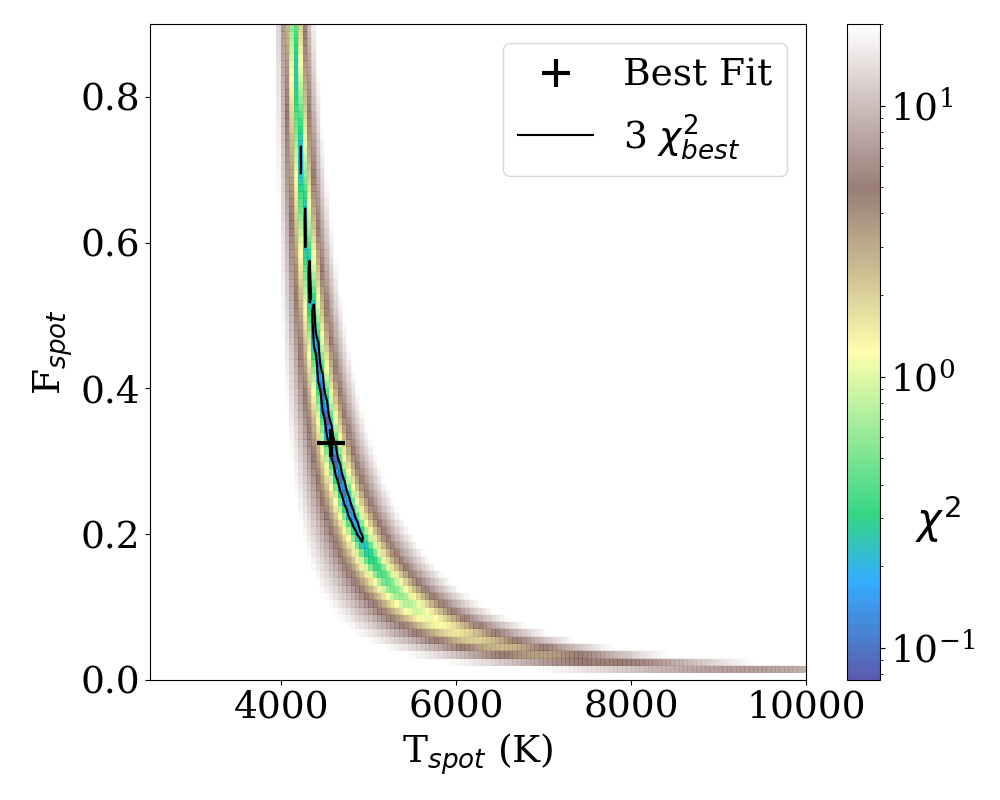} &
\includegraphics[width=4.3cm]{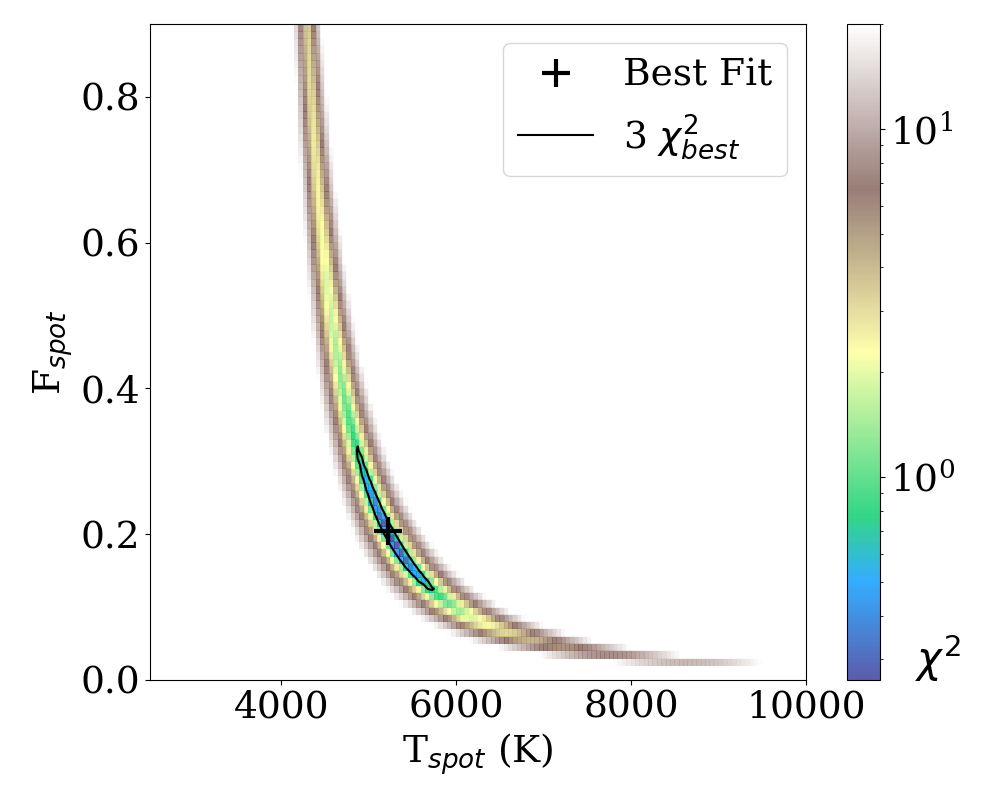} &
\includegraphics[width=4.3cm]{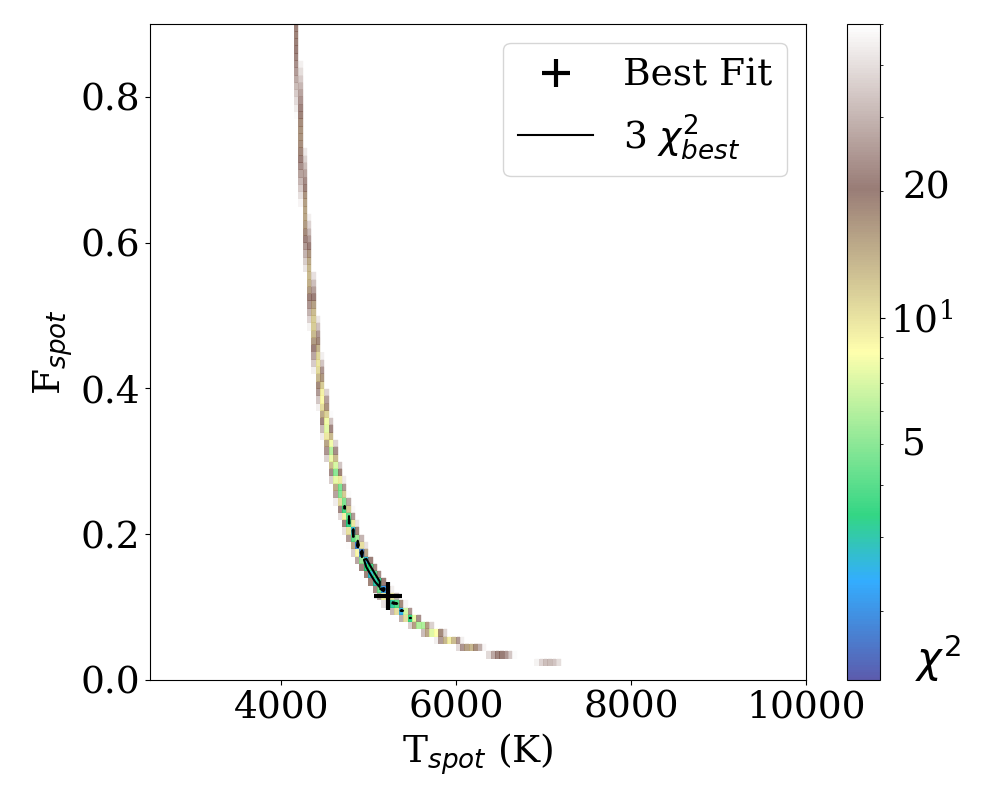} \\
\end{tabular}
\vskip 0.5truecm
{\bf \large{TW Hya} } \\
\begin{tabular}{ccc}
{\bf Full Dataset} &  {\bf  JD 2459657} &  {\bf  JD 2459674} \\
\includegraphics[width=4.3cm]{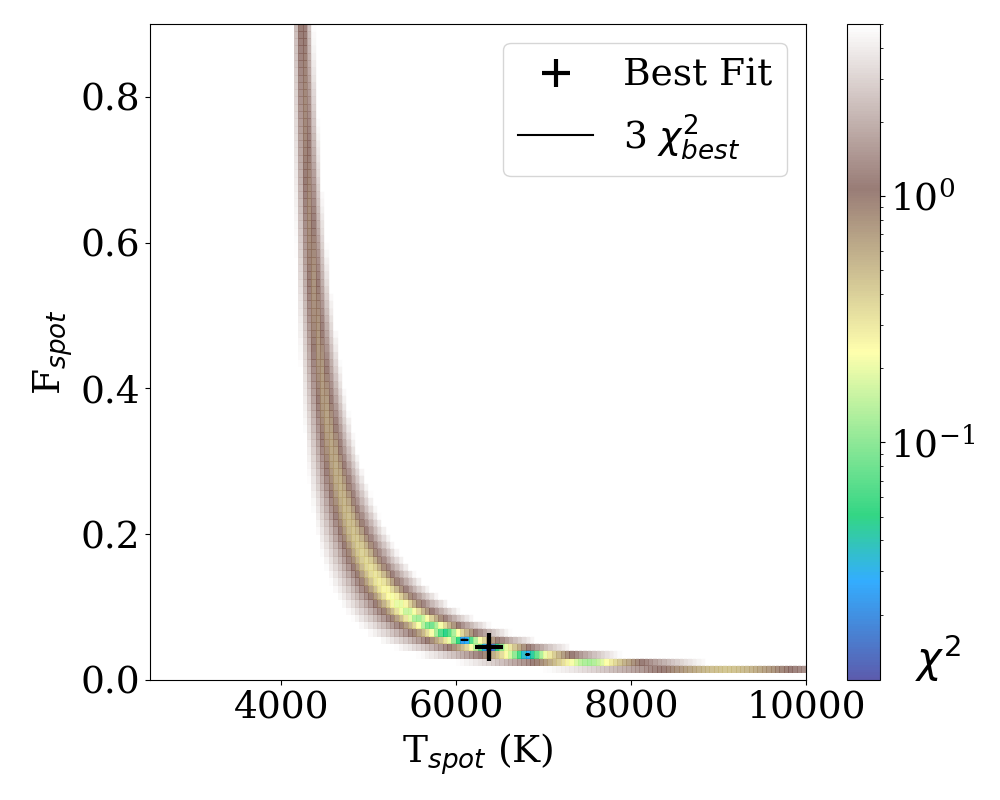} & 
\includegraphics[width=4.3cm]{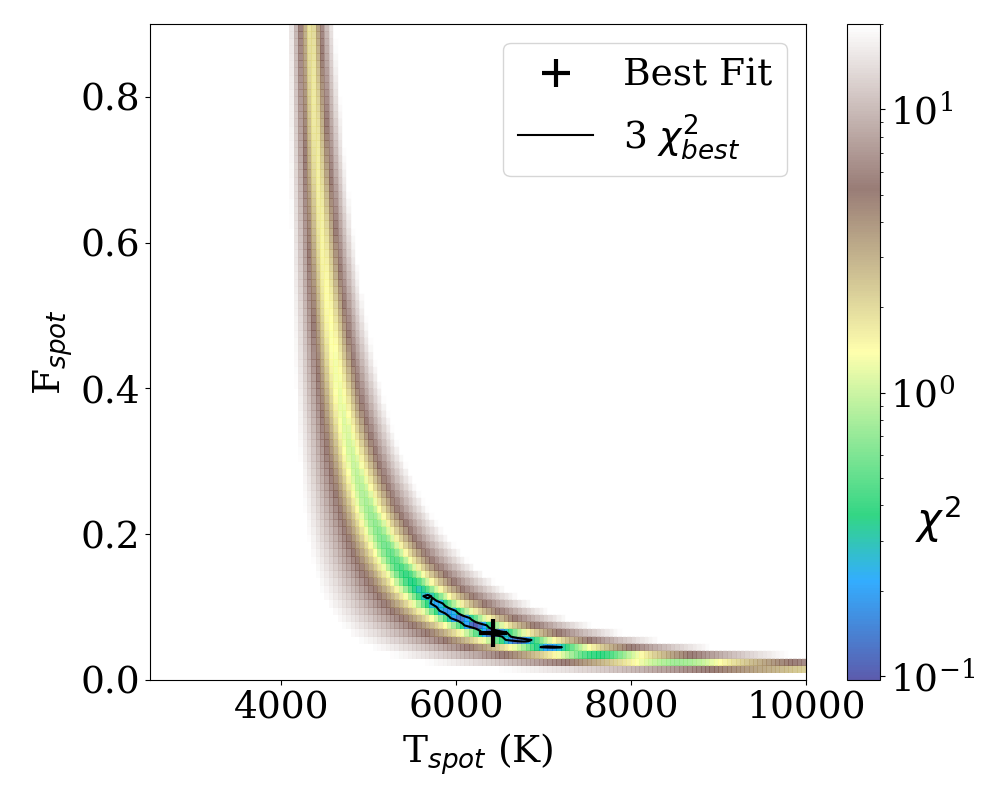} &
\includegraphics[width=4.3cm]{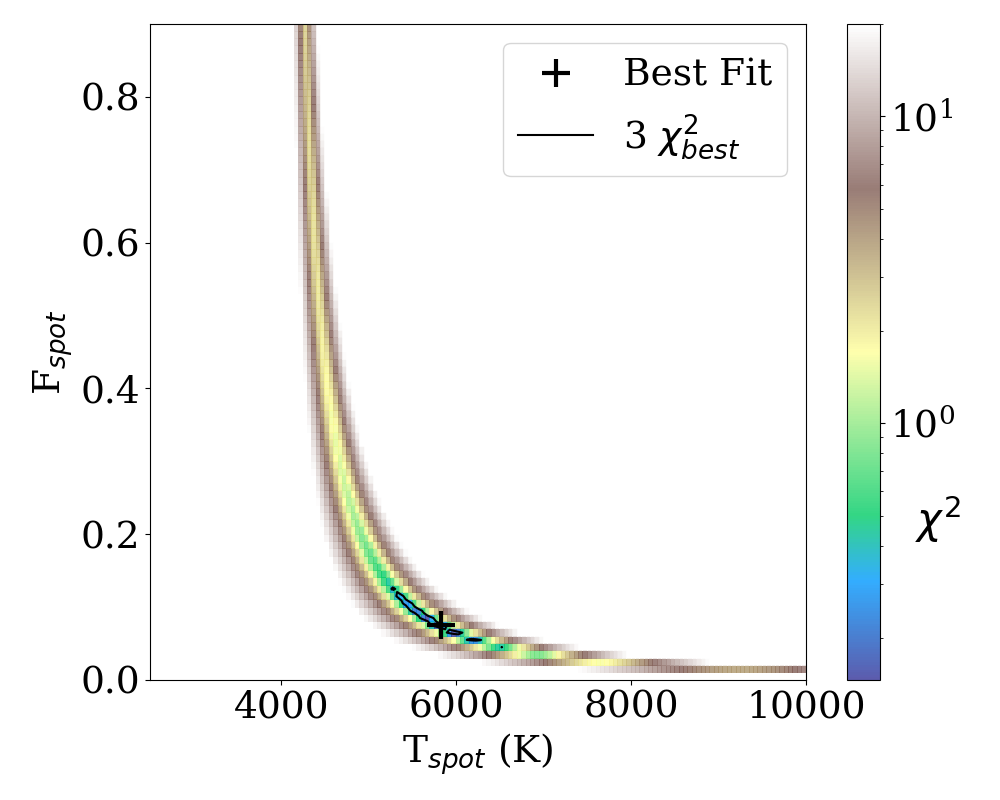}\\
\end{tabular}
\end{center}
\caption{Parameter space and $\chi^2$ for the simple spot models for EX Lupi (top and middle rows) and TW Hya (bottom). The individual plots are labeled accordingly, and they represent the different epochs (average JD of observations) following the description in Table \ref{exlupitwhya-simplespot}. The large cross marks the best-fit model, and the black contours surround the parameter space for values with $\chi^2$ up to 3 times the minimum (best-fit) value. Note that the high $\chi^2$ values appear in white to highlight the best-fit regions. For EX Lupi, we show the cold spot model for comparison in quiescence (which does not provide a good result) as well as hot spot models for quiescence and the various outburst epochs. For TW Hya, only the hot spot models are shown here, since the cold spot attempts do not provide good fits for any parameter combination (see text).}
\label{spot-params} 
\end{figure*}

In this appendix, we display the parameter space covered in the simple spot models, to highlight the best-fitting models and the area subtended by low values of the $\chi^2$. Figure \ref{spot-params} shows the results. As discussed in the text, although there is some degree of degeneracy, the best models for all hot spot cases are very well-defined. For EX Lupi, results such as the decrease in size of the hot spot as the outburst fades are extremely robust. The change of temperature, albeit somewhat less constrained, also agrees with the increase in temperature towards the end of the outburst.

\end{document}